\documentclass[10pt, nolongbibliography, groupedaddress, showpacs, showkeys, amsmath, amssymb, eqsecnum, aps, prd, nofootinbib]{revtex4-2}

\usepackage{lmodern}

\usepackage{amsmath}
\usepackage{tensor, mathtools, bm} 

\usepackage[]{graphicx}

\usepackage{epsfig}                                                 
\usepackage{tabularray}
\UseTblrLibrary{booktabs}

\usepackage[mathscr]{euscript}
\usepackage{hyperref, bookmark}

\begin{document}

\author{F. T. Brandt}  
\email{fbrandt@usp.br}
\affiliation{Instituto de F\'{\i}sica, Universidade de S\~ao Paulo, S\~ao Paulo, SP 05508-090, Brazil}

\author{J. Frenkel}
\email{jfrenkel@if.usp.br}
\affiliation{Instituto de F\'{\i}sica, Universidade de S\~ao Paulo, S\~ao Paulo, SP 05508-090, Brazil}

\author{S. Martins-Filho}    
\email{sergiomartinsfilho@usp.br}
\affiliation{Instituto de F\'{\i}sica, Universidade de S\~ao Paulo, S\~ao Paulo, SP 05508-090, Brazil}

\author{D. G. C. McKeon}
\email{dgmckeo2@uwo.ca}
\affiliation{
Department of Applied Mathematics, The University of Western Ontario, London, Ontario N6A 5B7, Canada}
\affiliation{Department of Mathematics and Computer Science, Algoma University, Sault Ste.~Marie, Ontario P6A 2G4, Canada}

\title{Renormalization of the Einstein-Cartan Theory in First-Order Form}

\date{\today}

\begin{abstract}
We examine the Einstein-Cartan (EC) theory in first-order form, which has a diffeomorphism as well as a local Lorentz invariance. We study the renormalizability of this theory in the framework of the Batalin-Vilkovisky formalism, which allows for a gauge invariant renormalization. Using the background field method, we discuss the gauge invariance of the background effective action and analyze the Ward identities which reflect the symmetries of the EC theory. As an application, we compute,  in a general background gauge, the self-energy of the tetrad field at one-loop order. 
\end{abstract}

\keywords{Einstein-Cartan theory, first-order form, renormalization}

\maketitle
\allowdisplaybreaks

\section{Introduction}\label{section:intro}

The EC theory is an extension of general relativity that describes a spacetime with curvature and torsion, arising from the spin of elementary particles. Due to this feature, the EC theory may solve the Big Bang singularity problem that occurs in cosmological models based on general relativity (GR) \cite{Poplawski:2011jz, Huang:2024ujj}. In GR, the fundamental entity is the metric $ g_{\mu \nu} (x)$ \cite{Buchbinder:2021wzv}. The EC theory differs from GR by being formulated in a Riemann-Cartan geometry that encodes, in addition to diffeomorphism, local Lorentz symmetry. It may be regarded as a gauge theory consisting of local translations and local Lorentz transformations \cite{Hehl:1976kj, Hehl:2023khc, Shapiro:2001rz}. The first-order form of the EC theory involves a tetrad field $ \tensor{e}{^a_{\mu}} (x) $ as well as an independent spin-connection field $ \omega_{\mu ab} (x)$. These are gauge fields that implement the diffeomorphism and the local Lorentz invariance of this theory. In a previous paper, the quantization of the EC theory in first-order form has been systematically discussed \cite{Brandt:2024rsy}. It was shown that the gauge algebra for both these gauge transformations is closed, and the BRST invariance of this theory was established.

The purpose of this paper is to show that this theory is renormalizable in the sense that the ultraviolet divergences are controlled by the gauge symmetries, so that there are counterterms available to cancel all such divergences. To this end, we will employ the Batalin-Vilkovisky (BV) formalism \cite{Batalin:1981jr, Batalin:1983ggl}, which is useful for maintaining gauge invariance during the renormalization process. We shall also use the background field method \cite{tHooft:1974toh, Abbott:1980hw, abbott:1982, Frenkel:2018xup, Barvinsky:2017zlx, Brandt:2022und}, which is convenient for preserving the gauge symmetries of the divergent part of the effective action of the EC theory in first-order form.

The paper is organized as follows. In Sec. II, we reproduce some properties of the EC theory in first-order form together with the background field method. In Sec. III, we formulate this theory in the framework of the BV formalism. In Sec. IV, we apply this approach to derive the Zinn-Justin equation \cite{Zinn-Justin:1984tfs} for the effective action and to provide a recursive proof of the renormalizability of this theory. Using this approach, all subdivergences at any order of the perturbative expansion are eliminated through the renormalization procedure. In Sec. V, we examine the gauge invariance of the background effective action, which is applied to study the ultraviolet divergences of the tetrad and spin connection self-energies at one-loop order. A brief discussion of the results is presented in Sec. VI. Some relevant details of the calculations are given in the appendices.

\section{EC theory in the background field gauge}\label{section:ECtheory}

In this section, we will review some results obtained in the first order form of the EC theory \cite{Brandt:2024rsy}, but will omit, for simplicity, the matter fields. The pure EC action involves the tetrad field $ \tensor{e}{^a_{\mu}} $ that is related to the metric  as 
\begin{equation}\label{eq:ccf21}
    g_{\mu\nu}(x) = \eta_{ab} \tensor{e}{^a_\mu}(x) \tensor{e}{^b_\nu}(x);\quad (\eta_{ab}=\text{diag}(+---))
\end{equation}
as well as the spin-connection field $ \omega_{\mu ab} $. It may be written in the form 
\begin{equation}\label{eq:ccf22}
    S_{\text{EC}}= -\frac{1}{ \kappa^{2} }\int \mathop{d^4x} e R(e, \omega),
\end{equation}
where $ \kappa^{2} = 16 \pi G$ ($G$ is Newton's constant), $e \equiv  \det \tensor{e}{^a_\mu} = \sqrt{-\det g_{\mu\nu}}$
and
\begin{equation}\label{e7}
R(e , \omega) = e^{a\mu} e^{b\nu} R_{\mu\nu ab}(\omega)
\end{equation}
with
\begin{equation}\label{e8o}
R_{\mu\nu ab} ( \omega)=  \omega_{\nu ab, \mu } -  \omega_{\mu ab, \nu } 
+ \omega_{\mu ap} \omega_{\nu\;b}^{\; p\;} -\omega_{\nu ap} \omega_{\mu\;b}^{\; p\;} .
\end{equation}

In the above equations, $ \tensor{e}{^a_{\mu}} $ and $ \omega_{\mu ab} $ are treated as independent fields. One can show that the solution to the classical equation of motion for $ \omega_{\mu ab}$ that follows from Eq.~\eqref{eq:ccf22} results in \cite{Kiriushcheva:2009tg}
\begin{equation}\label{e3}
    \omega_{\mu ab} = \frac{1}{2} \tensor{e}{_a^\nu}\left( e_{b\nu, \mu } -  e_{b\mu, \nu}\right) + 
    \frac{1}{4} \tensor{e}{_a^\sigma}  \tensor{e}{_b^\lambda} \left( e_{c\sigma , \lambda } -  e_{c\lambda , \sigma }\right) \tensor{e}{^c_\mu}
- (a\leftrightarrow b)
+ K_{\mu ab},
\end{equation}
where $ K_{\mu ab} $ is the contorsion tensor. 
Because the torsion does not affect the renormalizability of the EC theory, it will be disregarded, for simplicity, in the following discussion.

We assume the  metricity condition $ g_{\mu \nu ; \lambda} = 0$, which requires the tetrad to satisfy the compatibility condition 
\begin{equation}\label{eq:ccf26a}
\tensor{e}{^{a}_{\mu ; \lambda}} = \tensor{e}{^{a}_{\mu ; \lambda}} + \tensor{\omega}{_{\lambda}^{a}_{b}} \tensor{e}{^{b}_{\mu}} = \tensor{e}{^{a}_{\mu , \lambda}} - \tensor{\Gamma}{_{\lambda \mu}^{\sigma}} \tensor{e}{^{a}_{\sigma} } + \tensor{\omega}{_{\lambda}^{a}_{b}} \tensor{e}{^{b}_{\mu}}
 =0
\end{equation}
where comma denotes partial derivatives, semicolon denotes the usual covariant derivatives and bar denotes the total covariant derivative. Eq.~\eqref{e3} can also be derived from Eq.~\eqref{eq:ccf26a}.  

The action \eqref{eq:ccf22} is invariant under the  diffeomorphism and local Lorentz transformations
\begin{subequations}\label{eq:ccf26}
    \begin{align}
        \delta \tensor{e}{^a_{\mu}} ={}& - \kappa \epsilon^{\alpha} \tensor{e}{^{a}_{\mu , \alpha}} - \kappa \tensor{e}{^{a}_{\alpha}} \epsilon^{\alpha}_{, \mu} + \tensor{\lambda}{^{a}_{b}}\tensor{e}{^{b}_{\mu}} \intertext{and}
        \delta \omega_{\mu ab} ={}& - \kappa \epsilon^{\alpha} \omega_{\mu ab , \alpha} - \kappa \omega_{\alpha ab} \epsilon^{\alpha}_{, \mu} - \lambda_{ab , \mu } +  \tensor{\lambda}{_{a}^{p}} \omega_{\mu pb} +  \tensor{\lambda}{_{b}^{p}} \omega_{\mu ap},
    \end{align}
\end{subequations}
where $ \epsilon^{\mu} (x) $ and $ \lambda_{ab} (x) $ are  arbitrary (infinitesimal)  functions .

In the background field method\footnote{Note that, the background field method requires an invertible background metric ($ \det \bar{g}_{\mu \nu} \neq 0$) \cite{Brandt:2024rsy}. However, it is interesting to mention that the first order formalism contains another phase with a non-invertible metric, which was first studied in Ref. \cite{Tseytlin:1981ks}.}, the tetrad and spin-connection fields are replaced by the sums
\begin{equation}\label{eq:bm1}
    \tensor{e}{^a_{\mu}} = \tensor{\bar{e}}{^{a}_{\mu}} + \kappa \tensor{q}{^{a}_{\mu}} =\delta_{\mu}^{a} + \kappa \tensor{\bar{q}}{^{a}_{\mu}} + \kappa \tensor{q}{^{a}_{\mu}}
    \quad \text{and} \quad  \omega_{\mu ab} = \bar{\omega}_{\mu ab} + \kappa Q_{\mu ab},
\end{equation}
where $ \tensor{\bar{q}}{^{a}_{\mu}} $ ($ \tensor{\bar{e}}{^{a}_{ \mu }}$) and $ \bar{\omega}_{\mu ab} $ are the background fields, while $\tensor{q}{^{a}_{\mu}} $ and $ Q_{\mu ab} $ denote the quantum fields.           

We will next  choose gauge-fixing terms that respect the symmetries of Eqs.~\eqref{eq:ccf26} for the background fields $ \tensor{\bar{e}}{^{a}_{\mu}} $ and $ \bar{\omega}_{\mu ab} $, but break the symmetries
\begin{subequations}\label{eq:ccf28}
    \begin{align}\label{eq:ccf28a}
        \delta \tensor{\bar{e}}{^{a}_{\mu}} ={}&0, \\ \label{eq:ccf28b}
        \delta \tensor{q}{^{a}_{\mu}} ={}& - \epsilon^{\alpha} \tensor{e}{^{a}_{\mu , \alpha}} - \tensor{e}{^{a}_{\alpha}} \epsilon^{\alpha}_{, \mu} + \frac{1}{\kappa} \tensor{\lambda}{^{a}_{b}} \tensor{e}{^{b}_{\mu}} 
    \end{align}
\end{subequations}
and
\begin{subequations}\label{eq:ccf29}
    \begin{align}\label{eq:ccf29a}
        \delta \bar{\omega}_{\mu ab} ={}&0, \\\label{eq:ccf29b}
        \delta Q_{\mu ab} ={}& - \epsilon^{\alpha} \omega_{\mu ab , \alpha} - \omega_{\alpha ab} \epsilon^{\alpha}_{, \mu} - \frac{1}{\kappa} \lambda_{ab , \mu } + \frac{1}{\kappa} \tensor{\lambda}{_{a}^{p}} \omega_{\mu pb} + \frac{1}{\kappa} \tensor{\lambda}{_{b}^{p}} \omega_{\mu ap}.
    \end{align}
\end{subequations}

To fix the gauge transformations of Eqs.~\eqref{eq:ccf28} and \eqref{eq:ccf29}, we choose the gauge-fixing Lagrangian
\begin{equation}\label{eq:ccf210}
    \mathcal{L}_{\text{gf}} =   
    \frac{\xi}{2} b^{\nu} b_{\nu} - b^{\nu}  \eta_{ab} [ (\tensor{\bar{e}}{^{a}_{\mu}} \tensor{q}{^{b}_{\nu}} + \tensor{q}{^{a}_{\mu}} \tensor{\bar{e}}{^{b}_{\nu}})^{; \bar{\mu}} -2 \sigma (\bar{e}^{b \alpha } \tensor{q}{^{a}_{\alpha}} )_{; \bar{\nu} }] + \frac{\zeta}{2} B^{ab} B_{ab} - B^{ab} \tensor{Q}{_{\mu ab}^{; \bar{\mu}}}  ,
\end{equation}
where $ b^{\nu} $ and $ B^{ab} $ are the Nakanishi-Lautrup auxiliary fields \cite{Nakanishi:1966zz, lautrup:1967}.  In the above equation, $ \sigma $ denotes a general parameter such that for $ \sigma = 1/2$ we recover the linear part in the quantum fields $ \tensor{q}{^{a}_{\mu}} $ of the de Donder gauge \cite{harrison2013einstein} (gauge fixings involving quadratic quantum fields may hinder the renormalizability of the theory \cite{Das:1980qe, Das:1982rz}). With these choices, we do not break the diffeomorphism invariance in the background fields provided the indices are raised and lowered using  $ \bar{g}_{\mu \nu} = \eta_{ab}  \tensor{\bar{e}}{^{a}_{\mu}} \tensor{\bar{e}}{^{b}_{\nu}} $.

After some calculations, the corresponding ghost Lagrangian is found to be \cite{Brandt:2024rsy}
\begin{equation}\label{eq:ccf211}
    \begin{split}
    \mathcal{L}_{\text{gh}} ={}&
    c^{\star \nu}    \eta_{ab}   \big \{ \left[ \tensor{\bar{e}}{^{a}_{\mu}} \left ( - c^{\alpha} \tensor{e}{^{b}_{\nu ; \bar{\alpha}}} - \tensor{e}{^{b}_{\alpha}} c_{\  ; \bar{\nu}}^{\alpha} \right ) + \tensor{\bar{e}}{^{b}_{\nu}} \left( - c^{\alpha} \tensor{e}{^{a}_{\mu ; \bar{\alpha}}} - \tensor{e}{^{a}_{\alpha}} c_{\  ; \bar{\mu}}^{\alpha}    \right)\right]^{; \bar{\mu}} 
                        - 2 \sigma   [ \tensor{\bar{e}}{^{a \mu}}  ( - c^{\alpha} \tensor{e}{^{b}_{\mu ; \bar{\alpha } } } - \tensor{e}{^{b}_{\alpha}} c^{\alpha}_{; \bar{\mu} } ) ]_{; \bar{\nu}}  \big \} 
                            \\ & 
                            + \frac{1}{\kappa} c^{\star \nu}  \eta_{ab}\big \{  \left [ \tensor{\bar{e}}{^{a}_{\mu}} \left ( \tensor{C}{^{b}_{p}}\tensor{e}{^{p}_{\nu}} \right )+ \tensor{\bar{e}}{^{b}_{\nu}} \left ( \tensor{C}{^{a}_{p}}\tensor{e}{^{p}_{\mu}}\right )\right ]^{; \bar{\mu}} - 2 \sigma \left [\tensor{\bar{e}}{^{a \mu}} \left ( \tensor{C}{^{b}_{p}}\tensor{e}{^{p}_{ \mu }}\right)  \right ]_{; \bar{\nu}}  
                             \big\} 
                             + \kappa  C^{\star ab}  \left( - c^{\alpha} \omega_{\mu ab ; \bar{\alpha}} - \omega_{\alpha ab} c^{\alpha}_{\  ; \bar{\mu}} \right)^{; \bar{\mu}} 
                               \\ & +
                               C^{\star ab} \left [ -  C_{ab ; \bar{\mu}} +  \tensor{C}{_{a}^{p}} \omega_{\mu pb} +   \tensor{C}{_{b}^{p}} \omega_{\mu ap}\right ]^{; \bar{\mu} } .
\end{split}
\end{equation}
We note  that the two pairs of ghosts, $ ( c_{\mu} , {c}^{\star}_{\mu}) $ and $( C_{ab} , {C}^{\star}_{ab} )$, associated respectively with the diffeomorphisms and local Lorentz transformations, generally couple among themselves. However, for the terms quadratic in the ghost fields, the two types of ghosts decouple. Moreover, as shown in the appendices, there occurs a significant simplification of the calculations for $ \sigma = 1/2 $.

Setting $ \epsilon_{\mu} = c_{\mu} \eta $ and $ \lambda_{ab} = C_{ab} \eta$, where $ \eta $ is an infinitesimal Grassmann quantity with ghost number $-1$, one may verify that the Faddeev-Popov Lagrangian 
\begin{equation}\label{eq:ccf212}
    \mathcal{L}_{\text{FP}} = \mathcal{L}_{\text{EC}} + \bar{e} \mathcal{L}_{\text{gf}} + \bar{e} \mathcal{L}_{\text{gh}} 
\end{equation}
is invariant under the BRST transformations given in Eqs.~\eqref{eq:ccf28} and \eqref{eq:ccf29} together with 
\begin{equation}\label{eq:ccf213}
    \delta b_{\mu} = \delta B_{ab} =0,
\end{equation}
\begin{equation}\label{eq:ccf214}
    \delta {c}^{\star}_{\mu} = - b_{\mu} \eta, \quad \delta {C}^{\star}_{ab} = - \frac{1}{\kappa} B_{ab} \eta;
\end{equation}
and 
\begin{equation}\label{eq:ccf215}
    \delta c_{\mu} = \kappa c^{\alpha} c_{\mu, \alpha} \eta, \quad \delta C_{ab} = \left(-  \tensor{C}{_{a}^{k}}C_{kb} + \kappa c^{\alpha} C_{ab, \alpha} \right) \eta.
\end{equation}

It has been shown in Ref.~\cite{Brandt:2024rsy} that  the full BRST generator 
\begin{equation}\label{eq:ccf216}
    \delta = \delta_{\text{diff}} + \delta_{\text{LL} },
\end{equation}
where $ \delta_{\text{diff} }$ and $ \delta_{\text{LL}} $ generate respectively diffeomorphisms and local Lorentz transformations, is nilpotent due to the nilpotency of these generators as well as to the  anticommutation relation 
\begin{equation}\label{eq:ccf217}
    \{ \delta_{\text{diff}} , \delta_{\text{LL}}\} =0.
\end{equation}

It is usual to define the BRST  transformations of a functional $F$ of the quantum fields by  
\begin{equation}\label{eq:ccf218}
    \delta F = (s F) \eta,
\end{equation}
where $ s$ is the BRST generator. We will use the notation $ \mathop{\text{gh}}(F) $ for the ghost number of $ F $.
Since $ \mathop{ \text{gh} } (\eta ) = -1 $, it follows that the BRST generator $s$ increases the ghost number of $F$ by one unit, that is, $ \mathop{\text{gh}} (sF) = \mathop{\text{gh}} (F) +1$.

\section{BV formalism for the EC theory in the first-order form}\label{section:BVEC}

The purpose of the BV formalism is to control gauge invariance during the perturbative renormalization of gauge theories. In the present case, such an approach must be adapted due the fact  that we are dealing with two distinct gauge transformations.
To this end, it is convenient to use a compact notation where we denote the quantum fields $ ( \tensor{e}{^a_{\mu}} , \omega_{\mu ab} )$ by $ ( \phi_{i} , \Phi_{I} )$, $( b_{\mu} , B_{ab} )$ by $(b_{i} , B_{I} )$, $(c_{\mu} , {c}^{\star}_{\mu} , C_{ab} , {C}^{\star}_{ab} )$ by $(c_{i} , {c}^{\star}_{i} , C_{I} , {C}^{\star}_{I} )$, and the BRST transformations following from the gauge transformations \eqref{eq:ccf28b} and \eqref{eq:ccf29b} by  
\begin{equation}\label{eq:ccf31}
    s \phi_{i} = r_{ij} ( \phi ) c_{j} + r_{iJ} ( \phi ) C_{J}  
\end{equation}
and 
\begin{equation}\label{eq:ccf32}
    s \Phi_{I} = R_{Ij} ( \Phi ) c_{j} + R_{IJ} ( \Phi ) C_{J}. 
\end{equation}
Moreover, we have that $\mathop{\text{gh}}( \phi_{i} , \Phi_{I} ) = \mathop{\text{gh}}(b_{i} , B_{I} ) = 0 $, $\mathop{\text{gh}}(c_{i} , C_{I} ) = - \mathop{\text{gh}} ( {c}^{\star}_{i} , {C}^{\star}_{I} ) = 1$.

We will next introduce a fermionic potential $  \Psi $, which is a local functional of the quantum fields, with ghost number $-1$, as 
\begin{equation}\label{eq:ccf33}
    \Psi = \int \mathop{d^{4} x} \bar{e} \left [ {c}^{\star}_{i} \left ( f_{ij} \phi_{j} - \frac{\xi}{2} b_{i}\right ) + {C}^{\star}_{I} \left ( F_{IJ} \Phi_{J} - \frac{\zeta}{2} B_{I}\right )\right ],
\end{equation}
where the gauge-fixing functions $ f_{ij} $ and $ F_{IJ} $ may be identified with the corresponding functions given in Eq.~\eqref{eq:ccf210}.

Then, using the relations \eqref{eq:ccf31}, \eqref{eq:ccf32} and \eqref{eq:ccf33} together with the Eqs.~\eqref{eq:ccf213}, \eqref{eq:ccf214}, \eqref{eq:ccf215}, one may verify that the Faddeev-Popov action corresponding to \eqref{eq:ccf212}, may be written in the simple form 
\begin{equation}\label{eq:ccf34}
    \begin{split}
        S_{\text{FP}} ={}& S_{\text{EC}} + s \Psi \\
        ={}& S_{\text{EC}} 
        + \int \mathop{d^{4} x} \bar{e} \bigg\{-b_{i} \left ( f_{ij} \phi_{j} - \frac{\xi}{2} b_{i} \right ) -B_{I} \left( F_{IJ} \Phi_{J} - \frac{\zeta}{2} B_{I} \right) 
        \\ & \quad + {c}^{\star}_{i} f_{ij} \left [ r_{jk} ( \phi ) c_{k} + r_{jK} ( \phi ) C_{K}\right ] + {C}^{\star}_{I} F_{IJ} \left ]  R_{Jk} (\Phi ) c_{k} + R_{JK} ( \Phi ) C_{K}\right ]\bigg\},
\end{split}
\end{equation}
which is equivalent to the one arising from the Eq.~\eqref{eq:ccf22} and from the integration $ \int \mathop{d^{4} x} \bar{e} $ of Eqs.~\eqref{eq:ccf210} and \eqref{eq:ccf211}. 

In the BV-formalism, it turns out to be convenient to introduce as well gauge-invariant antifields associated with all quantum fields of the theory \cite{weinberg:1995a}. Thus, if we denote the quantum fields in our theory: $ \phi_{i} , \Phi_{I} , b_{i} , B_{I} , c_{i} , {c}^{\star}_{i} , C_{I} , {C}^{\star}_{I} $ collectively by $ \chi_{n} $, one adds a new gauge-invariant action (due to the nilpotency of the BRST generator $s$) given by 
\begin{equation}\label{eq:ccf35}
    \int \mathop{d^{4} x}  \chi_{n}^{\ddagger} s \chi_{n} 
\end{equation}
which should have a zero ghost number. To ensure this, the antifield $ \chi_{n}^{\ddagger} $ must have a ghost number equal to $ - \mathop{\text{gh}} ( \chi_{n} ) -1 $ because the BRST generator $s$ increases the ghost number by one unit. These antifields act as sources for the BRST transformations and are important to maintain the gauge invariance during renormalization.

Thus, the complete action 
\begin{equation}\label{eq:ccf36}
    S_{\Psi}  ( \bar{\chi}, \chi, \chi^{\ddagger} ) = S_{\text{EC}} + s \Psi + \int \mathop{d^{4} x}   \chi_{n}^{\ddagger} s \chi_{n},
\end{equation}
where the $ \bar{\chi} $ denotes background fields $ \tensor{\bar{e}}{^{a}_{\mu}} $ and $ \bar{\omega}_{\mu ab} $,
will be invariant under the BRST transformations. This invariance may be expressed in the form
\begin{equation}\label{eq:ccf37}
    \int \mathop{d^{4} x} \frac{\delta_{r} S_{\Psi} }{\delta \chi_{n} } \frac{\delta_{l} S_{\Psi} }{\delta \chi_{n}^{\ddagger}} = \int \mathop{d^{4} x} \frac{\delta_{r} S_{\Psi} }{\delta \chi_{n}} s \chi_{n} = 0,
\end{equation}
where the $r$/$l$ superscripts on the functional derivatives denote that these are taken from the right or the left, respectively. 

A main element in the BV formalism is the introduction of an antibracket defined for two general functionals $F$ and $G$ as 
\begin{equation}\label{eq:ccf38}
    (F, G) \equiv \int \mathop{d^{4} x}   \left ( \frac{\delta_{r} F }{\delta \chi_{n}} \frac{\delta_{l} G }{\delta \chi_{n}^{\ddagger}} - \frac{\delta_{r} F }{\delta \chi_{n}^{\ddagger}} \frac{\delta_{l} G }{\delta \chi_{n}}\right ).
\end{equation}
Replacing the functionals $F$ and $G$ by the bosonic action $ S_{\Psi} $, one can easily verify that 
\begin{equation}\label{eq:ccf39}
    \int \mathop{d^{4} x} \frac{\delta_{r} S_{\Psi} }{\delta \chi_{n} } \frac{\delta_{l} S_{\Psi} }{\delta \chi_{n}^{\ddagger}} = \frac{1}{2} (S_{\Psi} , S_{\Psi} ).
\end{equation}
Thus, using the relation \eqref{eq:ccf37}, one can see that this action  satisfies the master equation 
\begin{equation}\label{eq:ccf310}
    ( S_{\Psi} , S_{\Psi} ) = 0
\end{equation}
which expresses the BRST invariance of the action $ S_{\Psi}  ( \bar{\chi}, \chi, \chi^{\ddagger} )$. It is important to mention that since the EC theory is of Yang-Mills type, the Faddeev-Popov gauge-fixed action coincides with the BV action.

\section{Gauge-invariant renormalization of EC theory}\label{GIREC}

BRST invariance of the action leads to relations between the Green functions which are more directly expressed in terms of generating functionals. To this end we start from the functional: 
\begin{equation}\label{eq:ccf41}
    Z (J, \bar{\chi} , \chi^{\ddagger} ) = 
    \int \mathop{\mathcal{D} \chi} \exp i \left( S_{\Psi}  ( \bar{\chi}, \chi, \chi^{\ddagger} ) + \int \mathop{d^{4} x}  J_{n} \chi_{n}\right).
\end{equation}
Making a gauge transformation of the fields and using the gauge invariance of $Z$, leads to 
\begin{equation}\label{eq:ccf42}
    J_{m} 
    \int \mathop{\mathcal{D} \chi} s \chi_{m} \exp i \left( S_{\Psi}  ( \bar{\chi}, \chi, \chi^{\ddagger} ) + \int \mathop{d^{4} x}    J_{n} \chi_{n}\right)
    = J_{m} \left \langle s \chi_{m}\right\rangle |_{\chi^{\ddagger} , J}=0.
\end{equation}
One now defines a generating functional for connected Green functions by 
\begin{equation}\label{eq:ccf43}
    \exp i W(J, \bar{\chi} , \chi^{\ddagger} ) = \frac{Z(J, \bar{\chi},  \chi^{\ddagger} )}{Z(0, 0)}.
\end{equation}
Then, the relation \eqref{eq:ccf42} can be written as 
\begin{equation}\label{eq:ccf44}
    J_{m} \frac{\delta_{l} W }{\delta \chi_{m}^{\ddagger}} =0.
\end{equation}

Next, one performs a Legendre transformation with respect to $J$ 
\begin{equation}\label{eq:ccf45}
    \tilde{\chi}_{n} \equiv \langle \chi_{n} \rangle_{J, \tilde{\chi}^{\ddagger} } =  \frac{\delta_{l} W }{\delta J_{n}} 
\end{equation}
and define the quantum effective action for one-particle irreducible Green functions as 
\begin{equation}\label{eq:ccf46}
    \Gamma  ( \bar{\chi}, \chi, \chi^{\ddagger} ) = W ( {J} , \bar{\chi} , {\chi}^{\ddagger} ) - \int \mathop{d^{4} x}   {J}_{n} {\chi}_{n},
\end{equation}
where we dropped, for simplicity of notation, the tilde denoting the mean fields and antifields. The fact that the effective action is defined by a Legendre transform implies the relations 
\begin{equation}\label{eq:ccf47}
    \frac{\delta_{r} \Gamma }{\delta {\chi}_{n}} = -  {J}_{n}, \quad \frac{\delta_{l} \Gamma }{\delta {\chi}_{n}^{\ddagger}} = \frac{\delta_{l} W }{\delta {\chi}_{n}^{\ddagger}}. 
\end{equation}
Then, in terms of the effective action $ \Gamma $, the identity \eqref{eq:ccf44} takes the form  
\begin{equation}\label{eq:ccf48}
    \int \mathop{d^{4} x}  \frac{\delta_{r} \Gamma }{\delta {\chi}_{n}} \frac{\delta_{l}\Gamma  }{\delta {\chi}_{n}^{\ddagger}} = \frac{1}{2} ( \Gamma , \Gamma )_{{\chi} , {\chi}^{\ddagger}} =0.
\end{equation}
This important relation is called the Zinn-Justin master equation \cite{Zinn-Justin:1984tfs}.

We now remark that the master equations \eqref{eq:ccf310} and \eqref{eq:ccf48} are given in terms of unrenormalized quantities. In the framework of BV approach, one must show that a similar result may be obtained if we replace them by renormalized quantities. Thus, the required equations take the form 
\begin{equation}\label{eq:ccf49}
    (S_{\text{R}} , S_{\text{R}} ) = 0, \quad ( \Gamma_{\text{R}} , \Gamma_{\text{R}} ) =0. 
\end{equation}
BRST invariance is encoded in the above master equations. These are crucial to prove, order by order in a loop expansion, the renormalizability of the EC theory in first-order form. We note that BRST quantization involves local gauge transformations. Thus, in order to implement a BRST invariant renormalization, the counterterms must be local functions. Such a program works because the ultraviolet divergences, which occurs at very short distances, can be described by local functionals.
To evaluate these divergences, we will use dimensional regularization which preserves gauge invariance and henceforth follow the thorough discussion given in \cite{Lavrov:2019nuz} (for the renormalization of a gauge-affine gravity, see ref. \cite{Lee:1990xq}). 

Let us first consider the one-loop approximation for the effective action, which may be written as 
\begin{equation}\label{eq:ccf410}
    \Gamma = \Gamma^{(1)} + O ( \hbar^{2} )= S + \hbar [ \Gamma_{\text{div}}^{(1)} + \Gamma_{\text{fin}}^{(1)} ] + O ( \hbar^{2} ),
\end{equation}
where $ \hbar $ keeps track of the loop order in a perturbative expansion. Here, $ S = S_{\Psi}  ( \bar{\chi}, \chi, \chi^{\ddagger} )$ is the tree level action given in Eq.~\eqref{eq:ccf36}, and $ \Gamma_{\text{div}}^{(1)} $, $ \Gamma_{\text{fin}}^{(1)}  $ are respectively the divergent and finite parts of $ \Gamma $ in the one-loop approximation. This divergence fixes the form of the counterterm of the one-loop renormalized action $ S_{\text{1R}} $, such that 
\begin{equation}\label{eq:ccf411}
    S_{\text{1R}} = S - \hbar \Gamma^{(1)}_{\text{div}}.
\end{equation}

Let us now consider 
\begin{equation}\label{eq:ccf412}
(S_{\text{1R}} , S_{\text{1R}} ) = (S, S) - 2 \hbar (S, \Gamma^{(1)}_{\text{div}} ) + \hbar^{2} ( \Gamma_{\text{div}}^{(1)}, \Gamma_{\text{div}}^{(1)} ).
\end{equation}
To evaluate this quantity,  we substitute the expression \eqref{eq:ccf410} in the Eq.~\eqref{eq:ccf48}, which yields 
\begin{equation}\label{eq:ccf413}
    \begin{split}
        ( \Gamma , \Gamma ) ={}& (S, S) 
        + 2 \hbar ( S, \Gamma_{\text{div}}^{(1)} ) + 2 \hbar ( S , \Gamma_{\text{fin}}^{(1)} ) + O ( \hbar^{2} ) \\
        ={}& 
         2 \hbar ( S, \Gamma_{\text{div}}^{(1)} ) + 2 \hbar ( S , \Gamma_{\text{fin}}^{(1)} ) + O ( \hbar^{2} ) = 0,
    \end{split}
\end{equation}
where we used the master equation \eqref{eq:ccf310}. To first order in $ \hbar $, we have a sum of two terms which must vanish independently, since one of them is infinite. Thus, we get 
\begin{equation}\label{eq:ccf414}
    ( S , \Gamma_{\text{div}}^{(1 )} ) =0.
\end{equation}

Using this result in Eq.~\eqref{eq:ccf412}, together with Eq.~\eqref{eq:ccf310}, we obtain 
\begin{equation}\label{eq:ccf415}
    ( S_{\text{1R}}, S_{\text{1R}}) = \hbar^{2} 
    ( \Gamma_{\text{div}}^{(1)}, \Gamma_{\text{div}}^{(1)} ) .
\end{equation}
Thus, we have shown that the renormalized action $ S_{1R} $ satisfies the master equation \eqref{eq:ccf49} up to terms of order $ \hbar^{2} $. 

The renormalized effective action to one loop order $ \Gamma_{\text{1R}} $ can now be constructed by replacing in Eq.~\eqref{eq:ccf410},  $S$ by $S_{\text{1R}} $, which leads to the relation 
\begin{equation}\label{eq:ccf416}
    \Gamma_{\text{1R}} = S_{\text{1R}} + \hbar [ \Gamma_{\text{div}}^{(1)} + \Gamma_{\text{fin}}^{(1)} ] = S + \hbar \Gamma_{\text{fin}}^{(1)}.
\end{equation}
Thus, $ \Gamma_{\text{1R}} $ is finite at one-loop order and satisfies, by using Eq.~\eqref{eq:ccf415}, the relation 
\begin{equation}\label{eq:ccf417}
    ( \Gamma_{\text{1R}} , \Gamma_{\text{1R}} ) = \hbar^{2}  
    ( \Gamma_{\text{div}}^{(1)}, \Gamma_{\text{div}}^{(1)} ). 
\end{equation}

Let us now proceed by considering the two loop renormalized action $S_{\text{2R}} $ defined by 
\begin{equation}\label{eq:ccf418}
    S_{\text{2R}} = S_{\text{1R}} - \hbar^{2} \Gamma_{1, \text{div}}^{(2)}= S - \hbar \Gamma_{\text{div}}^{(1)} - \hbar^{2} \Gamma_{1 , \text{div}}^{(2)},
\end{equation}
where $ \Gamma_{1, \text{div}}^{(2)} $ is the divergent part at two-loop order constructed on the basis $ S_{\text{1R}} $.  
One can verify that this obeys the equation 
\begin{equation}\label{eq:ccf419}
    (S, \Gamma_{1, \text{div}}^{(2)} ) = \frac{1}{2}  
    ( \Gamma_{\text{div}}^{(1)}, \Gamma_{\text{div}}^{(1)} ). 
\end{equation}
Using this relation, we find that $ S_{\text{2R}} $ satisfies, up to terms of order $ \hbar^{3} $, the master equation 
\begin{equation}\label{eq:ccf420}
    ( S_{\text{2R}}, S_{\text{2R}}) = 2h^{3} ( \Gamma_{\text{div}}^{(1)} , \Gamma_{1, \text{div}}^{(2)} ) + O ( \hbar^{4} ). 
\end{equation}
The effective action $ \Gamma_{\text{2R}} $ is obtained from $ S_{\text{2R}} $ by a similar procedure to that used earlier (compare with Eq.~\eqref{eq:ccf416})
\begin{equation}\label{eq:ccf421}
    \begin{split}
        \Gamma_{\text{2R}} ={}& S_{\text{2R}} + \hbar [\Gamma_{\text{div}}^{(1)} + \Gamma^{(1)}_{\text{fin}}] + \hbar^{2} [\Gamma_{1, \text{div}}^{(2)} + \Gamma_{1, \text{fin}}^{(2)}] 
        \\
        ={}&S + \hbar \Gamma_{\text{fin}}^{(1)} + \hbar^{2} \Gamma_{1 , \text{fin}}^{(2)} 
    \end{split}
\end{equation}
and satisfies the equation 
\begin{equation}\label{eq:ccf422}
( \Gamma_{\text{2R}} , \Gamma_{\text{2R}} ) = 2h^{3} ( \Gamma_{\text{div}}^{(1)} , \Gamma_{1 , \text{div}}^{(2)} ) + O (  \hbar^{4}) .
\end{equation}

Using mathematical induction, we assume that the effective action $ \Gamma_{m-1 \, \text{R}} $ constructed on the basis $ S_{m-1 \,  \text{R}} $ is finite up to terms of order $ \hbar^{m-1} $. Then, one finds that the full renormalized action $S_{\text{R}} $ given by 
\begin{equation}\label{eq:ccf423}
    S_{\text{R}} = S - \sum_{m =1}^{\infty} \hbar^{m} \Gamma_{m-1, \text{div}}^{(m)} 
\end{equation}
satisfies exactly the master equation \eqref{eq:ccf49}. 
Moreover, the renormalized effective action $ \Gamma_{\text{R}} $ is finite to each order in perturbation theory: 
\begin{equation}\label{eq:ccf424}
    \Gamma_{\text{R}} = S + \sum_{m = 1}^{\infty} \hbar^{m} \Gamma_{m-1, \text{fin}}^{(m)} 
\end{equation}
and satisfies as well the master equation \eqref{eq:ccf49}.

Thus, we have shown that the theory is renormalizable in the sense that all infinites can be eliminated by an appropriate renormalization procedure. The renormalized action $ S_{\text{R}} $ and the effective action $ \Gamma_{\text{R}} $ preserve the BRST gauge invariance to all orders in the loop expansion. 

\section{Gauge invariance of the background effective action}\label{section:GIBEA}

Of special interest is the background effective action $ \bar{\Gamma} ( \bar{\chi} )$ defined as the effective action $ \Gamma ( \bar{\chi} , \chi , \chi^{\ddagger} )$ evaluated  for vanishing mean quantum fields and antifields. The Ward identities satisfied by this
action may be obtained in a similar way to that used in Sec. IV, by starting with the generating  functional $ \bar{Z} ( J, \bar{\chi} )$ 
 \begin{equation}\label{eq:ccf51}
     \bar{Z} (J, \bar{\chi} ) = \int \mathop{\mathcal{D} \chi} 
    \exp i \left ( S_{\text{FP}} ( \chi , \bar{\chi} ) + \int \mathop{d^{4} x} {J}_{n} \chi_{n} \right ),
\end{equation}                 
where $ \bar{\chi} $ denotes the background fields $ \tensor{\bar{q}}{^{a}_{\mu}} $ and $ \bar{\omega}_{\mu ab} $ (see Eq.~\eqref{eq:bm1}) and $ \chi $ denotes the quantum fields.

Making the background gauge transformations 
\begin{subequations}\label{eq:ccf52}
    \begin{align} \label{eq:ccf52a}
        {\Delta} \tensor{\bar{q}}{^{a}_{\mu}} 
        ={}& - \kappa \epsilon^{\alpha} \bar{q}_{\mu , \alpha}^{a} - \kappa \tensor{\bar{q}}{^{a}_{\alpha}} \epsilon^{\alpha}_{, \mu} + \tensor{\lambda}{^{a}_{b}}\tensor{\bar{q}}{^{b}_{\mu}} \intertext{and}
{\Delta} \bar{\omega}_{\mu ab} ={}&
        - \kappa \epsilon^{\alpha} \bar{\omega}_{\mu ab , \alpha} - \bar{\omega}_{\alpha ab} \epsilon^{\alpha}_{, \mu} - \kappa \lambda_{ab , \mu } + \tensor{\lambda}{_{a}^{p}} \bar{\omega}_{\mu pb} + \tensor{\lambda}{_{b}^{p}} \bar{\omega}_{\mu ap},
    \end{align}
\end{subequations}
and using the gauge invariance of $ \bar{Z} (J, \bar{\chi} )$ and of $S_{\text{FP}} ( \chi , \bar{\chi} )$, we obtain that 
\begin{equation}\label{eq:ccf53}
    J_{m} \int \mathop{\mathcal{D} \chi} {\Delta} \chi_{m} 
    \exp i \left ( S_{\text{FP}} (  \chi, \bar{\chi} ) + \int \mathop{d^{4} x} J_{n} {\chi }_{n} \right ) =0,
\end{equation}
where $ \Delta \chi_{m} $ denote the corresponding background gauge transformations of the quantum fields. For example, 
\begin{align}\label{eq:ccf54alt}
    \Delta \tensor{q}{^{a}_{\mu}} ={}&- \kappa \epsilon^{\alpha} \tensor{q}{^{a}_{\mu , \alpha}}- \kappa \tensor{q}{^{a}_{\alpha}} \epsilon^{\alpha}_{, \mu} + \tensor{\lambda}{^{a}_{b}} \tensor{q}{^{b}_{\mu}} \intertext{and}
    \Delta \tensor{Q}{_{\mu ab}} ={}& - \kappa \epsilon^{\alpha} \tensor{Q}{_{\mu ab, \alpha}}- \kappa \tensor{Q}{_{\alpha ab }} \epsilon^{\alpha}_{, \mu} + \tensor{\lambda}{_{a}^{p}} Q_{\mu pb} + \tensor{\lambda}{_{b}^{p}}Q_{\mu ap} . 
\end{align}

It is now convenient to write the field transformations \eqref{eq:ccf52} in the compact form used in Sec. III, as 
$
    \Delta \chi_{m} = \mathcal{R}_{ml} ( \chi ) \theta_{l} 
    $,
where $ \theta_{1} = \epsilon_{\mu} $ and $ \theta_{2} = \lambda_{ab} $.
Then, in terms of the generating functional  $ \bar{W} ( J , \bar{\chi}  )$ defined in Eq.~\eqref{eq:ccf43}, the Eq.~\eqref{eq:ccf53} becomes 
\begin{equation}\label{eq:ccf54}
    J_{m} \mathcal{R}_{ml} \left ( \frac{\delta }{\delta J}\right ) \bar{W} ( J, \bar{\chi}  ) \theta_{l} =0. 
\end{equation}

The background effective action $ \bar{\Gamma} ( \bar{\chi} ) $ is obtained in the standard way, by using a Legendre transform like
that  given in Eq.~\eqref{eq:ccf46}. Proceeding in a similar manner to that used previously, we get the result
\begin{equation}\label{eq:ccf55}
    {\Delta} \bar{\Gamma} ( \bar{\chi} ) = {\Delta} \bar{\chi}_{m} \frac{\delta \bar{\Gamma} }{\delta \bar{\chi}_{m}} =0
\end{equation}
which exhibits the diffeomorphism invariance of the background effective action $ \bar{\Gamma} ( \bar{\chi} )$ under the background gauge transformations of the background fields     $ \tensor{\bar{q}}{^{a}_{\mu}} $ and $ \bar{\omega}_{\mu ab} $.

This relation leads to Ward identities that reflect the background gauge symmetry of this action.
Such  identities are obtained by taking functional derivatives of Eq.~\eqref{eq:ccf55} with respect to the background fields  $ \tensor{\bar{q}}{^{a}_{\mu}} $ and $ \bar{\omega}_{\mu ab} $, evaluated at vanishing values of these fields. But  taking just a single functional derivative and using the transformations \eqref{eq:ccf52} of these fields, one gets trivial identities because the tadpoles
$ \delta \bar{\Gamma} / \delta \tensor{\bar{q}}{^{a}_{\mu}} $ and  $ \delta \bar{\Gamma} / \delta \bar{\omega}_{\mu ab} $ 
vanish in our theory.

On the other hand, by taking two functional derivatives of Eq.~\eqref{eq:ccf55} with respect to the background tetrad or spin connection fields, one obtains non-trivial relations which should be satisfied by the corresponding Green functions. 
For instance, taking two derivatives with respect to the tetrad fields, it is  straightforward to derive the following equations
\begin{equation}\label{eq:ccf56}
     \partial_{\alpha} \frac{\delta^{2} \bar{\Gamma} }{\delta \tensor{\bar{q}}{^{a}_{\alpha}} \delta \tensor{\bar{q}}{^{b}_{\nu}}} \bigg |_{ \bar{q} =   \bar{\omega} =0} 
    =0,
\end{equation}
where we have used the fact that $ \epsilon_{\mu} $ is an arbitrary independent parameter. 
The  Ward identity~\eqref{eq:ccf56} implies that the background tetrad self-energy should be a transverse function. 

\begin{figure}[ht]
    \includegraphics[scale=0.9]{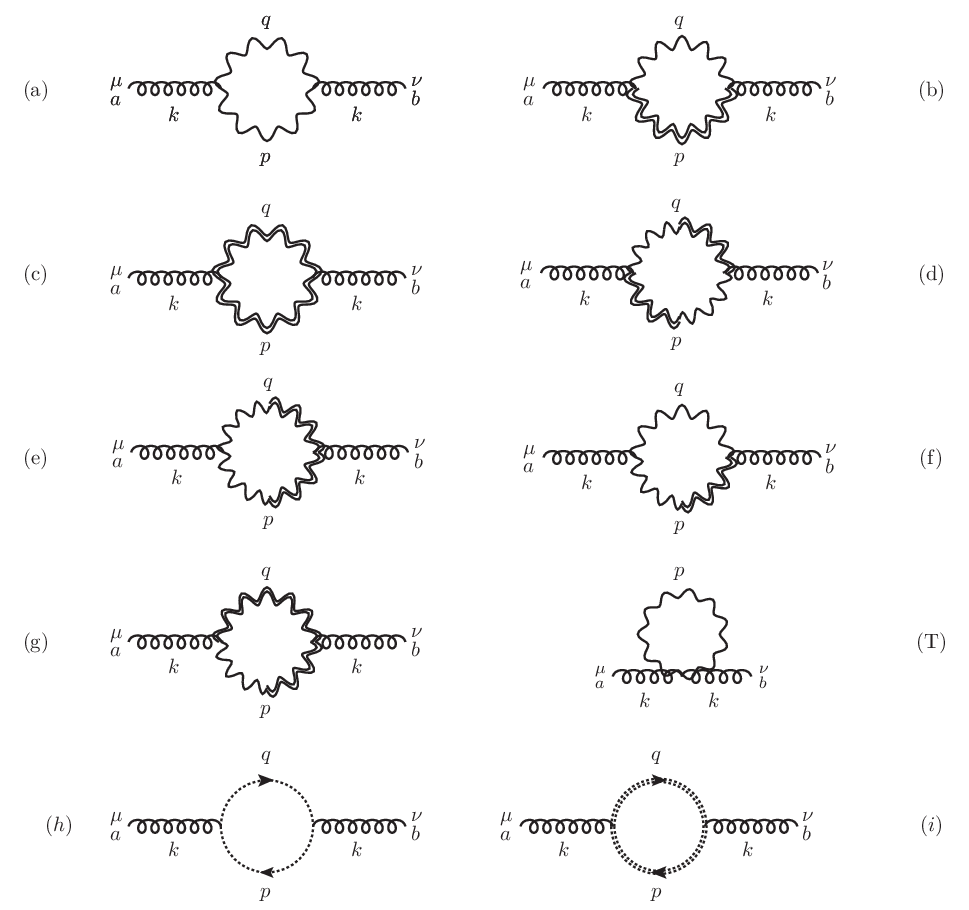}
    \caption{One-loop contributions to the self-energy $ \bar{\Pi}_{ab}^{\mu \nu} (k) $. The curly lines denote the background tetrad field, while the wavy and double wavy lines are associated, respectively, with the quantum fields $ \tensor{q}{^{a}_{\mu}} $ and $ Q_{\mu ab} $. The ghost fields $ {c}^{\star}_{\mu} $, $ c_{\mu} $ and $ {C}^{\star}_{ab} $, $ C_{ab} $ are denoted by dotted and double dotted lines. Diagrams like (1T) vanish in dimensional regularization.}\label{fig:diag1}
\end{figure}
The Feynman diagrams contributing at one loop to the tetrad self-energy are shown in Fig.~\ref{fig:diag1}.
The Feynman rules of the first-order form of the EC theory are derived in any space-time $D$ in appendix A.
As shown in appendix C (Eqs.~\eqref{eq:appV13} and \eqref{eq:appV11}), in $D=4 -2 \epsilon $ dimensions, the divergent part of this function has the transverse form 
\begin{equation}\label{eq:ccf510}
    \frac{i \kappa^{2} k^{2}}{16 \pi^{2}  \epsilon}\left[C_{1} \left ( \delta_{a}^{\mu} - \frac{k_{a} k^{\mu}}{k^{2}}\right ) \left ( \delta_{b}^{\nu} - \frac{k_{b} k^{\nu}}{k^{2}}\right ) + C_{2} \left ( \delta_{a}^{\nu} - \frac{k_{a} k^{\nu}}{k^{2}}\right ) \left ( \delta_{b}^{\mu} - \frac{k_{b} k^{\mu}}{k^{2}}\right ) +  \left ( C_{3} \eta_{ab} + C_{5} \frac{k_{a} k_{b}}{k^{2}}\right ) \left ( \eta^{\mu \nu} - \frac{k^{\mu} k^{\nu}}{k^{2}}\right )\right],
\end{equation}
where $C_{i} $ are linear functions of the gauge parameters $ \xi $ and $ \zeta $ introduced in \eqref{eq:ccf210}.

Similarly, the spin connection self-energy is also ultraviolet divergent. It may be verified that this obeys the Ward identity 
\begin{equation}\label{eq:ccf511}
        \partial_{\mu} \frac{\delta^{2} \bar{\Gamma} }{\delta \bar{\omega}_{\mu ab} \delta \bar{\omega}_{\nu cd}} \bigg |_{\bar{q} =  \bar{\omega} =0} =0.
\end{equation}
which shows that the spin connection self-energy is also a transverse function. 



\section{Discussion}\label{section:final}

We have examined the renormalizability of the EC theory in the first-order form, where the basic elements are the independent tetrad and spin connection fields. This theory possesses two BRST symmetries, namely, diffeomorphism and local Lorentz invariance. To this end, we used the BV formalism in conjunction with the background field method, which preserves gauge invariance during perturbative renormalization. We have argued that the EC theory in the first-order form is renormalizable, meaning that all the ultraviolet divergences can be subtracted in a gauge-invariant way by local covariant counterterms.

We have studied the background effective action obtained after switching off the quantum fields and antifields. It is important that the background field transformations do not change the form of the gauge-fixing terms for the quantum fields nor that of the ghost terms, which appear in the Faddeev-Popov action. The invariances of the background effective action lead to Ward identities that reflect the gauge symmetries of the first-order form of the EC theory. Such Ward identities relate the Green's functions and are very useful for simplifying perturbative calculations in this theory (exemplified in the appendices), which are generally quite involved. As an illustration, we have computed the divergent part of the background tetrad self-energy at one-loop order. Apart from the tetrad self-energy, there also occurs a divergent spin-connection as well as mixed tetrad-spin connection self-energy. Upon diagonalizing such a matrix, one would obtain a diagonal self-energy matrix whose elements correspond to the tetrad and spin connection normal modes of the EC theory. This is currently under consideration.

In previous works, it was shown, by using a Lagrange multiplier field that ensures the classical equations of motion, that all contributions beyond one-loop order may be eliminated in quantum gravity. It has also been argued that this important feature leads to a renormalizable and unitary theory \cite{Brandt:2020gms, Brandt:2021qgh}. It may be worth considering an extension of such an approach to the EC theory, particularly in the formalism recently formulated in \cite{Brandt:2022kjo, McKeon:2024psy}.


\begin{acknowledgments}
F. T. B., J. F. and S. M.-F. would like to thank CNPq (Brazil) for financial support.  D. G. C. M. is indebted to Ann Alsoy for enlightening conversations.
\end{acknowledgments}

\appendix

\section{Feynman Rules}
In this appendix, we will derive the propagators of the EC theory in first-order form that comes from the action~\eqref{eq:ccf212}. We will also present the vertices which are relevant to the computation of the background self-energy of the tetrad field at one-loop order. Throughout the appendices we have used FeynCalc \cite{Mertig:1990an,Shtabovenko:2020gxv}. 

\subsection{Propagators}

 Using Eq.~\eqref{eq:bm1} in Eq.~\eqref{eq:ccf22}, we can collect the bilinear terms in the quantum fields obtaining 
 \begin{equation}\label{eq:bm3}
     \frac{1}{2} \begin{pmatrix}
         \tensor{q}{^{a}_{\mu}} & Q_{\sigma  cd}
     \end{pmatrix}
     \begin{pmatrix}
         A^{\mu \nu}_{ab} & B^{\mu \, \rho lm}_{a} \\
         C^{\sigma cd \, \nu}_{b}  & D^{\sigma cd \, \rho lm}
     \end{pmatrix}
     \begin{pmatrix}
         \tensor{q}{^{b}_{\nu}} \\
         Q_{\rho lm}.
     \end{pmatrix}
 \end{equation}
We will also include the gauge fixing terms due to the gauges:
\begin{subequations}\label{eq:bm13}
    \begin{align} \label{eq:36aa}
        F_{\nu}(q)  \equiv \eta_{ab} [(\tensor{\bar{e}}{^{a}_{\mu}} \tensor{q}{^{b}_{\nu}} + \tensor{q}{^{a}_{\mu}} \tensor{\bar{e}}{^{b}_{\nu}})^{; \bar{\mu}} -2 \sigma (\bar{e} _{\alpha}^{b} q^{a \, \alpha} )_{; \bar{\nu} }]
        ={}&0 
        \\ \label{eq:36bb}
        F_{ab} (Q)\equiv \tensor{Q}{_{\mu ab}^{; \bar{\mu}}} ={}&0.
    \end{align}
\end{subequations}
When $ \sigma = 1/2$, the gauge in Eq.~\eqref{eq:36aa} is equal to the linear part of the de Donder gauge $ h_{\mu \nu}^{; \bar{\mu}} - \frac{1}{2} h_{; \bar{\nu}} $, where \begin{equation}\label{eq:cm2}
    h_{\mu \nu} (x) = \eta_{ab} \left [ \tensor{(\eta + \bar{q})}{^{a}_{\mu}} \tensor{q}{^{b}_{\nu}} + \tensor{(\eta + \bar{q})}{^{b}_{\nu}} \tensor{q}{^{a}_{\mu}} + \kappa \tensor{q}{^{a}_{\mu}} \tensor{q}{^{b}_{\nu}} \right ],
\end{equation}
in the quantum field $ \tensor{q}{^{a}_{\mu}} $. 

Integrating out the Nakanishi-Lautrup fields in Eq.~\eqref{eq:ccf210} leads to
\begin{equation}\label{eq:gfixingout}
    \mathcal{L}_{\text{gf}} =  - \frac{1}{2 \xi} F^{\mu} F_{\mu} - \frac{1}{2 \zeta}  F^{ab} F_{ab}. 
\end{equation}
Thus, we obtain that, in momentum space ($ \partial \to + i p $),
\begin{subequations}\label{eq:bm14}
    \begin{align}
        A^{\mu \nu}_{ab} ={}&
-\frac{4 \sigma ^2 {p}^{2} \delta^{\mu}_{a} \delta^{\nu}_{b}}{\xi }+\frac{4 \sigma {p}_a {p}^{\mu} \delta^{\nu}_{b}}{\xi }-\left(\frac{4 \sigma {p}_b {p}^{\nu } \delta^{\mu}_{a}}{\xi }
+\frac{{p}_a {p}_b \delta^{\mu \nu }}{\xi }+\frac{{p}_b {p}^{\mu } \delta^{\nu}_{a}}{ \xi }+\frac{{p}_a {p}^{\nu} \delta^{\mu}_{b}}{ \xi}+\frac{{p}^{\mu } {p}^{\nu } \eta_{ab}}{\xi }\right),
        \\
        B^{\mu \, \rho lm}_{a} ={}{}& 
-i ({p}^l \delta_{a}^{\mu} \eta^{m\rho}-{p}_{a } \eta^{l \mu} \eta^{m\rho}-{p}^m \delta_{a}^{\mu} \eta^{l \rho}-{p}^l \eta^{m \mu} \delta_a^{\rho}+{p}^m \eta^{l \mu} \delta_{a}^{\rho}+{p}_{a } \eta^{m\mu} \eta^{l\rho}),
                                         & \\
        C^{\sigma cd \, \nu }_{b} ={}& - B^{\nu \, \sigma cd}_{b}, \\
        D^{\sigma cd \, \rho lm} ={} &
        -2   \tensor{I}{^{a b \sigma  \rho}} (\tensor{I}{_{ap}^{cd}} \tensor{I}{^{p}_{b}^{lm}} - \tensor{I}{_{ap}^{lm} } \tensor{I}{^{p}_{b}^{cd}}) - \frac{1}{\zeta} p^{\sigma} p^{\rho} \tensor{I}{^{cd}^{lm}},
    \end{align}
\end{subequations}
where $ I_{ab c d} = ( \eta_{ac} \eta_{bd} - \eta_{ad} \eta_{b c} )/2$.

In order to derive the propagator, we will invert the $2 \times 2 $ matrix in Eq.~\eqref{eq:bm3} using 
\begin{equation}\label{eq:matinv}
\begin{pmatrix}
    \bm{A} & \bm{B} \\
    \bm{C} & \bm{D}
\end{pmatrix}^{-1} = \begin{pmatrix}
    \bm{X}^{-1} & - \bm{X}^{-1} \bm{B} \bm{D}^{-1} \\
    - \bm{D}^{-1} \bm{C} \bm{X}^{-1}      & \bm{D}^{-1} + \bm{D}^{-1} \bm{C} \bm{X} \bm{B} \bm{D}^{-1}
\end{pmatrix},
\end{equation}
where $ \bm{X} = \bm{A} - \bm{B} \bm{D}^{-1} \bm{C}$ is the Schur complement. 

First, we will compute the  inverse of $ \bm{D} $. We introduce the tensor basis: \begin{subequations}\label{eq:tensorbasisD}
    \begin{align}
        (T_{QQ}^{(1)} )^{\mu ab \, \nu cd}   ={}&  \frac{1}{2} \eta^{ac} \eta^{bd} \eta^{\mu \nu }-\frac{1}{2} \eta^{ad} \eta^{bc} \eta^{\mu \nu } = I^{\mu ab \, \nu cd}
, \\ (T_{QQ}^{(2)})^{\mu ab \, \nu cd} ={}&
 \frac{1}{4} \eta^{a\mu } \eta^{bd} \eta^{c \nu }-\frac{1}{4} \eta^{ad} \eta^{b\mu 
} \eta^{c \nu }-\frac{1}{4} \eta^{a\mu } \eta^{bc} \eta^{d \nu }+\frac{1}{4}
\eta^{ac} \eta^{b\mu
} \eta^{d \nu } 
, \\ (T_{QQ}^{(3)})^{\mu ab \, \nu cd} ={}&
 \frac{1}{4} \eta^{a \nu } \eta^{bd} \eta^{c\mu }-\frac{1}{4} \eta^{ad} \eta^{b \nu } \eta^{c\mu }-\frac{1}{4} \eta^{a \nu } \eta^{bc} \eta^{d\mu }+\frac{1}{4}
\eta^{ac} \eta^{b \nu
} \eta^{d\mu } 
, \\ (T_{QQ}^{(4)})^{\mu ab \, \nu cd} ={}&
 \frac{p^{\mu } p^{\nu } \eta^{ac} \eta^{bd}}{2 p^2}-\frac{p^{\mu } p^{\nu}
 \eta^{ad} \eta^{bc}}{2 p^2} 
, \\ (T_{QQ}^{(5)})^{\mu ab \, \nu cd} ={}&
 -\frac{p^c p^{\mu } \eta^{a \nu } \eta^{bd}}{4 p^2}+\frac{p^c p^{\mu }
\eta^{ad} \eta^{b \nu }}{4 p^2}-\frac{p^d p^{\mu } \eta^{a \nu } \eta^{bc}}{4
p^2}+\frac{p^d
p^{\mu } \eta^{ac} \eta^{b \nu }}{4 p^2} 
\\
                                         & 
-\frac{{p}^a {p}^{\nu } {g}^{bd}
   {g}^{c\mu }}{4
   p^{2}}+\frac{{p}^b
   {p}^{\nu } {g}^{ad} {g}^{c\mu }}{4
   p^{2}}+\frac{{p}^a
   {p}^{\nu } {g}^{bc} {g}^{d\mu }}{4
   p^{2}}-\frac{{p}^b
   {p}^{\nu } {g}^{ac} {g}^{d\mu }}{4
   p^{2}}
, \\ (T_{QQ}^{(6)})^{\mu ab \, \nu cd} ={}&
 -\frac{p^c p^{\nu } \eta^{a\mu } \eta^{bd}}{4 p^2}+\frac{p^c p^{\nu }
\eta^{ad} \eta^{b\mu }}{4 p^2}-\frac{p^d p^{\nu } \eta^{a\mu } \eta^{bc}}{4
p^2}+\frac{p^d
p^{\nu } \eta^{ac} \eta^{b\mu }}{4 p^2} 
, \\ (T_{QQ}^{(7)})^{\mu ab \, \nu cd} ={}&
 \frac{p^a p^c \eta^{b \nu } \eta^{d\mu }}{4 p^2}-\frac{p^b p^c \eta^{a \nu
} \eta^{d\mu }}{4 p^2}-\frac{p^a p^d \eta^{b \nu } \eta^{c\mu }}{4
p^2}+\frac{p^b p^d \eta^{a \nu
} \eta^{c\mu }}{4 p^2} 
, \\ (T_{QQ}^{(8)})^{\mu ab \, \nu cd} ={}&
 \frac{p^a p^c \eta^{b\mu } \eta^{d \nu }}{4 p^2}-\frac{p^b p^c \eta^{a\mu }
\eta^{d \nu }}{4 p^2}-\frac{p^a p^d \eta^{b\mu } \eta^{c \nu }}{4
p^2}+\frac{p^b p^d \eta^{a\mu
} \eta^{c \nu }}{4 p^2} 
, \\ (T_{QQ}^{(9)})^{\mu ab \, \nu cd} ={}& 
\frac{{p}^a {p}^c {g}^{bd}
   {g}^{\mu \nu }}{4
   p^{2}}-\frac{{p}^b {p}^c
   {g}^{ad} {g}^{\mu \nu }}{4
   p^{2}}-\frac{{p}^a {p}^d
   {g}^{bc} {g}^{\mu \nu }}{4
   p^{2}}+\frac{{p}^b {p}^d
   {g}^{ac} {g}^{\mu \nu }}{4
p^{2}} 
, \\ (T_{QQ}^{(10)})^{\mu ab \, \nu cd} ={}&
 \frac{p^a p^c p^{\mu } p^{\nu } \eta^{bd}}{4 p^4}-\frac{p^b p^c
p^{\mu } p^{\nu } \eta^{ad}}{4 p^4}-\frac{p^a p^d p^{\mu } p^{\nu }
\eta^{bc}}{4 p^4}+\frac{p^b
p^d p^{\mu } p^{\nu } \eta^{ac}}{4 p^4}.
    \end{align}
\end{subequations}
We obtain 
\begin{equation}\label{eq:bm15a}
    (D^{-1})^{\sigma cd \, \rho lm} =  \sum_{i=1}^{10} y_{(i)} (T^{(i)}_{QQ} )^{\sigma cd \, \rho lm} ,
\end{equation}
where
\begin{subequations}\label{eq:bm15}
\begin{align}
    y_{(1)}&= -\frac{1}{2} & 
 y_{(2)}&= \frac{2 \left(\zeta +p^2\right)}{(D-2) \zeta +(D-3) p^2} \\
 y_{(3)}&= -1 & 
 y_{(4)}&= \frac{1}{2}-\frac{\zeta }{2 \zeta +p^2} \\
 y_{(5)}&= \frac{2 \zeta }{2 \zeta +p^2}-1 & 
 y_{(6)}&= \frac{2 p^2}{(D-2) \zeta +(D-3) p^2} \\
 y_{(7)}&= \frac{2 \zeta }{2 \zeta +p^2}-1 & 
 y_{(8)}&= -\frac{2 p^2}{(D-2) \left((D-2) \zeta +(D-3) p^2\right)} \\
 y_{(9)}&= 1-\frac{2 \zeta }{2 \zeta +p^2} & 
 y_{(10)}&= -\frac{2 (D-4) p^4}{\left(2 \zeta +p^2\right) \left((D-2) \zeta +(D-3) p^2\right)} 
\end{align}
\end{subequations}
and $ D $ is the spacetime dimension. 

Then, using Eq.~\eqref{eq:bm14} and the result in Eq.~\eqref{eq:bm15a}, we obtain the Schur complement (for simplicity, in $ D=4$):
\begin{equation}\label{eq:bm17}
    \begin{split}
        X^{a \mu \, b \nu} ={}& A^{a \mu \, b \nu} - \tensor{B}{^{a \mu}_{\sigma cd}} ( D^{-1})^{ \sigma cd \rho lm} \tensor{C}{_{\rho lm}^{b \nu}}\\
        ={}& 
        -\frac{4 \sigma ^2 {p}^2 {\eta}^{a \mu } {\eta}^{b \nu }}{\xi }+\frac{4 \sigma  {p}^a {p}^{\mu } {\eta}^{b \nu }}{\xi }+\frac{4 \sigma  {p}^b {p}^{\nu } {\eta}^{a \mu }}{\xi }-\frac{{p}^a {p}^{\nu } {\eta}^{b \mu }}{\xi }-\frac{{p}^a {p}^b {\eta}^{\mu  \nu }}{\xi }-\frac{{p}^b {p}^{\mu } {\eta}^{a \nu }}{\xi }-\frac{{p}^{\mu } {p}^{\nu } {\eta}^{a b}}{\xi } \\ & \quad  
        +3 {p}^a {p}^b {\eta}^{\mu  \nu }-{p}^a {p}^{\nu } {\eta}^{b \mu }-{p}^a {p}^{\mu } {\eta}^{b \nu }-{p}^b {p}^{\nu } {\eta}^{a \mu }-{p}^b {p}^{\mu } {\eta}^{a \nu }+{p}^2 {\eta}^{a \nu } {\eta}^{b \mu }+{p}^2 {\eta}^{a \mu } {\eta}^{b \nu } \\ & \quad  
+\frac{5 {p}^a {p}^b {p}^{\mu } {p}^{\nu }}{{p}^2+2 \zeta }+\frac{2 \zeta  {p}^a {p}^b {p}^{\mu } {p}^{\nu }}{{p}^2 \left({p}^2+2 \zeta \right)}-\frac{{p}^a {p}^b {p}^{\mu } {p}^{\nu }}{{p}^2}
        \frac{6 \zeta  {p}^a {p}^{\mu } {\eta}^{b \nu }}{{p}^2+2 \zeta }-\frac{8 \zeta  {p}^a {p}^b {\eta}^{\mu  \nu }}{{p}^2+2 \zeta }+\frac{6 \zeta  {p}^b {p}^{\nu } {\eta}^{a \mu }}{{p}^2+2 \zeta } \\ & \quad -\frac{2 \zeta  {p}^{\mu } {p}^{\nu } {\eta}^{a b}}{{p}^2+2 \zeta }+\frac{{p}^4 {\eta}^{a \nu } {\eta}^{b \mu }}{{p}^2+2 \zeta }-\frac{3 {p}^4 {\eta}^{a \mu } {\eta}^{b \nu }}{{p}^2+2 \zeta }+\frac{3 {p}^2 {p}^a {p}^{\mu } {\eta}^{b \nu }}{{p}^2+2 \zeta }+\frac{3 {p}^2 {p}^b {p}^{\nu } {\eta}^{a \mu }}{{p}^2+2 \zeta }+\frac{2 \zeta  {p}^2 {\eta}^{a b} {\eta}^{\mu  \nu }}{{p}^2+2 \zeta }\\ & \quad -\frac{{p}^2 {p}^a {p}^{\nu } {\eta}^{b \mu }}{{p}^2+2 \zeta }-\frac{{p}^2 {p}^b {p}^{\mu } {\eta}^{a \nu }}{{p}^2+2 \zeta }-\frac{6 \zeta  {p}^2 {\eta}^{a \mu } {\eta}^{b \nu }}{{p}^2+2 \zeta }-\frac{7 {p}^2 {p}^a {p}^b {\eta}^{\mu  \nu }}{{p}^2+2 \zeta }
        .
\end{split}
\end{equation}

\subsection{Propagator $qq$}\label{app:A2}
The propagator $ (P_{qq})_{\mu \nu}^{ab} \equiv  \langle 0|T \tensor{e}{^a_{\mu}} \tensor{e}{^{b}_{\nu}}| 0 \rangle $ of the tetrad field  is given by $ i (\bm{X}^{-1})_{\mu \nu}^{ab}  $. To invert $ \bm{X}$, we use the basis:
\begin{subequations}\label{eq:b18}
    \begin{align}
        {(T_{qq}^{(1)} )}^{a \mu \, b \nu} ={}&   \eta^{a\mu} \eta^{b\nu}, \\
{(T_{qq}^{(2)} )}^{a \mu \, b \nu} ={}&   \eta^{a\nu} \eta^{b\mu}, \\
{(T_{qq}^{(3)} )}^{a \mu \, b \nu} ={}&   \eta^{ab} \eta^{\mu\nu} \equiv I^{a \mu b \nu} , \\
{(T_{qq}^{(4)} )}^{a \mu \, b \nu} ={}&   \frac{p^\mu p^\nu \eta^{ab}}{p^2}, \\
{(T_{qq}^{(5)} )}^{a \mu \, b \nu} ={}&   \frac{p^a p^b \eta^{\mu\nu}}{p^2}, \\
{(T_{qq}^{(6)} )}^{a \mu \, b \nu} ={}&   \frac{p^a p^\mu \eta^{b\nu}}{p^2}+
\frac{p^b p^\nu \eta^{a\mu}}{p^2}, \\
{(T_{qq}^{(7)} )}^{a \mu \, b \nu} ={}&   \frac{p^a p^\nu \eta^{b\mu}}{p^2} +\frac{p^b p^\mu \eta^{a\nu}}{p^2}, \\
{(T_{qq}^{(8)} )}^{a \mu \, b \nu} ={}&  \frac{p^a p^b p^\mu p^\nu}{p^4}.
    \end{align}
\end{subequations}
One can verify that  
\begin{equation}\label{eq:bm19}
    ( \bm{X}^{-1})^{a \mu \, b \nu} = -\sum_{i=1}^{8} x_{(i)} ( T_{qq}^{(i)} )^{a \mu \, b \nu}, 
\end{equation}
where 
\begin{subequations}\label{eq:bm20}
    \begin{align}
        x_{(1)}&= \frac{1}{2 (D-2) p^2} & 
        x_{(2)}&= -\frac{\zeta +p^2}{2 p^4} \\
        x_{(3)}&= \frac{\zeta }{2 p^4} & 
        x_{(4)}&= \frac{-4 \xi +D(2 \xi  +1)-3}{2 (D-2) p^2} \\
        x_{(5)}&= \frac{D-3}{2 (D-2) p^2} & 
        x_{(6)}&= \frac{1-2 \sigma }{2 (D-2) p^2 (\sigma -1)} \\
 x_{(7)}&= \frac{1}{2 (D-2) p^2} 
        & 
 x_{(8)}&= \frac{(1-2 \sigma) (D (\xi  (2 \sigma -3)-2)
 +2 (\xi +1) (2 \sigma -3))}{4 (D-2) p^2 (\sigma -1)^2}.
\end{align}
\end{subequations}
We have a pole when $ \sigma = 1$ which is associated to the vanishing of the Faddeev-Popov determinant \cite{harrison2013einstein}. 
The gauge~\eqref{eq:36aa} and the de Donder gauge are equivalent in the lowest order in $ \kappa $. Thus, the propagators in either gauge should be the same.

\subsection{Propagators $QQ$, $Qq$ and $qQ$}
We proceed, and compute the $QQ$ propagator $ P_{QQ} $ which is given by $ i(\bm{D}^{-1} + \bm{D}^{-1} \bm{C} \bm{X} \bm{B} \bm{D}^{-1})$. We obtain that 
\begin{equation}\label{eq:bm22}
    (P_{QQ})_{\mu ab \, \nu cd} =  i \sum_{i =1}^{10} z_{(i)} ( T_{QQ}^{(i)} )_{\mu ab \, \nu cd}, 
\end{equation}
where 
\begin{subequations}\label{eq:bm23}
\begin{align}
 z_{(1)}={}& -\frac{1}{2} & 
 z_{(2)}={}& \frac{2}{D-2} & 
 z_{(3)}={}&- 1 & 
 z_{(4)}={}& \frac{1}{2}-\frac{\zeta }{p^2}& 
 z_{(5)}={}&- 1 \\ 
 z_{(6)}={}& \frac{2}{D-2} & 
 z_{(7)}={}& 1 & 
 z_{(8)}={}&- \frac{2}{D-2} & 
 z_{(9)}={}& 1 & 
 z_{(10)}={}& -\frac{2(D-3)}{D-2}.
\end{align}
\end{subequations}

To compute the mixed propagators, we  introduce another basis:
\begin{subequations}\label{eq:bm24}
    \begin{align}
        (T_{Qq}^{(1)})^{\mu ab \, c \nu } ={}& \frac{{p}^b \eta^{a c} \eta^{\mu  \nu }}{2 {p}^2}-\frac{{p}^a \eta^{b c} \eta^{\mu  \nu }}{2 {p}^2} , \\
        (T_{Qq}^{(2)})^{\mu ab \, c \nu }={}&\frac{{p}^b \eta^{a \nu } \eta^{c \mu }}{2 {p}^2}-\frac{{p}^a \eta^{b \nu } \eta^{c \mu }}{2 {p}^2} , \\
(T_{Qq}^{(3)})^{\mu ab \, c \nu }={}&\frac{{p}^b \eta^{a \mu } \eta^{c \nu }}{2 {p}^2}-\frac{{p}^a \eta^{b \mu } \eta^{c \nu }}{2 {p}^2} , \\
(T_{Qq}^{(4)})^{\mu ab \, c \nu }={}&\frac{{p}^{\nu } \eta^{a \mu } \eta^{b c}}{2 {p}^2}-\frac{{p}^{\nu } \eta^{a c} \eta^{b \mu }}{2 {p}^2} , \\
(T_{Qq}^{(5)})^{\mu ab \, c \nu }={}&\frac{{p}^{\mu } \eta^{a \nu } \eta^{b c}}{2 {p}^2}-\frac{{p}^{\mu } \eta^{a c} \eta^{b \nu }}{2 {p}^2} , \\
(T_{Qq}^{(6)})^{\mu ab \, c \nu }={}&\frac{{p}^c \eta^{a \mu } \eta^{b \nu }}{2 {p}^2}-\frac{{p}^c \eta^{a \nu } \eta^{b \mu }}{2 {p}^2} , \\
(T_{Qq}^{(7)})^{\mu ab \, c \nu }={}&\frac{{p}^a {p}^c {p}^{\mu } \eta^{b \nu }}{2 {p}^2}-\frac{{p}^b {p}^c {p}^{\mu } \eta^{a \nu }}{2 {p}^2} , \\
(T_{Qq}^{(8)})^{\mu ab \, c \nu }={}&\frac{{p}^a {p}^c {p}^{\nu } \eta^{b \mu }}{2 {p}^2}-\frac{{p}^b {p}^c {p}^{\nu } \eta^{a \mu }}{2 {p}^2} , \\
(T_{Qq}^{(9)})^{\mu ab \, c \nu }={}&\frac{{p}^a {p}^{\mu } {p}^{\nu } \eta^{b c}}{2 {p}^2}-\frac{{p}^b {p}^{\mu } {p}^{\nu } \eta^{a c}}{2 {p}^2}.
\end{align}
\end{subequations}
The $Qq$ propagator $P_{Qq} $ is given by $ -i \bm{D}^{-1} \bm{C} \bm{X}^{-1}$. Computing and projecting it in basis~\eqref{eq:bm24} yields 
\begin{equation}\label{eq:bm25}
    (P_{Qq} )_{\mu ab \, c\nu } =    i \sum_{i=1}^{9} w_{(i)} (T_{Qq}^{(i)})_{\mu ab \,  c\nu },
\end{equation}
where \begin{subequations}\label{eq:bm26}
    \begin{align}
        w_{(1)}&= 0 & 
        w_{(2)}&= 1 & 
 w_{(3)}&= \frac{1}{2-D} \\
 w_{(4)}&= \frac{1}{2-D} & 
 w_{(5)}&= -\frac{\zeta }{p^2} & 
 w_{(6)}&= \frac{1}{D-2} \\
 w_{(7)}&= \frac{D-3}{(D-2) p^2} & 
 w_{(8)}&= \frac{1-2 \sigma }{(D-2) (\sigma -1) p^{2} } & 
 w_{(9)}&= \frac{1}{(D-2) p^2} 
\end{align}
\end{subequations}

One can verify that $ (P_{q Q})_{c \nu \, \mu ab} = - (P_{Qq} )_{\mu ab \, c \nu} $. Note that, the propagators $QQ$, $Qq$ and $qQ$ have no dependency on the gauge parameter $ \xi $ because 
\begin{equation}\label{eq:bm27}
    \bm{C} \bm{X}^{-1} \bm{A} = 0.
\end{equation}

\subsection{Ghost sector}\label{section:ghsector}

The bilinear terms in $ {c}^{\star}_{\mu} $, $ c_{\mu} $ and $ {C}^{\star}_{ab} $, $ C_{ab} $ from the ghost Lagrangian~\eqref{eq:ccf211} reads 
\begin{equation}\label{eq:ccf211bilinear}
    \begin{split}
     & -c^{\star \nu}    \eta_{ab}   \big \{ \left[ \delta_{\mu}^{a} \left (  \delta_{\alpha}^{b} c_{, \bar{\nu}}^{\alpha}   \right ) + \delta_{\nu}^{b} \left(  \delta_{\alpha}^{a} c_{, \bar{\mu}}^{\alpha} \right)\right]^{, \bar{\mu}} 
     -2 \sigma   [ \delta_{\mu}^{a}  (  \delta_{\alpha}^{b} c^{\alpha , \bar{\mu} } ) ]_{, \bar{\nu}}  \big \} 
  \\ & \quad  
+ c^{\star \nu}  \big \{ \eta_{ab} \left [ \delta_{\mu}^{a} \left ( \tensor{C}{^{b}_{p}}\delta_{\nu}^{p} \right )+ \delta_{\nu}^{b} \left ( \tensor{C}{^{a}_{p}}\delta_{\mu}^{p}\right )\right ]^{, \bar{\mu}} - 2 \sigma \left [\delta_{\mu}^{a} \left ( \tensor{C}{^{b}_{p}}e^{p \, \mu }\right)  \right ]_{, \bar{\nu}}  \big\} 
-C^{\star ab}  \tensor{C}{_{ab , \mu }^{, \bar{\mu} }}.
    \end{split}
\end{equation}
However, note that the second term in above equation can be simplified to
\begin{equation}\label{eq:appDm2}
    c^{\star \nu}  \big \{  \left [ \delta_{\mu}^{a} \left ( \tensor{C}{_{ap}}\delta_{\nu}^{p} \right )+ \delta_{\nu}^{b} \left ( \tensor{C}{_{bp}}\delta_{\mu}^{p}\right )\right ]^{, \bar{\mu}} - 2 \sigma  \left [\delta_{\mu}^{a} \left ( \tensor{C}{_{ap}}e^{p \, \mu }\right)  \right ]_{, \bar{\nu}}  \big\} = c^{\star \nu} ( \tensor{C}{_{\mu \nu}^{, \mu}} + \tensor{C}{_{\nu \mu}^{, \mu}} - 2 \sigma \tensor{C}{_{\mu}^{\mu}_{, \nu}})
\end{equation}
which vanishes, since $ C_{ab} = - C_{ba} $. Then, we see that, in this order, the ghost fields decouple. There is no mixed ghost propagators. 

Using Eq.~\eqref{eq:appDm2} in Eq.~\eqref{eq:ccf211bilinear}, we have 
\begin{equation}\label{eq:appDm3}
    \begin{split}
     & -c^{\star \nu}    
     \big [   \eta_{\mu \nu}  \partial^{\pi} \partial_{\pi} +(1- 2 \sigma)     \partial_{\nu} \partial_{\mu} \big ] 
     c^{\mu} 
     -C^{\star ab}  \partial_{\pi} \partial^{\pi} I_{ablm} \tensor{C}{^{lm}}.
\end{split}
\end{equation}
The propagator $ (P_{{C}^{\star} C})_{ab lm}  \equiv \left \langle {C}^{\star}_{ab} C_{lm}\right\rangle $ of the ghost field $ {C}_{ab} $ is simply given by 
\begin{equation}\label{eq:appDM4}
    -\frac{i}{p^{2}} I_{ablm}.
\end{equation}
While the propagator $ (P_{{c}^{\star} c})_{\mu \nu}  \equiv  \left \langle {c}^{\star}_{\mu} c_{\nu}\right\rangle $  is 
\begin{equation}\label{eq:appDM5}
    -\frac{i}{p^{2}} \left ( \eta_{\mu \nu} + \frac{1}{2} \frac{1- 2 \sigma}{\sigma -1} \frac{p_{\mu} p_{\nu}}{p^{2}}\right ).
\end{equation}
When $ \sigma =1/2$ (in de Donder-like gauge), it simplifies to $ - i \eta_{\mu \nu} / p^{2} $.

\subsection{Vertices}\label{section:vertices}

To derive the vertices, we need to consider the expansion \eqref{eq:bm1} in several terms. In the determinant $ e$, we obtain that
\begin{equation}\label{eq:appV01}
    e = \bar{e} \left[ 1 + \frac{\kappa}{2} \tensor{h}{_{\nu}^{\nu}} + O ( \kappa^{2} )\right], 
\end{equation}
where 
\begin{equation}\label{eq:appV02}
    \tensor{h}{_{\nu}^{\nu}}= h_{\mu \nu} \bar{g}^{\mu \nu} = \eta_{ab} ( \tensor{\bar{e}}{^{a}_{\mu}} \tensor{q}{^{b}_{\nu}} + \tensor{\bar{e}}{^{b}_{\nu}} \tensor{q}{^{a}_{\mu}} + \kappa \tensor{q}{^{a}_{\mu}} \tensor{q}{^{b}_{\nu}} ) ( \eta^{\mu \nu} - \kappa \bar{h}^{\mu \nu} + O (\kappa^{2} )).
\end{equation}
Thus, 
\begin{equation}\label{eq:appV03}
    \begin{split}
        e ={}&\bar{e} \left[ 1 +  \kappa \eta_{ab} \tensor{\bar{e}}{^{a}_{\mu}} q^{b}_{\ \nu} \bar{g}^{\mu \nu}  + O ( \kappa^{2} )\right] \\
        ={}&  1 +  \kappa \eta_{ab} \delta_{\mu}^{a} \tensor{\bar{q}}{^{b \mu}}   +  \kappa \eta_{ab} \delta_{\mu}^{a} \tensor{q}{^{b}_{\nu}} \eta^{\mu \nu} 
        - \frac{\kappa^{2}}{2} \eta_{ab} ( \delta^{a}_{c} \tensor{q}{^{b}_{\nu}} \bar{q}^{\nu \, c } + \delta_{\mu}^{a} \delta_{c}^{\nu} \bar{q}^{\mu \, c}\tensor{q}{^{b}_{\nu}} + \delta_{\nu}^{b} \delta_{c}^{\mu} \tensor{q}{^{a}_{\mu}} \bar{q}^{\nu \, c} + \delta_{c}^{b} \tensor{q}{^{a}_{\mu}} \bar{q}^{\mu \, c  } ) 
        \\ & 
        + \frac{\kappa^{2}}{2} \eta_{ab} \eta^{\mu \nu} ( 
 \tensor{\bar{q}}{^{a}_{\mu}} \tensor{q}{^{b}_{\nu}} + \tensor{\bar{q}}{^{b}_{\nu}} \tensor{q}{^{a}_{\mu}} ) 
 + \frac{\kappa^{2} }{4} \eta_{ab} \eta_{cd} \eta^{\mu \nu} \eta^{\alpha \beta} (
 \tensor{\bar{q}}{^{a}_{\mu}} {\delta }{^{b}_{\nu}} + \tensor{\bar{q}}{^{b}_{\nu}} {\delta}{^{a}_{\mu}} 
 )(
 {\delta }{^{c}_{\alpha}} \tensor{q}{^{d}_{\beta}} + {\delta}{^{d}_{\beta}} \tensor{q}{^{c}_{\alpha}} 
 ) 
        + O ( \kappa^{3} ).
\end{split}
\end{equation}
We have omitted nonlinear terms on the quantum field $ q^{a \mu} $. 

Moreover, using the relation $ \tensor{e}{^{a}_{\mu}} \tensor{e}{_{a}^{\nu}} = \delta_{\mu}^{\nu} $, we find that the inverse tetrad $ \tensor{e}{_{a}^{\mu}}$ is given by 
\begin{equation}\label{eq:appV6}
    \tensor{e}{_{a}^{\mu}} = \tensor{\bar{e}}{_a^{\mu}} - \kappa \tensor{q}{^{\mu}_a} + \cdots = \delta^{\mu}_{a} - \kappa \tensor{\bar{q}}{^{\mu}_{a}} - \kappa \tensor{q}{_{\nu}_a} \eta^{\mu \nu}  + \kappa^{2} \tensor{( \bar{q} + q )}{^{\mu}_\gamma } \tensor{(\bar{q} + q)}{^{\gamma}_a}  + O (\kappa^{3} ),
\end{equation}
which appears in the action \eqref{e7}.

In the following derivation, we assume that all the external momenta are inward. We also remember that curly lines represents the background tetrad field $ \bar{q} $. The wavy and double wave lines represent respectively the quantum fields $q$ and $Q$. The ghost fields $c$, $C$ by dotted and double dotted lines. 

\subsubsection{Vertex $ qq \bar{q} $}\label{section:vertexqqbarq}

Using Eq.~\eqref{eq:appV03}, we obtain the first contribution to the vertex $ qq \bar{q} $ that comes from the gauge fixing term 
\begin{equation}\label{eq:}
    - \bar{e} \frac{\kappa}{2\xi}
    \bar{g}^{\nu \beta} \eta_{ab} \eta_{cd} 
    \bar{g}^{\mu \gamma} \bar{g}^{\alpha \omega}  [ (\tensor{\bar{e}}{^{a}_{\mu}} \tensor{q}{^{b}_{\nu}} + \tensor{q}{^{a}_{\mu}} \tensor{\bar{e}}{^{b}_{\nu}})_{; \bar{\gamma}} -2 \sigma (\tensor{\bar{e}}{^{b}_{\gamma} } \tensor{q}{^{a}_{\mu}} )_{; \bar{\nu} }]
    [ (\tensor{\bar{e}}{^{c}_{\alpha}} \tensor{q}{^{d}_{\beta}} + \tensor{q}{^{c}_{\alpha}} \tensor{\bar{e}}{^{d}_{\beta}})_{; \bar{\omega}} -2 \sigma (\tensor{\bar{e}}{^{d}_{\omega}} \tensor{q}{^{c}_{\alpha}} )_{; \bar{\beta} }]
     ,
\end{equation}
which is
\begin{equation}\label{eq:ccf210exp}
    - \frac{\kappa}{2\xi}\eta_{uv} \delta_{\pi}^{u} \bar{q}^{v \, \pi} 
\eta^{\nu \beta} \eta_{ab} [ (\delta_{\mu}^{a} \tensor{q}{^{b}_{\nu}} + \tensor{q}{^{a}_{\mu}} \delta_{\nu}^{b})^{, {\mu}} - (\delta _{\tau}^{b} \tensor{q}{^{a}_{ \lambda}} \eta^{\lambda \tau}  )_{, {\nu} }] 
\eta_{cd} [ (\delta_{\alpha}^{c} \tensor{q}{^{d}_{\beta}} + \tensor{q}{^{c}_{\alpha}} \delta_{\beta}^{d})^{, {\alpha}} - (\delta _{\pi}^{d} q^{c}_{\omega} \eta^{\omega \pi}  )_{, {\beta} }], 
\end{equation}
where we have set $ \sigma =1/2$. 

Now, let us consider the contributions due to the expansion \eqref{eq:bm1} in
\begin{equation}\label{eq:ccf210exp2a}
    \begin{split}
        \mathcal{L}_{11\bar{1}}^{(1)}  ={}& 
    -  \frac{\kappa}{2\xi}
    \bar{g}^{\nu \beta} \eta_{ab} \eta_{cd} 
    \bar{g}^{\mu \gamma} \bar{g}^{\alpha \omega}  [ (\tensor{\bar{e}}{^{a}_{\mu}} \tensor{q}{^{b}_{\nu}} + \tensor{q}{^{a}_{\mu}} \tensor{\bar{e}}{^{b}_{\nu}})_{, \bar{\gamma}} -2 \sigma (\tensor{\bar{e}}{^{b}_{\gamma} } \tensor{q}{^{a}_{\mu}} )_{, \bar{\nu} }]
    [ (\tensor{\bar{e}}{^{c}_{\alpha}} \tensor{q}{^{d}_{\beta}} + \tensor{q}{^{c}_{\alpha}} \tensor{\bar{e}}{^{d}_{\beta}})_{, \bar{\omega}} -2 \sigma (\tensor{\bar{e}}{^{d}_{\omega}} \tensor{q}{^{c}_{\alpha}} )_{, \bar{\beta} }]
        \\
        ={}&
        - \frac{1}{2 \xi} \eta_{ab} \eta_{cd} \eta^{\nu \beta} 
    [ \tensor{\bar{e}}{^{a}_{\mu}^{, \mu}} \tensor{q}{^{b}_{\nu}} + \tensor{\bar{e}}{^{a}_{\mu}} \tensor{q}{^{b}_{\nu}^{, \mu}} + \tensor{q}{^{a}_{\mu}^{, \mu}} \tensor{\bar{e}}{^{b}_{\nu}} + \tensor{q}{^{a}_{\mu}} \tensor{\bar{e}}{^{b}_{\nu}^{, \mu}}  - \bar{e}_{\tau \, , \nu }^{a }  q^{b \, \tau } - \tensor{\bar{e}}{^{a}_{\tau}} q^{b \, \tau }_{,\nu}  ]
        \\ & \qquad \qquad \qquad \quad \times [
 \tensor{\bar{e}}{^{c}_{\alpha}^{, \alpha}} \tensor{q}{^{d}_{\beta}} + \tensor{\bar{e}}{^{c}_{\alpha}} \tensor{q}{^{d}_{\beta}^{, \alpha}} + \tensor{q}{^{c}_{\alpha}^{, \alpha}} \tensor{\bar{e}}{^{d}_{\beta}} + \tensor{q}{^{c}_{\alpha}} \tensor{\bar{e}}{^{d}_{\beta}^{, \alpha}}  - \bar{e}_{\pi \, , \beta }^{c }  q^{d \, \pi } - \tensor{\bar{e}}{^{c}_{\pi}} q^{d \, \pi }_{,\beta}  
        ]
        \\ & 
        + \frac{\kappa}{2 \xi} \eta_{ab} \eta_{cd} \eta_{ef} \eta^{\mu \gamma} \eta^{\alpha \omega}   [\delta^{e \nu} \bar{q}^{\beta f} + \delta^{f \beta} \bar{q}^{\nu e} ] 
[ (\delta_{\mu}^{a} \tensor{q}{^{b}_{\nu}} + \tensor{q}{^{a}_{\mu}} \delta_{\nu}^{b})_{, {\gamma}} - (\delta _{\gamma}^{b} \tensor{q}{^{a}_{ \mu}} )_{, {\nu} }]
[ (\delta_{\alpha}^{c} \tensor{q}{^{d}_{\beta}} + \tensor{q}{^{c}_{\alpha}} \delta_{\beta}^{d})_{, {\omega}} - (\delta _{\omega}^{d} \tensor{q}{^{c}_{ \alpha}}   )_{, {\beta} }]   
        \\
           & 
           + \frac{\kappa}{2 \xi} \eta_{ab} \eta_{cd} \eta_{ef} \eta^{\nu \beta} \eta^{\alpha \omega} [\delta^{e \mu} \bar{q}^{\gamma f} + \delta^{f \gamma} \bar{q}^{\mu e} ] 
[ (\delta_{\mu}^{a} \tensor{q}{^{b}_{\nu}} + \tensor{q}{^{a}_{\mu}} \delta_{\nu}^{b})_{, {\gamma}} - (\delta _{\gamma}^{b} \tensor{q}{^{a}_{ \mu}} )_{, {\nu} }]
[ (\delta_{\alpha}^{c} \tensor{q}{^{d}_{\beta}} + \tensor{q}{^{c}_{\alpha}} \delta_{\beta}^{d})_{, {\omega}} - (\delta _{\omega}^{d} \tensor{q}{^{c}_{ \alpha}}   )_{, {\beta} }]   
           \\
           & 
+ \frac{\kappa}{2 \xi} \eta_{ab} \eta_{cd} \eta_{ef} \eta^{\nu \beta} \eta^{\mu  \gamma} [\delta^{e \alpha} \bar{q}^{\omega f} + \delta^{f \omega } \bar{q}^{\alpha e} ] 
[ (\delta_{\mu}^{a} \tensor{q}{^{b}_{\nu}} + \tensor{q}{^{a}_{\mu}} \delta_{\nu}^{b})_{, {\gamma}} - (\delta _{\gamma}^{b} \tensor{q}{^{a}_{ \mu}} )_{, {\nu} }]
[ (\delta_{\alpha}^{c} \tensor{q}{^{d}_{\beta}} + \tensor{q}{^{c}_{\alpha}} \delta_{\beta}^{d})_{, {\omega}} - (\delta _{\omega}^{d} \tensor{q}{^{c}_{ \alpha}}   )_{, {\beta} }].
    \end{split}
\end{equation}

The part of the vertex $qq \bar{q} $ coming from Eq.~\eqref{eq:ccf210exp2a} is given by 
\begin{equation}\label{eq:neweq1}
    \begin{split}
        - \frac{1}{2 \xi} \eta_{ab} \eta_{cd} \eta^{\nu \beta} 
    \frac{\delta^{3}  \mathcal{L}_{1 1 \bar{1}}^{(1)}}{ \delta \tensor{q}{^{l}_{\rho}} (x_1)\delta \tensor{q}{^{m}_{\gamma}} (x_2) \delta \bar{q}^{v \, \theta} ( x_{3} )}   \Big  |_{\bar{q}=q =0} \end{split},
\end{equation}
which, in momentum space, becomes
\begin{equation}\label{eq:appV15a}
    \begin{split}
        & \frac{\kappa }{2 \xi} \eta_{ab} \eta_{cd} \eta^{\nu \beta} 
        \{ [p_{3}^{\mu} \eta_{\theta \mu} \delta_{v}^{a} \delta_{\nu}^{\gamma} \delta_{m}^{b} + p_{2}^{\mu} \eta_{\mu \theta} \delta_{v}^{a} \delta_{\nu}^{\gamma} \delta_{m}^{b} + p_{2}^{\mu} \delta_{\mu}^{\gamma} \delta_{m}^{a} \delta_{v}^{b} \delta_{\nu \theta} + p_{3}^{\mu} \delta_{\mu}^{\gamma} \delta_{m}^{a} \delta_{v}^{b} \eta_{\nu \theta} - p_{3 \nu } \eta_{\tau \theta} \delta_{v}^{a} \delta_{m}^{b} \eta^{\tau \gamma} - p_{2 \nu} \delta_{v}^{a} \eta_{\tau \theta} \delta_{m}^{b} \eta^{\tau \gamma }]
        \\ & \qquad \qquad \qquad \times 
        [p_{1}^{\alpha} \delta_{\alpha}^{c} \delta_{\beta}^{\rho} \delta_{l}^{d} + p_{1}^{\alpha} \delta_{\alpha}^{\rho} \delta_{l}^{c} \delta_{\beta}^{d} - p_{1 \beta} \delta_{\pi}^{c} \delta_{l}^{d} \eta^{\pi \rho} ]
        + 
        [ p_{2}^{\mu} \delta_{\mu}^{a} \delta_{\nu}^{\gamma} \delta_{m}^{b}  + p_{2}^{\mu} \delta_{\mu}^{\gamma} \delta_{m}^{a} \delta_{\nu}^{b} - p_{2 \nu} \delta_{\tau}^{a} \delta_{m}^{b} \eta^{\tau \gamma} ]
        \\ & 
        \qquad \qquad \qquad 
        \times 
        [
        p_{3}^{\alpha} \delta_{v}^{c} \eta_{\alpha \theta} \delta_{\beta}^{\rho} \delta_{l}^{d} + p_{1}^{\alpha} \delta_{v}^{c} \eta_{\alpha \theta} \delta_{\beta}^{\rho} \delta_{l}^{d} + p_{1}^{\alpha} \delta_{\alpha}^{\rho} \delta_{l}^{c} \delta_{v}^{d} \eta_{\beta \theta} + p_{3}^{\alpha} \delta_{\alpha}^{\rho} \delta_{l}^{c} \delta_{v}^{d} \eta_{\beta \theta} - p_{3 \beta} \delta_{v}^{c} \eta_{\pi \theta} \delta_{l}^{d} \eta^{\pi \rho} - p_{1 \beta} \delta_{v}^{c} \eta_{\pi \theta} \delta_{l}^{d} \eta^{\pi \rho}  
    ]\}
        \\ & 
        - \frac{\kappa}{2 \xi} \eta_{ab} \eta_{cd} \eta_{ef} \{ [ \delta^{e \nu} \delta_{v}^{\beta} \delta_{\theta}^{f} + \delta^{f \beta} \delta_{v}^{\nu} \delta_{\theta}^{e} ]
[p_{1}^{\alpha} \delta_{\alpha}^{c} \delta_{\beta}^{\rho} \delta_{l}^{d} 
        + p_{1}^{\alpha} \delta_{\alpha}^{\rho} \delta_{l}^{c} \delta_{\beta}^{d} - p_{1 \beta} \delta_{\pi}^{c} \delta_{l}^{d} \eta^{\pi \rho} 
        ]
[ p_{2}^{\mu} \delta_{\mu}^{a} \delta_{\nu}^{\gamma} \delta_{m}^{b}  + p_{2}^{\mu} \delta_{\mu}^{\gamma} \delta_{m}^{a} \delta_{\nu}^{b} - p_{2 \nu} \delta_{\tau}^{a} \delta_{m}^{b} \eta^{\tau \gamma} ]
\\
           & \quad + \eta^{\nu \beta} \eta^{\alpha \omega} (\delta^{e \mu} \delta_{v}^{\gamma } \delta_{\theta}^{f} + \delta^{f \gamma} \delta_{v}^{\mu} \delta_{\theta}^{e}  ) 
        [ p_{1}^{\alpha} ( \delta_{\alpha}^{c} \delta_{l}^{d} \delta_{\beta}^{\rho} + \delta_{l}^{c} \delta_{\alpha}^{\rho} \delta_{\beta}^{d} ) - p_{1 \beta} \delta_{\pi}^{d}  \delta_{l}^{c} \eta^{\pi \rho} ] 
     [ p_{2}^{\mu} ( \delta_{\mu}^{a} \delta_{m}^{b} \delta_{\nu}^{\gamma} + \delta_{m}^{a} \delta_{\mu}^{\gamma} \delta_{\nu}^{b} ) - p_{2 \nu} \delta_{\tau}^{b} \delta_{m}^{a} \eta^{\tau \gamma} ]  
        \\ &  \quad + \eta^{\nu \beta} \eta^{\mu \gamma} ( \delta^{e \alpha} \delta_{v}^{\omega} \delta_{\theta}^{f} + \delta^{f \omega}  \delta_{v}^{\alpha} \delta_{\theta}^{e} )
 [ p_{1}^{\alpha} ( \delta_{\alpha}^{c} \delta_{l}^{d} \delta_{\beta}^{\rho} + \delta_{l}^{c} \delta_{\alpha}^{\rho} \delta_{\beta}^{d} ) - p_{1 \beta} \delta_{\pi}^{d}  \delta_{l}^{c} \eta^{\pi \rho} ] 
[ p_{2}^{\mu} ( \delta_{\mu}^{a} \delta_{m}^{b} \delta_{\nu}^{\gamma} + \delta_{m}^{a} \delta_{\mu}^{\gamma} \delta_{\nu}^{b} ) - p_{2 \nu} \delta_{\tau}^{b} \delta_{m}^{a} \eta^{\tau \gamma} ]  \}
        \\ & + (p_1, l, \rho ) \leftrightarrow (p_2, m, \gamma )
        .
    \end{split}
\end{equation}

Finally, we have the interactions terms coming from the background covariant derivatives 
\begin{align}\label{eq:covdev}
    V_{\alpha ; \bar{\mu}}^{a }   ={}& V_{\alpha , \mu }^{a} -  \kappa \tensor{\bar{e}}{_{u}^{\beta}}    \partial_{\mu} \tensor{\bar{q}}{^{u}_{\alpha}} V_{\beta}^{a}  - \tensor{\bar{e}}{_{u}^{\beta}}  \tensor{\bar{e}}{^{v}_{\alpha}}    \tensor{\bar{\omega}}{_{\mu}^{u}_{v}} V_{\beta}^{a} 
    \end{align}
which lead to the contributions to the vertices:
\begin{equation}\label{eq:ccf210expCD}
    \begin{split}
 & - \frac{1}{2 \xi} \eta_{ab} \eta_{cd} \eta^{\nu \beta} 
          \frac{\delta^{3} }{ \delta \tensor{q}{^{l}_{\rho}} (x_1)\delta \tensor{q}{^{m}_{\gamma}} (x_2) \delta \bar{q}^{v \, \theta} ( x_{3} )}      
          [ \delta_{\mu}^{a \, ; \mu} \tensor{q}{^{b}_{\nu}} + \tensor{\bar{e}}{^{a}_{\mu}} \tensor{q}{^{b}_{\nu}^{: \mu}} + \tensor{q}{^{a}_{\mu}^{: \mu}} \tensor{\bar{e}}{^{b}_{\nu}} + \tensor{q}{^{a}_{\mu}} \delta_{\nu}^{b \, ; \mu}  - \delta_{\tau \, ; \nu }^{a }  q^{b \, \tau } - \tensor{\bar{e}}{^{a \tau}} \tensor{q}{^{b}_{ \tau }_{:\nu} } ]
        \\ & \qquad \qquad \qquad \qquad \qquad \qquad \qquad \qquad \qquad \times 
[
\tensor{\bar{e}}{^{c}_{\alpha}^{, \alpha}} \tensor{q}{^{d}_{\beta}} + \tensor{\bar{e}}{^{c}_{\alpha}} \tensor{q}{^{d}_{\beta}^{, \alpha}} + \tensor{q}{^{c}_{\alpha}^{, \alpha}} \tensor{\bar{e}}{^{d}_{\beta}} + \tensor{q}{^{c}_{\alpha}} \tensor{\bar{e}}{^{d}_{\beta}^{, \alpha}}  - \bar{e}_{\pi \, , \beta }^{c }  q^{d \, \pi } - \tensor{\bar{e}}{^{c}_{\pi}} q^{d \, \pi }_{,\beta} ]
 \Big |_{\bar{q} = q =0}
        \\ & 
- \frac{1}{2 \xi} \eta_{ab} \eta_{cd} \eta^{\nu \beta} 
          \frac{\delta^{3} }{ \delta \tensor{q}{^{l}_{\rho}} (x_1)\delta \tensor{q}{^{m}_{\gamma}} (x_2) \delta \bar{q}^{v \, \theta} ( x_{3} )}
        [ \tensor{\bar{e}}{^{a}_{\mu}^{, \mu}} \tensor{q}{^{b}_{\nu}} + \tensor{\bar{e}}{^{a}_{\mu}} \tensor{q}{^{b}_{\nu}^{, \mu}} + \tensor{q}{^{a}_{\mu}^{, \mu}} \tensor{\bar{e}}{^{b}_{\nu}} + \tensor{q}{^{a}_{\mu}} \tensor{\bar{e}}{^{b}_{\nu}^{, \mu}}  - \bar{e}_{\tau \, , \nu }^{a }  q^{b \, \tau } - \tensor{\bar{e}}{^{a}_{\tau}} q^{b \, \tau }_{,\nu}  ] 
        \\ & \qquad \qquad \qquad \qquad \qquad \qquad \qquad \qquad \qquad \times 
[
\delta_{\alpha}^{c \, ; \alpha} \tensor{q}{^{d}_{\beta}} + \tensor{\bar{e}}{^{c}_{\alpha}} \tensor{q}{^{d}_{\beta}^{: \alpha}} + \tensor{q}{^{c}_{\alpha}^{: \alpha}} \tensor{\bar{e}}{^{d}_{\beta}} + \tensor{q}{^{c}_{\alpha}} \delta_{\beta}^{d \, ;\alpha}  - \delta_{\pi \, ; \beta }^{c }  q^{d \, \pi } - \tensor{\bar{e}}{^{c \pi}} \tensor{q}{^{d}_{ \pi }_{:\beta}}  
 ] 
 \Big |_{\bar{q} = q =0}.
    \end{split}
\end{equation}
where $ q_{: \mu} \equiv  q_{; \mu} - q_{, \mu} $. Note that, $ \delta_{\alpha; \mu }^{a} $ does not vanishes. By omitting $ \bar{\omega}_{\mu ab} $ and taking the linear part in $ \tensor{\bar{q}}{^{a}_{\mu}} $, it is given by $ - \kappa \delta_{u}^{\beta} \partial_{\mu} \tensor{\bar{q}}{^{u}_{\alpha}} \delta_{\beta}^{a} = - \kappa \partial_{\mu} \tensor{\bar{q}}{^{a}_{\alpha}}  $. Using the same conditions, we get that $ q_{ \alpha : \mu}^{a} = - \kappa \delta_{u}^{\lambda} \partial_{\mu} \tensor{\bar{q}}{^{u}_{\alpha}} q_{\lambda }^{a}$. 

The contributions in Eq.~\eqref{eq:ccf210expCD}, in momentum space, reads
\begin{equation}\label{eq:verticecov110}
    \begin{split}
& - \frac{\kappa}{2 \xi} \eta_{ab} \eta_{cd} \eta^{\nu \beta}
 \{ [p_{3}^{\mu} \delta_{v}^{a} \eta_{\mu \theta} \delta_{\nu}^{\gamma} \delta_{m}^{b} + p_{3}^{\mu} \delta_{\mu}^{a} \delta_{u}^{\lambda} \eta_{\mu \theta} \delta_{v}^{u} \delta_{\lambda}^{\gamma } \delta_{m}^{b} + p_{3}^{\mu} \delta_{u}^{\lambda} \delta_{v}^{u} \eta_{\nu \theta} \delta_{m}^{a} \delta_{\lambda}^{ \gamma} \delta_{\nu}^{b} + p_{3}^{\mu} \delta_{\mu}^{\gamma} \delta_{m}^{a} \delta_{v}^{b} \eta_{\nu \theta} \\ 
           &  \quad \quad \quad \quad \quad \quad - p_{3 \nu} \delta_{v}^{a} \eta_{\tau \theta} \delta_{m}^{b} \eta^{\tau \gamma} 
           -p_{3 \nu} \delta_{\tau}^{a} \delta^{\tau}_{u} \delta_v^{u} \eta_{\lambda \theta} \delta_{m}^{a} \eta^{\lambda \gamma}  ]
        [p_{1}^{\alpha} \delta_{\alpha}^{c} \delta_{\beta}^{\rho} \delta_{l}^{d} 
        + p_{1}^{\alpha} \delta_{\alpha}^{\rho} \delta_{l}^{c} \delta_{\beta}^{d} - p_{1 \beta} \delta_{\pi}^{c} \delta_{l}^{d} \eta^{\pi \rho} 
        ]
        \\ & +  
        [ p_{2}^{\mu} \delta_{\mu}^{a} \delta_{\nu}^{\gamma} \delta_{m}^{b}  + p_{2}^{\mu} \delta_{\mu}^{\gamma} \delta_{m}^{a} \delta_{\nu}^{b} - p_{2 \nu} \delta_{\tau}^{a} \delta_{m}^{b} \eta^{\tau \gamma} ] [ p_{3}^{\alpha} \delta_{v}^{c} \eta_{\alpha \theta} \delta_{\beta}^{\rho} \delta_{l}^{d} + p_{3}^{\alpha} \delta_{\alpha}^{c} \delta_{u}^{\lambda} \delta_{v}^{u} \eta_{\beta \theta} \delta_{l}^{d} \delta_{\lambda}^{\rho} + p_{3}^{\alpha} \delta_{u}^{\lambda} \delta_{v}^{u} \eta_{\alpha \theta} \delta_{\lambda}^{\rho} \delta_{l}^{c} \delta_{\beta}^{d} 
        + p_{3}^{\alpha} \delta_{\alpha}^{\rho} \delta_{l}^{c} \delta_{v}^{d} \eta_{\beta \theta}
        \\ & 
     \quad \quad    
 - p_{3 \beta} \delta_{v}^{c} \eta_{\pi \theta} \delta_{l}^{d} \eta^{\pi \rho} - p_{3 \beta} \delta_{\pi}^{c} \delta_{u}^{\pi}  \delta_{v}^{u} \eta_{\lambda \theta} \delta_{l}^{d} \eta^{\lambda \rho}]\}
        \\ & + (p_1, l, \rho ) \leftrightarrow (p_2, m, \gamma )
     . 
    \end{split}
\end{equation}
end{equation}

Thus, collecting all contributions, one can find that the vertex $qq \bar{q}$ is given by
\begin{equation}
    \label{fig:FRqqq}
    \vcenter{\hbox{\includegraphics[scale=1]{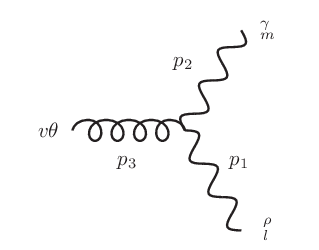}}} 
    \quad \quad 
i\tensor{[V_{11 \bar{1}}]}{_{l}^{\rho }_{m}^{\gamma}_{v \theta }} 
(p_1,p_2,p_3),
\end{equation}
where 
\begin{equation}\label{eq:appV08}
    \begin{split}
    \allowdisplaybreaks
        \tensor{[V_{11 \bar{1}}]}{_{l}^{\rho }_{m}^{\gamma}_{v \theta }} 
(p_1,p_2,p_3)
       ={}& 
       \frac{\kappa}{\xi}
       [
-{p}_1{}_l {p}_2{}_m \delta_{v}^{\rho} \delta^{\gamma}_{\theta}-{p}_3{}_l
   {p}_2{}_m \delta_{v}^{\rho} \delta^{\gamma}_{\theta}+\delta_{l}^{\rho} {p}_2{}_m
   {p}_1{}_v \delta^{\gamma}_{\theta}-{p}_1{}_l \delta_{m}^{\rho} {p}_2{}_v
   \delta^{\gamma}_{\theta}
       \\ & +\delta_{l}^{\rho} {p}_1{}_m {p}_2{}_v \delta^{\gamma \theta
   }
    -\delta^{lm} {p}_2{}_v {p}_1{}^{\rho } \delta^{\gamma}_{\theta}+{p}_1{}_l
   {p}_2{}_m \eta_{v \theta} \eta^{\gamma \rho}-\delta_{l \theta} {p}_2{}_m
   {p}_1{}_v \eta^{\gamma \rho}-{p}_1{}_l \delta_{m \theta} {p}_2{}_v
   \eta^{\gamma \rho} 
   \\ & -\delta_{l}^{\rho} {p}_2{}_m \eta_{v \theta} {p}_1{}^{\gamma
   }+\delta_{l \theta} {p}_2{}_m \delta_{v}^{\rho} {p}_1{}^{\gamma }+\delta_{l}^{\rho}
   \delta_{m \theta} {p}_2{}_v {p}_1{}^{\gamma }+{p}_1{}_l \delta_{m}^{\rho}
   \eta_{v \theta} {p}_2{}^{\gamma }-\delta_{l}^{\rho} {p}_1{}_m \eta_{v \theta}
   {p}_2{}^{\gamma } 
   \\ & -{p}_1{}_l \delta_{m \theta} \delta_{v}^{\rho} {p}_2{}^{\gamma
   }-{p}_3{}_l \delta_{m \theta} \delta_{v}^{\rho} {p}_2{}^{\gamma }+\delta_{l \theta}
   {p}_1{}_m \delta_{v}^{\rho} {p}_2{}^{\gamma }+\delta_{l \theta} {p}_3{}_m
   \delta_{v}^{\rho} {p}_2{}^{\gamma }
    +\delta_{l}^{\rho} \delta_{m \theta} {p}_1{}_v
    {p}_2{}^{\gamma } \\ & -\delta_{l \theta} \delta_{m}^{\rho} {p}_1{}_v {p}_2{}^{\gamma
   }+\delta_{l \theta} {p}_2{}_m \delta_{v}^{\rho} {p}_3{}^{\gamma }-{p}_1{}_l
   {p}_2{}_m \delta_{v}^{\gamma} \delta^{\rho}_{\theta}-{p}_1{}_l {p}_3{}_m
   \delta_{v}^{\gamma} \delta^{\rho}_{\theta}  +{p}_2{}_l \delta_{m}^{\gamma} {p}_1{}_v
   \delta^{\rho}_{\theta} \\ & -\delta_{l}^{\gamma} {p}_2{}_m {p}_1{}_v \delta^{\theta \rho
   }+{p}_1{}_l \delta_{m}^{\gamma} {p}_2{}_v \delta^{\rho}_{\theta}-\delta^{lm}
   {p}_1{}_v {p}_2{}^{\gamma } \delta^{\rho}_{\theta} +\delta_{l}^{\rho} {p}_2{}_m
   \delta_{v}^{\gamma} {p}_1{}_{\theta}+\delta_{l}^{\rho} {p}_3{}_m \delta_{v}^{\gamma}
   {p}_1{}_{\theta} \\ & +{p}_2{}_l \delta_{m}^{\gamma} \delta_{v}^{\rho} {p}_1{}^{\theta
   }-\delta_{l}^{\gamma} {p}_2{}_m \delta_{v}^{\rho} {p}_1{}_{\theta}-\delta_{l}^{\rho}
   \delta_{m}^{\gamma} {p}_2{}_v {p}_1{}_{\theta}-\delta^{lm} \delta_{v}^{\rho}
   {p}_2{}^{\gamma } {p}_1{}_{\theta}   -{p}_1{}_l \delta_{m}^{\rho} \delta_{v}^{\gamma}
   {p}_2{}_{\theta}\\ & +\delta_{l}^{\rho} {p}_1{}_m \delta_{v}^{\gamma} {p}_2{}^{\theta
   }+{p}_1{}_l \delta_{m}^{\gamma} \delta_{v}^{\rho} {p}_2{}_{\theta}+{p}_3{}_l
   \delta_{m}^{\gamma} \delta_{v}^{\rho} {p}_2{}_{\theta}-\delta_{l}^{\rho} \delta_{m}^{\gamma}
   {p}_1{}_v {p}_2{}_{\theta}-{p}_1{}_l \delta_{m}^{\rho} \delta_{v}^{\gamma}
   {p}_3{}_{\theta} 
   \\ & +\delta_{l}^{\rho} {p}_1{}_m \delta_{v}^{\gamma} {p}_3{}^{\theta
   }+{p}_2{}_l \delta_{m}^{\gamma} \delta_{v}^{\rho} {p}_3{}_{\theta}-\delta_{l}^{\gamma}
   {p}_2{}_m \delta_{v}^{\rho} {p}_3{}_{\theta}-\delta^{lm} \delta_{v}^{\rho}
   {p}_2{}^{\gamma } {p}_3{}_{\theta}+{p}_2{}_l \delta_{m \theta} \delta_{v}^{\gamma}
   {p}_1{}^{\rho }
   \\ & +{p}_3{}_l \delta_{m \theta} \delta_{v}^{\gamma} {p}_1{}^{\rho
   }-\delta_{l \theta} {p}_2{}_m \delta_{v}^{\gamma} {p}_1{}^{\rho }-\delta_{l \theta}
   {p}_3{}_m \delta_{v}^{\gamma} {p}_1{}^{\rho }-{p}_2{}_l \delta_{m}^{\gamma}
   \eta_{v \theta} {p}_1{}^{\rho }+\delta_{l}^{\gamma} {p}_2{}_m \eta_{v \theta}
   {p}_1{}^{\rho }\\ & +\delta_{l \theta} \delta_{m}^{\gamma} {p}_2{}_v {p}_1{}^{\rho
   }-\delta_{l}^{\gamma} \delta_{m \theta} {p}_2{}_v {p}_1{}^{\rho }+\delta^{lm}
   \eta_{v \theta} {p}_2{}^{\gamma } {p}_1{}^{\rho }-\delta^{lm} \delta_{v}^{\gamma}
   {p}_2{}_{\theta} {p}_1{}^{\rho }-\delta^{lm} \delta_{v}^{\gamma} {p}_3{}_{\theta}
   {p}_1{}^{\rho }
   \\ & +{p}_1{}_l \delta_{m \theta} \delta_{v}^{\gamma} {p}_2{}^{\rho
   }
    -{p}_1{}_l \delta_{m}^{\gamma} \eta_{v \theta} {p}_2{}^{\rho }
    +\delta_{l \theta}
   \delta_{m}^{\gamma} {p}_1{}_v {p}_2{}^{\rho }+{p}_1{}_l \delta_{m \theta}
   \delta_{v}^{\gamma} {p}_3{}^{\rho }
   -\delta_{l}^{\rho} \delta_{m \theta} \delta_{v}^{\gamma}
   \left({p}_1\cdot {p}_2\right)
   \\ & +\delta_{l}^{\rho} \delta_{m}^{\gamma} \eta_{v \theta}
   \left({p}_1\cdot {p}_2\right)-\delta_{l \theta} \delta_{m}^{\gamma} \delta_{v}^{\rho}
   \left({p}_1\cdot {p}_2\right)-\delta_{l}^{\rho} \delta_{m \theta} \delta_{v}^{\gamma}
   \left({p}_1\cdot {p}_3\right)-\delta_{l \theta} \delta_{m}^{\gamma} \delta_{v}^{\rho}
   \left({p}_2\cdot {p}_3\right)
       ],
    \end{split}
\end{equation}
where $ p \cdot q \equiv p_{\mu} q^{\mu}  $.

\subsubsection{Vertex $Qq \bar{q} $}\label{section:vertexQqbarq}

Using Eq.~\eqref{eq:appV03}, we can obtain the contribution to the vertex $ Qq \bar{q} $ which comes from the determinant: 
\begin{equation}\label{eq:appV06}
    \begin{split}
&   \frac{\kappa}{2} \eta_{ab} 
 ( \delta_{u}^{a} \tensor{q}{^{b}_{\nu}} \bar{q}^{\nu u} + \delta_{\mu}^{a} \delta_{u}^{\nu} \bar{q}^{\mu u}\tensor{q}{^{b}_{\nu}} + \delta_{\nu}^{b} \delta_{u}^{\mu} \tensor{q}{^{a}_{\mu}} \bar{q}^{\nu u} + \delta_{u}^{b} \tensor{q}{^{a}_{\mu}} \bar{q}^{\mu u} ) 
 (\delta^{c \alpha} \delta^{d \beta} - \delta^{c \beta} \delta^{d \alpha} )Q_{\beta cd, \alpha} 
 \\ & 
 - \frac{\kappa}{2} [ \eta_{lb} \eta^{\gamma \nu} ( 
 \tensor{\bar{q}}{^{l}_{\gamma}} \tensor{q}{^{b}_{\nu}} + \tensor{\bar{q}}{^{b}_{\nu}} \tensor{q}{^{l}_{\gamma}} ) 
 + \frac{1}{2} \eta_{lb} \eta_{cd} \eta^{\gamma \nu} \eta^{\lambda \omega} (
 \tensor{\bar{q}}{^{l}_{\gamma}} {\delta }{^{b}_{\nu}} + \tensor{\bar{q}}{^{b}_{\nu}} {\delta}{^{l}_{\gamma}} 
 )(
 {\delta }{^{c}_{\lambda}} \tensor{q}{^{d}_{\omega}} + {\delta}{^{d}_{\omega}} \tensor{q}{^{c}_{\lambda}} 
 )] 
(\delta^{u \alpha} \delta^{v \beta} - \delta^{u \beta} \delta^{v \alpha} ) Q_{\beta uv, \alpha} 
     .
    \end{split}
\end{equation}
Now, using Eq.~\eqref{eq:appV6}, we obtain the following interaction terms 
\begin{equation}\label{eq:appV14ant}
    \begin{split}
    & -\kappa \bar{q}^{\lambda c} \tensor{q}{_{\sigma}^b} \eta^{\sigma \nu}  (Q_{\nu cb , \lambda} - Q_{\lambda cb , \nu} )
    + \kappa \tensor{q}{_\pi^c} \eta^{\pi \lambda} \bar{q}^{\nu b} 
(Q_{\nu cb , \lambda} - Q_{\lambda cb , \nu} )
\\ & \quad 
- \kappa \delta^{b \nu} ( \bar{q}^{\lambda u} \tensor{q}{_{u}^{c}} - q^{\lambda u} \tensor{\bar{q}}{_{u}^{c}})
        (Q_{\nu cb , \lambda} - Q_{\lambda cb , \nu} ) 
+ \kappa \delta^{c \nu } ( \bar{q}^{\nu u} \tensor{q}{_{u}^{b}} - q^{\nu u} \tensor{\bar{q}}{_{u}^{b}})
(Q_{\nu cb , \lambda} - Q_{\lambda cb , \nu} )
    \end{split} 
\end{equation}
which yields the following vertex contribution
\begin{equation}\label{eq:appV14}
    \begin{split}
     - \frac{1}{\kappa} 
    \frac{\delta^{3} }{\delta Q_{\rho lm} (x_1) \delta \tensor{q}{^{a}_{\mu}}(x_2) \bar{q}^{v \pi} (x_3)} 
   &  [ \delta^{c \lambda} - \kappa \bar{q}^{\lambda \, c} - \kappa  \tensor{q}{_{\pi}^c} \eta^{\lambda \pi} + \kappa^{2}  ( \bar{q} + q )^{\lambda \, u} \tensor{( \bar{q} + q)}{_{u}^{ c}}]
    \\ & \times 
    [ \delta^{b \nu} - \kappa \bar{q}^{\nu \, b} - \kappa  \tensor{q}{_{\sigma}^b} \eta^{\sigma \nu} + \kappa^{2}  ( \bar{q} + q )^{\nu \, u} \tensor{( \bar{q} + q)}{_{u}^{ b}} ] (Q_{\nu cb , \lambda} - Q_{\lambda cb , \nu} )\big |_{q = \bar{q} =Q=0}.
    \end{split} 
 \end{equation}
 In momentum space, it becomes 
\begin{equation}\label{eq:appV15}
\begin{split}
    & -i \kappa (p_{1 \lambda} \delta_{\nu}^{\rho} - p_{1 \nu} \delta_{\lambda}^{\rho} )\tensor{I}{_{cb}^{lm}  } \\ & \quad \quad \times
    [ \delta_{v}^{\lambda} \delta_{\pi}^{c} \delta_{a}^{\nu} \eta^{\mu b}  + \delta_{a}^{\lambda} \eta^{\mu c} \delta_{v}^{\nu} \delta_{\pi}^{b} + \delta^{b \nu } \left ( 
            \delta_{v}^{\lambda} \delta_{\pi}^{u} \eta_{a u} \eta^{c \mu} + \eta_{v u} \eta_{c \pi} \delta_{a}^{\lambda} \eta^{c \mu}  
        \right )  
        + \delta^{c \lambda } 
        \left ( 
        \delta_{v}^{\nu} \delta_{\pi}^{u} \eta_{a u} \eta^{b \mu} + \eta_{v u} \delta^{b}_{ \pi} \delta^{ \nu}_a \eta^{u \mu}  
        \right ) 
    ].
\end{split}
\end{equation}

We also have the following interactions terms due to the expansion in EC action and the determinant:
\begin{equation}\label{eq:appV16ant}
  -     [\eta_{uj} \delta_{\tau}^{u} \bar{q}^{j \, \tau} +   \eta_{uj} \delta_{\tau}^{u} \eta^{\tau \gamma } \tensor{q}{^{j}_{ \gamma}}]
[  \delta^{c \lambda} - \kappa \bar{q}^{\lambda \, c} - \kappa  \tensor{q}{_{\pi}^c} \eta^{\lambda \pi}  ]
[   \delta^{b \nu} - \kappa \bar{q}^{\nu \, b} - \kappa   \tensor{q}{_{\sigma}^b} \eta^{\sigma \nu}  ] 
(Q_{\nu cb , \lambda} - Q_{\lambda cb, \nu} )
.
\end{equation}
The terms that contributes to the $ Q q \bar{q} $ vertex are 
\begin{equation}\label{eq:appV16}
    \kappa \eta_{u j} \delta_{r}^{u} \bar{q}^{j \tau} ( \delta^{c \lambda} \tensor{q}{_{\sigma}^b} \eta^{\sigma \nu} + \delta^{b \nu} \tensor{q}{_{\pi}^c} \eta^{\pi \lambda} )
    (Q_{\nu cb , \lambda} - Q_{\lambda cb, \nu} ) 
    + \kappa \eta_{uj} \delta_{\tau}^{u} \eta^{\tau \gamma} \tensor{q}{^{j}_{\gamma}}( \delta^{c \lambda} \bar{q}^{\nu b} + \delta^{b \nu} \bar{q}^{\lambda c} )
    (Q_{\nu cb , \lambda} - Q_{\lambda cb, \nu} ) 
\end{equation}
which leads to the contribution, in momentum space,
\begin{equation}\label{eq:appV17}
    i \kappa (p_{1 \lambda} \delta_{\nu}^{\rho} - p_{1 \nu} \delta_{\lambda}^{\rho} )\tensor{I}{_{cb}^{lm}}[\eta_{uj} \delta_{\tau}^{u} \delta_{v}^{j} \delta_{\pi}^{\tau} ( \delta_{a}^{\lambda} \eta^{\mu c} \delta^{b \nu}  + \delta^{c \lambda} \delta_{a}^{\nu} \eta^{\mu b} ) + \eta_{u j} \delta_{\tau}^{u} \eta^{\tau \gamma} \delta_{a}^{j} \delta_{\gamma}^{\mu} (\delta^{c \lambda} \delta_{v}^{\nu} \delta_{\pi}^{b} + \delta_{v}^{\lambda} \delta_{\pi}^{c} \delta^{b \nu} )].
\end{equation}

Collecting all the interaction terms above, in momentum space ($ \partial \to +i p $), we get 
\begin{equation}
    \label{fig:FRQqq}
    \vcenter{\hbox{\includegraphics[scale=1]{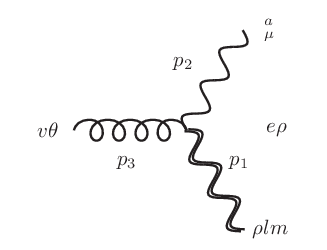}}} 
    \quad \quad 
        \tensor{i[V_{21 \bar{1}} ]}{^{\rho lm}_{a}^{\mu}_{v \pi}}  (p_1,p_2,p_3),
    \end{equation}
    where
\begin{equation}\label{eq:appV07}
    \begin{split}
        \tensor{[V_{21 \bar{1}} ]}{^{\rho lm}_{a}^{\mu}_{v \pi}}  (p_1,p_2,p_3) 
        ={}& 
  i\kappa \eta_{db}
( \delta^{d}_{c} \delta_{\nu}^{\mu} \delta_{a}^{b} \delta_{v}^{\nu} \delta_{\pi}^{c}   
+ \delta_{\beta}^{d} \delta_{c}^{\nu} \delta_{\nu}^{\mu} \delta_{a}^{b} \delta_{v}^{\beta} \delta_{\pi}^{c}  
+ \delta_{\nu}^{b} \delta_{c}^{\beta} \delta_{\beta}^{\mu} \delta_{a}^{d} \delta_{v}^{\nu} \delta^{c}_{\pi}    
+ \delta_{c}^{b} \delta_{\beta}^{\mu} \delta_{a}^{d} \delta_{v}^{\beta} \delta_{\pi}^{c}   )
I^{lm \alpha \rho}    p_{1 \alpha} 
  \\ & 
  -i \kappa  p_{1 \alpha} I^{cd \alpha \beta}  \delta_{\beta}^{\rho} \tensor{I}{_{cd}^{lm}} [\eta_{ub} \eta^{\gamma \nu}(  \delta_{v}^{u} \eta_{\pi \gamma} \delta_{\nu}^{\mu} \delta_{a}^{b} + \delta_{v}^{b} \eta_{\pi \nu } \delta_{a}^{u} \delta_{\gamma}^{\mu} ) 
  \\ & 
  - \frac{\eta_{ub} \eta_{cd} }{2} \eta^{\gamma \nu} \eta^{\lambda \omega }  (\delta_{v}^{u} \eta_{\gamma \pi} \delta_{\nu}^{b} + \delta_{v}^{b} \eta_{\nu \pi} \delta_{\gamma}^{u}  )
  (\delta_{\lambda}^{c} \delta_{a}^{d} \delta_{\omega}^{\mu} + \delta_{\omega}^{d} \delta_{a}^{c} \delta_{\lambda}^{\mu} )] 
\\ & 
    +i \kappa (p_{1 \lambda} \delta_{\nu}^{\rho} - p_{1 \nu} \delta_{\lambda}^{\rho} )\tensor{I}{_{cb}^{lm}}
    \\  & \quad \quad \times [\delta_{v}^{\lambda} \delta_{\pi}^{c} \delta_{a}^{\nu} \eta^{\mu b} + \delta_{a}^{\lambda} \eta^{\mu c} \delta_{v}^{\nu} \delta_{\pi}^{b} - \delta^{b \nu } \left ( 
            \delta_{v}^{\lambda} \delta_{\pi}^{u} \eta_{a u} \eta^{c \mu} + \eta_{v u} \eta_{c \pi} \delta_{a}^{\lambda} \eta^{c \mu}  
        \right )  
        \\ & \quad \quad \quad \quad 
        - \delta^{c \lambda } 
        \left ( 
        \delta_{v}^{\nu} \delta_{\pi}^{u} \eta_{a u} \eta^{b \mu} + \eta_{v u} \delta^{b}_{ \pi} \delta^{ \nu}_a \eta^{u \mu}  
        \right ) 
    -
    \eta_{uj} \delta_{\tau}^{u} \delta_{v}^{j} \delta_{\pi}^{\tau} ( \delta_{a}^{\lambda} \eta^{\mu c} \delta^{b \nu}  + \delta^{c \lambda} \delta_{a}^{\nu} \eta^{\mu b} ) 
     \\ & \quad \quad \quad \quad  \quad + \eta_{u j} \delta_{\tau}^{u} \eta^{\tau \gamma} \delta_{a}^{j} \delta_{\gamma}^{\mu} (\delta^{c \lambda} \delta_{v}^{\nu} \delta_{\pi}^{b} + \delta_{v}^{\lambda} \delta_{\pi}^{c} \delta^{b \nu} )].
  \end{split}
\end{equation}

\subsubsection{Vertex $ Q Q \bar{q} $}\label{section:vertexQQbarq}

The first contribution comes from the gauge fixing term 
\begin{equation}\label{eq:appV1}
    - \bar{e} \frac{1}{2 \zeta } \tensor{Q}{_{\mu ab}^{; \bar{\mu}}} \tensor{Q}{_{\nu}^{ ab}^{; \bar{\nu}} }.
\end{equation}
Using Eq.~\eqref{eq:appV03}, we get the following contribution to the vertex $ QQ \bar{q} $
\begin{equation}\label{eq:appV2alt}
    - \frac{\kappa }{ 2\zeta } \eta_{uv} \delta_{\alpha}^{u} \bar{q}^{v \, \alpha} \partial^{\mu} \tensor{Q}{_{\mu ab}} \partial_{\nu} \tensor{Q}{^{\nu ab} }.
\end{equation}
We also have one from the EC action: 
\begin{equation}\label{eq:appV2alta}
    -\kappa \eta_{uv} \delta_{\pi}^{u} \bar{q}^{v \, \pi} 
    \delta^{a \mu} \delta^{b \nu}  ( Q_{\mu ap} Q_{\nu\;b}^{\; p\;} -Q_{\nu ap} Q_{\mu\;b}^{\; p\;} ).
\end{equation}

Now, we will consider the interaction terms coming from the covariant derivatives
\begin{equation}\label{eq:appV2}
    \begin{split}
    V_{\alpha ; \bar{\mu}}^{ab}   ={}& V_{\alpha , \mu }^{ab} -  \kappa \tensor{\bar{e}}{_{u}^{\beta}}    \partial_{\mu} \tensor{\bar{q}}{^{u}_{\alpha}} V_{\beta}^{ab}  - \tensor{\bar{e}}{_{u}^{\beta}}  \tensor{\bar{e}}{^{v}_{\alpha}}    \tensor{\bar{\omega}}{_{\mu}^{u}_{v}} V_{\beta}^{ab}.
\end{split} 
\end{equation}

Using Eq.~\eqref{eq:appV2}, we have that
\begin{equation}\label{eq:appV4}
\tensor{Q}{_{\nu}^{ ab}^{; \bar{\nu} } }  = \tensor{Q}{_{\nu}^{ ab}_{, \omega}} 
[\eta^{\omega \nu} - \kappa \eta_{ef} ( \delta^{e \omega} \bar{q}^{ \nu f} + \delta^{f \nu} \bar{q}^{ \omega e} )] 
- 
{\delta}{_{l}^{\alpha}} \partial^{\nu} \tensor{\bar{q}}{^{l}_{\nu}} \tensor{Q}{_{\alpha}^{ ab}} - \tensor{\bar{e}}{_{l}^{\alpha}} \tensor{\bar{e}}{^{m}_{\nu}} \tensor{\bar{\omega}}{^{\nu}^{l}_{m}} \tensor{Q}{_{\alpha}^{ ab}} + O( \bar{q}^{2} ) 
.
\end{equation}
This leads to the following interaction terms 
\begin{equation}\label{eq:appV5}
    \begin{split}
& \frac{\kappa}{2\zeta} 
\delta_{l}^{\alpha} \partial^{\mu} \tensor{\bar{q}}{^{l}_{\mu}} \tensor{Q}{_{\alpha}^{ab}}
\tensor{Q}{_{\nu}^{ ab}_{, \nu }} 
+ \frac{\kappa}{2\zeta} \tensor{Q}{_{\mu}^{ ab}^{, \mu }} \delta_{l}^{\alpha} \partial^{\nu} \tensor{\bar{q}}{^{l}_{\nu}} \tensor{Q}{_{\alpha}^{ab}}
\\ & 
+ \frac{\kappa}{2 \zeta} Q_{\pi ab}^{, \pi} \tensor{Q}{_{\tau}^{ab}_{, \gamma }} \eta_{ef} ( \delta^{e \gamma} \bar{q}^{ \tau f} + \delta^{f \tau} \bar{q}^{ \gamma e} )
+ \frac{\kappa}{2 \zeta} Q_{\pi ab \, , \gamma} \tensor{Q}{_{\tau}^{ab}^{, \tau }} \eta_{ef} ( \delta^{e \gamma} \bar{q}^{ \pi f} + \delta^{f \pi} \bar{q}^{ \gamma e} )
    .
    \end{split}
\end{equation}

The remaining contribution comes from the following term in the action $ S_{\text{EC}} $ (without $ \bar{e} $): 
\begin{equation}\label{eq:appV7}
    + \kappa (\delta^{a \mu} \bar{q}^{\nu \, b}  + \delta^{b \nu} \bar{q}^{\nu \, a} )( Q_{\mu ap} Q_{\nu\;b}^{\; p\;} -Q_{\nu ap} Q_{\mu\;b}^{\; p\;} ).
\end{equation}
Summing all the interaction terms $QQ \bar{q} $, we obtain
\begin{equation}\label{eq:appV8}
    \begin{split}
        & 
- \kappa \eta_{uv} \delta_{\pi}^{u} \bar{q}^{v \, \pi} 
    \delta^{a \mu} \delta^{b \nu}  ( Q_{\mu ap} Q_{\nu\;b}^{\; p\;} -Q_{\nu ap} Q_{\mu\;b}^{\; p\;} )
    + 
    2 \kappa \delta^{a \mu} \bar{q}^{\nu \, b}  ( Q_{\mu ap} Q_{\nu\;b}^{\; p\;} -Q_{\nu ap} Q_{\mu\;b}^{\; p\;} )
                                 \\ & 
                                 + 
\frac{\kappa}{2\zeta} 
\delta_{l}^{\alpha} \partial^{\mu} \tensor{\bar{q}}{^{l}_{\mu}} \tensor{Q}{_{\alpha}^{ab}}
\tensor{Q}{_{\nu}^{ ab}^{, \nu }} 
+ \frac{\kappa}{2\zeta} \tensor{Q}{_{\mu}^{ ab}^{, \mu }} \delta_{l}^{\alpha} \partial^{\nu} \tensor{\bar{q}}{^{l}_{\nu}} \tensor{Q}{_{\alpha}^{ab}}
+ \frac{\kappa}{2 \zeta} Q_{\pi ab}^{, \pi} \tensor{Q}{_{\tau}^{ab}_{, \gamma }} \eta_{ef} ( \delta^{e \gamma} \bar{q}^{ \tau f} + \delta^{f \tau} \bar{q}^{ \gamma e} )
                                 \\ & 
+ \frac{\kappa}{2 \zeta} Q_{\pi ab \, , \gamma} \tensor{Q}{_{\tau}^{ab}^{, \tau }} \eta_{ef} ( \delta^{e \gamma} \bar{q}^{ \pi f} + \delta^{f \pi} \bar{q}^{ \gamma e} )
    - \frac{\kappa }{ 2\zeta } \eta_{uv} \delta_{\alpha}^{u} \bar{q}^{v \, \alpha} \partial^{\mu} \tensor{Q}{_{\mu ab}} \partial_{\nu} \tensor{Q}{^{\nu ab} }
    \end{split}
\end{equation}
which, in momentum space $ \partial \to -i p$, gives us the vertex: 
\begin{equation}
    \label{fig:FRQQq}
    \vcenter{\hbox{\includegraphics[scale=1]{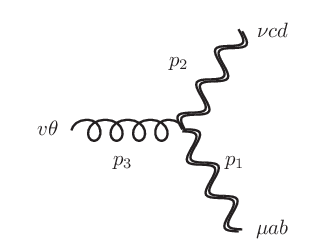}}} 
    \quad \quad 
        \tensor{i[V_{22\bar{1}}]}{_{\mu ab, \nu cd}^{ v \theta }} (p_{1} , p_{2} , p_{3}) ,
    \end{equation}
    where
\begin{equation}\label{eq:appV9}
    \begin{split}
        \tensor{[V_{22\bar{1}}]}{_{\mu ab, \nu cd}^{ v \theta }} (p_{1} , p_{2} , p_{3}) 
        ={}&
 -\kappa  \delta^{v \, \theta }  
\delta^{l}_{\mu} \delta^{m}_{\nu} ( 
    \tensor{I}{_{lp ab} } \tensor{I}{^{p}_{mcd}} 
- 
    \tensor{I}{_{lp cd} } \tensor{I}{^{p}_{mab}} 
    )
+2 \kappa \delta^{l}_{\mu } \eta_{ \nu}^{v  } \eta^{\theta m}    (
        \tensor{I}{_{lp ab}} \tensor{I}{^{p}_{m cd}}  
        - 
        \tensor{I}{_{lp cd}} \tensor{I}{^{p}_{m ab}}  
        )
                                                                                              \\ & 
        - \frac{\kappa}{2 \zeta} I_{ab cd}\left(
            p_{3}^{\theta} p_{2 \nu}  \delta_{\mu}^{v}  
            + 
            p_{3}^{\theta} p_{1 \mu}  \delta_{\nu}^{v}  
 \right)
-\frac{\kappa}{2 \zeta} \eta_{lm} p_{2}^{\pi} p_{1 \gamma} I_{ab cd} \eta_{\mu \pi} \eta_{\nu \tau} ( \delta^{l \gamma} \eta^{\theta  m } \eta^{v \tau} + \delta^{m \tau} \eta^{l \theta } \eta^{\gamma v} ) 
\\ & 
-\frac{\kappa}{2 \zeta} I_{ab cd} \eta_{lm} p_{2 \gamma} p_{1}^{\tau} \eta_{\pi \mu} \eta_{\tau \nu} ( \delta^{l \gamma} \eta^{m \theta } \eta^{\pi v} + \delta^{m \pi} \delta^{l \theta } \delta^{v \gamma} )
+ \frac{\kappa}{2 \zeta} I_{ab cd} \eta^{v \theta} p_{1 \mu} p_{2 \nu} 
\\ & 
+ (p_1 , \mu ab) \leftrightarrow ( p_{2} , \nu cd).
    \end{split}
\end{equation}

\subsubsection{Vertices $ {C}^{\star} C \bar{q} $ and $ {c}^{\star} c \bar{q} $}\label{section:vertexghostbarq}

Consider the $ C_{ab}^{\star}  $ ghost Lagrangian 
\begin{equation}\label{eq:appV24}
     \bar{e} C^{\star ab} \left [ -  C_{ab ; \bar{\mu}} +  \tensor{C}{_{a}^{p}} \omega_{\mu pb} +  \tensor{C}{_{b}^{p}} \omega_{\mu ap}\right ]_{; \bar{\nu} } \bar{g}^{\nu \mu}.
\end{equation}
Using the expansion \eqref{eq:bm1} in the determinant yields 
\begin{equation}\label{eq:appV24a1}
 -\kappa I_{ablm}\eta_{uv} \delta_{\alpha}^{u} \bar{q}^{v \, \alpha}C^{\star ab}   {\partial_{\pi}}\partial^{\pi}  \tensor{C}{^{lm}},
\end{equation}
and the expansion in $ \bar{g}^{\nu \mu} $: 
\begin{equation}\label{eq:V24a2}
    +\kappa C^{\star ab}  \left [ -  C_{ab, \mu \nu } \right ] \eta_{ef} ( \delta^{e \mu} \bar{q}^{\nu f} + \delta^{\nu f} \bar{q}^{\mu e} ).
\end{equation}

We also have a interaction term (at order $ \kappa $) due to the covariant derivative 
\begin{equation}\label{eq:V24b1}
    C_{ab ; \bar{\mu} ; \bar{\nu} } 
= C_{ab , \mu \nu} - \kappa \tensor{\bar{e}}{_{l}^{\beta}} \partial_{\nu} \tensor{\bar{q} }{^{l}_{\mu}} C_{ab, {\beta}} - \tensor{\bar{e}}{_{u}^{\beta}} \tensor{\bar{e}}{^{v}_{\mu}} \tensor{\bar{\omega}}{_{\nu}^{u}_{v}} C_{ab , {\beta} } ,
\end{equation}
which reads 
\begin{equation}\label{eq:V24b2}
    \kappa \eta^{\mu \nu} C^{\star ab} \delta_{a}^{\beta} \partial_{\nu} \tensor{\bar{q}}{^{a}_{\mu}}C_{ab , \beta}. 
\end{equation}

Therefore, the interaction terms $ C^{\star} C \bar{q} $ are given by 
\begin{equation}\label{eq:V24total}
 -\kappa I_{ablm}\eta_{uv} \delta_{\alpha}^{u} \bar{q}^{v \, \alpha}C^{\star ab}   {\partial_{\pi}}\partial^{\pi}  \tensor{C}{^{lm}}
    +\kappa C^{\star ab}  \left [ -  C_{ab, \mu \nu } \right ] \eta_{ef} ( \delta^{e \mu} \bar{q}^{\nu f} + \delta^{\nu f} \bar{q}^{\mu e} )
    +\kappa \eta^{\mu \nu} C^{\star ab} \delta_{a}^{\beta} \delta_{\nu} \tensor{\bar{q}}{^{a}_{\mu}}C_{ab , \beta}. 
\end{equation}

Now, we have the $ c_{\mu}^{\star} $ ghost Lagrangian 
\begin{align}\label{eq:appV22}
& 
\bar{e} c^{\star \nu }  \bar{g}^{\mu \gamma}  \eta_{ab}   
\left\{\left[ \tensor{\bar{e}}{^{a}_{\mu}} \left ( - c^{\alpha} \tensor{\bar{e}}{^{b}_{\nu , {\alpha}}} - \tensor{\bar{e}}{^{b}_{\alpha}} c_{\ ,{\nu}}^{\alpha} \right ) + \tensor{\bar{e}}{^{b}_{\nu}} \left( - c^{\alpha} \tensor{\bar{e}}{^{a}_{\mu , {\alpha}}} - \tensor{\bar{e}}{^{a}_{\alpha}} c_{\  , {\mu}}^{\alpha}  \right)\right]_{; \bar{\gamma}}  
-                          [ \tensor{\bar{e}}{^{a }_{\gamma }}  ( - c^{\alpha} \tensor{\bar{e}}{^{b}_{\mu , {\alpha } } } - \tensor{\bar{e}}{^{b}_{\alpha}} c^{\alpha}_{\ , {\mu} } ) ]_{; \bar{\nu}} \right\}
\end{align}
in which linear terms in the quantum tetrad field $ q $ were omitted.

From Eq.~\eqref{eq:appV01}, we obtain the first interaction term
\begin{equation}\label{eq:appV04}
      - \kappa \eta_{\mu \nu} \eta_{uv} \delta_{\alpha}^{u} \bar{q}^{v \, \alpha}  
      c^{\star \nu}    
          \partial^{\pi} \partial_{\pi} 
     c^{\mu} 
\end{equation}

Now, we have the terms coming from $ \bar{g}^{\mu \lambda} $:
\begin{equation}\label{eq:appV21}
    -\kappa c^{\star \nu}  \eta_{ab}\eta_{cd} ( \delta^{c \gamma} \bar{q}^{\mu d} + \delta^{d \mu} \bar{q}^{ \gamma c} )    
        \{ \left[ {\delta}{^{a}_{\mu}} \left ( - {\delta}{^{b}_{\alpha}} c_{, \nu }^{\alpha} \right ) + {\delta}{^{b}_{\nu}} \left( - {\delta}{^{a}_{\alpha}} c_{, \mu }^{\alpha}  \right)\right]_{, \gamma }
        - 
        [ \delta^{a}_{\gamma}  \delta_{\alpha}^{b} c_{, \mu}^{ \alpha} ]_{,\nu}
    \}.
\end{equation}

Moreover, the covariant derivatives $ \tensor{\delta}{^{a}_{\mu : \alpha} } = - \kappa \partial_{\alpha} \tensor{\bar{q}}{^{a}_{\mu} } + O ( \bar{\omega} )$ and \begin{equation}
    (c^{\alpha} \tensor{\bar{e}}{^{b}_{\nu , \alpha}} + \tensor{\bar{e}}{^{b}_{\alpha}} c^{\alpha}_{\ , \nu} )_{: \bar{\gamma}} = 
    - \kappa
    \tensor{e}{_{u}^{\beta}} \partial_{\gamma} \tensor{\bar{q}}{^{u}_{\nu}}
    (c^{\alpha} \tensor{\bar{e}}{^{b}_{\beta , \alpha}} + \tensor{\bar{e}}{^{b}_{\alpha}} c^{\alpha}_{\ , \beta}) + O ( \bar{\omega} ) 
\end{equation} 
also lead to interaction terms
\begin{equation}\label{eq:appV23}
    \begin{split}
 & 
 \kappa c^{\star \nu }   \eta^{ \mu \gamma} \eta_{ab}  
 [ 
 \partial_{\gamma} \tensor{\bar{q}}{^{a}_{\mu}}  \delta_{\alpha}^{b} c^{\alpha}_{\ , \nu}  + \partial_{\gamma} \tensor{\bar{q}}{^{b}_{\nu}} \delta_{\alpha}^{a} c^{\alpha}_{\ , \mu} - \partial_{\nu} \tensor{\bar{q}}{^{a}_{\gamma}} \delta_{\alpha}^{b} c^{\alpha}_{\ , \mu} ] 
 \\ & 
 +
 \kappa c^{\star \nu }   \eta^{ \mu \gamma} \eta_{ab}  
 (\delta_{\mu}^{a} \delta_{u}^{\beta} \partial_{\gamma} \tensor{\bar{q}}{^{u}_{\nu}}
 {\delta}{^{b}_{\alpha}} c^{\alpha}_{\ , \beta}
+ \delta_{\nu}^{b} \delta_{u}^{\beta} \partial_{\gamma} \tensor{\bar{q}}{^{u}_{\mu}} \delta{^{a}_{\alpha}} c^{\alpha}_{\ , \beta}
 - 
 \delta_{\gamma}^{a} 
 \delta_{u}^{\beta} \partial_{\nu} \tensor{\bar{q}}{^{u}_{\mu}} \delta {^{a}_{\alpha}} c^{\alpha}_{\ , \beta}
 ).
    \end{split}
\end{equation}

The remaining interactions terms arises from the expansion \eqref{eq:bm1} in
\begin{align}\label{eq:appV22remain}
c^{\star \nu }    \eta_{ab}   
\left\{\left[ \tensor{\bar{e}}{^{a}_{\mu}} \left ( - c^{\alpha} \tensor{\bar{e}}{^{b}_{\nu , {\alpha}}} - \tensor{\bar{e}}{^{b}_{\alpha}} c_{\ ,{\nu}}^{\alpha} \right ) + \tensor{\bar{e}}{^{b}_{\nu}} \left( - c^{\alpha} \tensor{\bar{e}}{^{a}_{\mu , {\alpha}}} - \tensor{\bar{e}}{^{a}_{\alpha}} c_{\  , {\mu}}^{\alpha}  \right)\right]_{, {\gamma}}  
-                          [ \tensor{\bar{e}}{^{a }_{\gamma }}  ( - c^{\alpha} \tensor{\bar{e}}{^{b}_{\mu , {\alpha } } } - \tensor{\bar{e}}{^{b}_{\alpha}} c^{\alpha}_{\ , {\mu} } ) ]_{, {\nu}} \right\}
\end{align}
which reads 
\begin{equation}\label{eq:appV22remainterms}
    \begin{split}
    & \kappa c^{\star \nu}     \eta_{ab}   
    \left[ \delta{^{a}_{\mu}} \left ( - c^{\alpha} \tensor{\bar{q}}{^{b}_{\nu , {\alpha}}} - \tensor{\bar{q}}{^{b}_{\alpha}} c_{\  , {\nu}}^{\alpha} \right ) + \delta{^{b}_{\nu}} \left( - c^{\alpha} \tensor{\bar{q}}{^{a}_{\mu , {\alpha}}} - \tensor{\bar{q}}{^{a}_{\alpha}} c_{\  , {\mu}}^{\alpha}  \right)\right]^{, \mu} 
     -\kappa c^{\star \nu}    \eta_{ab}   
     [ \delta{^{a \mu}}  ( - c^{\alpha} \tensor{\bar{q}}{^{b}_{\mu , {\alpha } } } - \tensor{ \bar{q}}{^{b}_{\alpha}} c^{\alpha}_{, {\mu}}  ) ]_{, \nu} 
 \\ & 
+\kappa c^{\star \nu }     \eta_{ab}   
\left[ \tensor{\bar{q}}{^{a}_{\mu}} \left ( - \delta{^{b}_{\alpha}} c_{\  , {\nu}}^{\alpha} \right ) + \tensor{\bar{q}}{^{b}_{\nu}} \left(  - \delta{^{a}_{\alpha}} c_{\  , {\mu}}^{\alpha}  \right)\right]^{, \mu} 
     - \kappa c^{\star \nu}     \eta_{ab}   
     [ \tensor{\bar{q}}{^{a \mu}}  (  - \delta{^{b}_{\alpha}} c^{\alpha}_{, {\mu}}  ) ]_{, \nu} .
    \end{split}
\end{equation}

Collecting all contributions, in the momentum space, we have the vertices: 
\begin{equation}
    \label{fig:FRccq}
    \vcenter{\hbox{\includegraphics[scale=1]{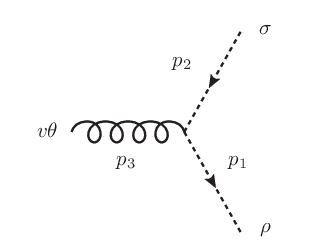}}} 
    \quad \quad 
    i[V_{3^{\star}  3 \bar{1}}]_{\rho , \sigma , v \theta} (p_{1} , p_{2} , p_{3} ) 
\end{equation}
and
\begin{equation}
    \label{fig:FRCCq}
    \vcenter{\hbox{\includegraphics[scale=1]{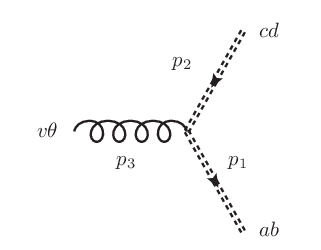}}} 
    \quad \quad 
    i[V_{{4}^{\star}  4 \bar{1}}]_{ab, cd, v \theta }  (p_{1} , p_{2} , p_{3} ) 
    ,
\end{equation}
where
\begin{equation}\label{eq:appV05}
    \begin{split}
        [V_{{3}^{\star}  3 \bar{1}}]^{\rho , \sigma , v \theta} (p_{1} , p_{2} , p_{3} ) 
        ={}& \kappa [ 
-2 {p}_2^{\theta } {p}_2^v \eta^{\rho \sigma }-{p}_3^{\theta }
   {p}_2^v \eta^{\rho \sigma }+{p}_2^{2} \eta^{\rho \sigma } \eta^{v\theta
   }+{p}_1^{\rho } {p}_2^v \eta^{\theta \sigma }+{p}_2^{\theta }
   {p}_1^{\rho } \eta^{v\sigma }-{p}_2^{\rho } {p}_1^v \eta^{\theta \sigma
   } \\ & -{p}_1^{\theta } {p}_2^{\rho } \eta^{v\sigma }-{p}_3^{\theta }
   {p}_2^{\rho } \eta^{v\sigma }+{p}_3^{\rho } {p}_2^v \eta^{\theta \sigma
   }+{p}_2^{\theta } {p}_3^{\rho } \eta^{v\sigma }-{p}_3^{\sigma }
   {p}_1^v \eta^{\theta \rho }  -{p}_3^{\sigma } {p}_2^v \eta^{\theta \rho
   } 
   \\ & -{p}_1^{\theta } {p}_3^{\sigma } \eta^{v\rho }+{p}_1^{\rho }
   {p}_3^{\sigma } \eta^{v\theta }-\left({p}_1\cdot {p}_2\right) \eta^{\theta
   \rho } \eta^{v\sigma }-\left({p}_1\cdot {p}_2\right) \eta^{\theta \sigma }
   \eta^{v\rho }-\left({p}_2\cdot {p}_3\right) \eta^{\theta \rho } \eta^{v\sigma }]
    \end{split} 
\end{equation}
and 
\begin{equation} \label{eq:appV05b}
    \begin{split}
        [V_{4^{\star}  4 \bar{1}}]_{ab, cd, v \theta }  (p_{1} , p_{2} , p_{3} ) 
        ={}&   \kappa I_{abcd}   \eta_{v\theta} p_{2}^{\pi} p_{2 \pi} 
        - \kappa I_{ab cd} \eta_{ef} p_{2 \nu} p_{2 \mu} ( \delta^{e \mu} \delta_{v}^{\nu} \delta_{\theta}^{f} + \delta^{\nu f} \delta_{v}^{\mu} \delta_{\theta}^{e} ) 
        - \kappa I_{ab cd}  p_{3 \theta} p_{2 v}   
    \end{split} 
\end{equation}

\section{Transversality of the tetrad self-energy}\label{section:TransSE}

One can express the tetrad self-energy in terms of the basis \eqref{eq:b18} as follows 
\begin{equation}\label{eq:frenkel71}
    \begin{split}
        \bar{\Pi}\tensor{}{_a^\mu_b^\nu} (k) ={}& c_{1} \delta_{a}^{\mu} \delta_{b}^{\nu} + c_{2} \delta_{a}^{\nu} \delta_{b}^{\mu} + c_{3} \eta^{\mu \nu} \eta_{ab} + c_{4}  \eta_{ab} \frac{k^{\mu} k^{\nu}}{k^{2}} + c_{5} \eta^{\mu \nu} \frac{k_{a} k_{b}}{k^{2}} \\ & + c_{6} \left ( \delta_{b}^{\nu} \frac{k_{a} k^{\mu}}{k^{2}} + \delta_{a}^{\mu} \frac{k_{b} k^{\nu}}{k^{2}} \right )  + c_{7} \left ( \delta_{b}^{\mu} \frac{k_{a} k^{\nu }}{k^{2}} + \delta_{a}^{\nu} \frac{k_{b} k^{\mu}}{k^{2}}\right ) + c_{8} \frac{k_{a} k_{b} k^{\mu} k^{\nu}}{k^{4}},  
    \end{split} 
\end{equation}
where $ c_{i} $ are scalar functions with mass dimension $2$.

The transversality condition \eqref{eq:ccf56} leads to the following constraints: 
\begin{equation}\label{eq:frenkel72}
    c_{1} + c_{6} = 0, \quad c_{2} + c_{7} =0, \quad c_{3} + c_{4} =0 \quad \text{and} \quad  c_{5} + c_{6} + c_{7} + c_{8} =0. 
\end{equation}
Using these relation in Eq.~\eqref{eq:frenkel71}, we obtain the transverse expression 
\begin{equation}\label{eq:frenkel73}
    \begin{split}
    & c_{1} ( \xi , \zeta ) \left ( \delta_{a}^{\mu} - \frac{k_{a} k^{\mu}}{k^{2}}\right ) \left ( \delta_{b}^{\nu} - \frac{k_{b} k^{\nu}}{k^{2}}\right ) + c_{2} ( \xi , \zeta )\left ( \delta_{a}^{\nu} - \frac{k_{a} k^{\nu}}{k^{2}}\right ) \left ( \delta_{b}^{\mu} - \frac{k_{b} k^{\mu}}{k^{2}}\right ) 
 \\ & \quad \quad +  \left ( c_{3} ( \xi , \zeta )\eta_{ab} + c_{5} ( \xi , \zeta )\frac{k_{a} k_{b}}{k^{2}}\right ) \left ( \eta^{\mu \nu} - \frac{k^{\mu} k^{\nu}}{k^{2}}\right ),
    \end{split}
\end{equation}
where $ \xi $ and $ \zeta $ are the gauge parameters introduced in Eq.~\eqref{eq:ccf211}.


\section{Self-energy of the background tetrad field at one-loop order}\label{section:SEtetrad}

To compute the diagrams in Fig.~\ref{fig:diag1}, we will use dimensional regularization and employ tensor decompositions as done in Refs. \cite{Passarino:1978jh, Brandt:2020vre}. After the loop momentum integration, the diagram (I) will have the following tensorial structure 
\begin{equation}\label{eq:appV13a}
    [\bar{\Pi}^{(\text{I})} ]\tensor{}{_a^\mu_b^\nu} (k) = \sum_{m=1}^{8} C^{(\text{I})}_{m} ( T_{qq}^{(m)} )\tensor{}{_a^\mu_b^\nu}(k),
\end{equation}
where $ (T_{qq}^{(m)})\tensor{}{_a^\mu_b^\nu}$ is the $m$th tensor of the basis \eqref{eq:b18} and $\text{I} = \text{a, b, c, d, e, f, g, h, i} $. 
We can obtain the coefficients $ C_{m}^{(I)} $ by solving the system of 8 linear equations: 
\begin{equation}\label{eq:appsystem}
    (T_{qq}^{(n)} )\tensor{}{^a_\mu^b_\nu} [ \bar{\Pi}^{(\text{I})}]\tensor{}{_a^\mu_b^\nu} = \sum_{m =1}^{8} C_{m}^{( \text{I} )} ( T_{qq}^{(m)} )\tensor{}{_a^\mu_b^\nu} ( T_{q q}^{(n)} )\tensor{}{^a_\mu^b_\nu},
\end{equation}
where $ n = 1, 2,\ldots,8$. 

Now, we have to compute the scalar integrals $
(T_{qq}^{(n)} )\tensor{}{^a_\mu^b_\nu} [ \bar{\Pi}^{(\text{I})}]\tensor{}{_a^\mu_b^\nu}$, which have the following form 
\begin{equation}\label{eq:appformofSI}
    \int \frac{\mathop{d^{D} p}}{(2 \pi)^D} s^{(\text{I})} (p,q,k),
\end{equation}
where $ p$ is taken to be the loop momentum, $ q = p +k$, $k$ is the external momentum and $ s^{( \text{I} )} (p,q,k)$ are scalar functions of the $ p \cdot k $, $q \cdot k $, $ p \cdot q $, $ p^{2} $, $ q^{2} $ and $ k^{2} $. We can simplify these scalar functions using the relations 
\begin{subequations}\label{eq:apprelationmomenta}
    \begin{align}
        p \cdot k ={}& \frac{q^{2} - p^{2} - k^{2}}{2}, \\
        q \cdot k ={}& \frac{q^{2} + k^{2} - p^{2}}{2}, \\
        p \cdot q ={}& \frac{p^{2} + q^{2} - k^{2}}{2}; 
    \end{align}
\end{subequations}
which reduces $ s^{(\text{I} )} (p,q,k)$ to combinations of powers of $ p^2$ and $ q^2$.  
In the end, the integrals in Eq.~\eqref{eq:appformofSI} becomes a combination of the simple integrals \cite{Brandt:2020vre}:
\begin{equation}\label{eq:appintegralsimple}
    I^{lm} \equiv \int \frac{\mathop{d^{D} p}}{( 2 \pi)^D}  \frac{1}{( p^{2} )^{l} (q^{2} )^{m}} =                i^{1+D}\frac{ (k^{2})^{D/2-l-m} }{(4 \pi )^{D/2}} 
               \frac{\Gamma (l+m-D/2)}{\Gamma (l) \Gamma (m)} 
               \frac{{\Gamma (D/2-m) \Gamma (D/2-l)} }{ \Gamma (D-m-l)}. 
\end{equation}
The only non-vanishing integrals are 
\begin{subequations}\label{eq:I11}
    \begin{align}
        I_{11}  ={}&  i^{1+D} \frac{ (k^{2})^{D/2-2}}{(4 \pi )^{D/2}} 
           \frac{\Gamma (2-D/2) \Gamma (D/2-1)^{2}  }{ \Gamma (D-2)}, \\
        I_{12} ={}& I_{21} = \frac{3-D}{k^{2}} I_{11 }, \\
        I_{22} ={}& \frac{(3-D)(6-D)}{k^{4}} I_{11 }.
    \end{align}
\end{subequations}
Other integrals comes from massless tadpole-like contributions that go to zero when dimensional regularization is used \cite{Leibbrandt:1975dj}. Note that, in the massive case, these contributions may not vanish. The procedure described here for the massless case can be modified accordingly to the massive case, which could be required to the diagonal formulation of the EC theory in first-order form.

In $ D =4 - 2 \epsilon $ dimensions, the ultraviolet pole $ 1 / \epsilon $ part of the integral $ I^{11} $ is given by 
\begin{equation}\label{eq:basisUVI11}
    I_{\text{UV}} = \frac{i}{16 \pi^{2} \epsilon} .
\end{equation}
From Eq.~\eqref{eq:I11}, we see that if the UV pole of the integrals in Eq.~\eqref{eq:appformofSI} vanish, then the entire integrals vanish. This implies that, if the background tetrad self-energy $ \bar{\Pi}_{\mu \nu}^{ab} $ is finite, then it must vanish. 

\subsection{Results in $ D = 4 - 2 \epsilon $ dimensions}

Using the method described above, we obtained the divergent part of the diagrams in Fig.~\ref{fig:diag1} in $ D = 4 - 2 \epsilon $ dimensions which are presented in Table 1 (for a general $ \xi $ and $ \zeta $-gauge).
\begin{table*}[ht]
    \centering
    \caption{The coefficients $  C_{m}^{\text{(I)}} $ (see Eq.~\eqref{eq:appV13a}), in units of $ \kappa^{2} k^{4} I_{\text{UV}} $, for the divergent part of the diagrams in Fig.~\ref{fig:diag1} decomposed in the basis \eqref{eq:b18}.}
\label{tab:1}
\begin{tblr}{l|rccrcrcrr}
    \toprule
    (I) & (a) & (b) & (c) & (d) & (e) & (f) & (g) & (h) & (i)\\
\midrule
    $C_1^{\text{(I)}}   $ & $ \dfrac{1}{15}  $ & $  \dfrac{\xi }{24}+\dfrac{1}{20}  $ & $  \dfrac{\zeta }{6 {k}^2}+\dfrac{33}{64}  $ & $  -\dfrac{43}{480}  $ & $  0  $ & $  \dfrac{1}{4}  $ & $  \dfrac{\zeta }{3 {k}^2}+\dfrac{7}{24}  $ & $  \dfrac{1}{60}  $ & $  -\dfrac{1}{10}  $ \\
    $C_2^{\text{(I)}}   $ & $ \dfrac{1}{15}  $ & $  \dfrac{\xi }{24}-\dfrac{13}{240}  $ & $  \dfrac{23}{192}-\dfrac{\zeta }{12 {k}^2}  $ & $  \dfrac{247}{480}  $ & $  0  $ & $  -\dfrac{3}{8}  $ & $  -\dfrac{\zeta }{6 {k}^2}-\dfrac{23}{96}  $ & $  -\dfrac{19}{60}  $ & $  -\dfrac{1}{10}  $ \\
    $C_3^{\text{(I)}}   $ & $ -\dfrac{1}{10}  $ & $  -\dfrac{\xi }{6}-\dfrac{13}{240}  $ & $  \dfrac{19}{96}-\dfrac{\zeta }{12 {k}^2}  $ & $  -\dfrac{1}{160}  $ & $  0  $ & $  -\dfrac{1}{24}  $ & $  -\dfrac{\zeta }{6 {k}^2}-\dfrac{23}{96}  $ & $  -\dfrac{1}{15}  $ & $  -\dfrac{1}{10}  $ \\
    $C_4^{\text{(I)}}   $ & $ \dfrac{1}{10}  $ & $  \dfrac{\xi }{6}+\dfrac{7}{60}  $ & $  \dfrac{\zeta }{12 {k}^2}-\dfrac{1}{16}  $ & $  \dfrac{19}{240}  $ & $  0  $ & $  \dfrac{1}{24}  $ & $  \dfrac{\zeta }{6 {k}^2}-\dfrac{1}{32}  $ & $  \dfrac{1}{15}  $ & $  \dfrac{1}{10}  $ \\
    $C_5^{\text{(I)}}   $ & $ -\dfrac{1}{15}  $ & $  \dfrac{7 \xi }{12}+\dfrac{7}{60}  $ & $  \dfrac{5}{96}-\dfrac{\zeta }{4 {k}^2}  $ & $  \dfrac{3}{80}  $ & $  0  $ & $  \dfrac{7}{24}  $ & $  \dfrac{\zeta }{2 {k}^2}-\dfrac{25}{96}  $ & $  \dfrac{37}{30}  $ & $  -\dfrac{2}{5}  $ \\
    $C_6^{\text{(I)}}   $ & $ -\dfrac{1}{15}  $ & $  -\dfrac{\xi }{24}-\dfrac{11}{120}  $ & $  \dfrac{\zeta }{6 {k}^2}-\dfrac{61}{192}  $ & $  \dfrac{1}{60}  $ & $  0  $ & $  -\dfrac{1}{4}  $ & $  -\dfrac{2 \zeta }{3 {k}^2}-\dfrac{3}{8}  $ & $  -\dfrac{1}{60}  $ & $  \dfrac{1}{10}  $ \\
    $C_7^{\text{(I)}}   $ & $ -\dfrac{1}{15}  $ & $  \dfrac{7}{60}-\dfrac{\xi }{24}  $ & $  \dfrac{25}{192}-\dfrac{\zeta }{12 {k}^2}  $ & $  -\dfrac{7}{30}  $ & $  0  $ & $  \dfrac{1}{6}  $ & $  \dfrac{\zeta }{3 {k}^2}-\dfrac{7}{48}  $ & $  \dfrac{19}{60}  $ & $  \dfrac{1}{10}  $ \\
    $C_8^{\text{(I)}}   $ & $ \dfrac{1}{5}  $ & $  -\dfrac{\xi }{2}-\dfrac{1}{10}  $ & $  -\dfrac{5}{48}  $ & $  \dfrac{7}{60}  $ & $  0  $ & $  0  $ & $  \dfrac{5}{6}  $ & $  -\dfrac{23}{15}  $ & $  \dfrac{1}{5}  $ \\
\bottomrule
\end{tblr}
\end{table*}

Summing all the contributions of the diagrams in Fig.~\ref{fig:diag1}, we find that the divergent part of the background tetrad self-energy  is given by (see Eq.~\eqref{eq:ccf510})
\begin{equation}\label{eq:appV13}
    \tensor{{\Pi}}{_{a}^{\mu}_b^{ \nu }} = \kappa^{2} k^2 I_{\text{UV}} \sum_{m=1}^{8} C_{m} \tensor{( T_{qq}^{(m)} )}{_{a}^{\mu}_b^{ \nu}},
\end{equation}
where 
\begin{equation}\label{eq:appV11}
    \begin{split}
 C_1 ={}&\left(\frac{\xi  }{24}+\frac{961 }{960}\right)k^2 +\frac{\zeta }{2} ,\\
 C_2 ={}&\left(\frac{\xi  }{24}-\frac{123 }{320}\right)k^2 -\frac{\zeta }{4} ,\\
 C_3 ={}&-\left(\frac{\xi}{6} +\frac{197 }{480}\right)k^2 -\frac{\zeta }{4} ,\\
 C_4 ={}& - C_3,\\
 C_5 ={}&\left(\frac{7 \xi }{12}+\frac{241 }{240}\right)k^2 +\frac{\zeta }{4} ,\\
 C_6 ={}&-C_1,\\
 C_7 ={}&-C_2,\\
 C_8 ={}&-C_5+C_1+C_2.
  \end{split}
\end{equation}
The above relations between the coefficients $C$ are exactly the transversality conditions derived in appendix B (see Eq.~\eqref{eq:frenkel72}). 

The gauge parameter $ \zeta $ does not appear in the coefficient $ C_{8} $, since non-local terms, as 
\begin{equation}\label{eq:appV26}
    \zeta \frac{k_{a} k_{b} k_{\mu} k_{\nu}}{ k^2} \xrightarrow[\text{space} ]{\text{position}} \zeta \frac{\partial_{a} \partial_{b} \partial_{m} \partial_{\nu}}{\partial^{2}} ,
\end{equation}
would arise. Such terms are forbidden due to the locality of gauge transformation and the corresponding BRST transformations.  

Moreover, the following relation holds
\begin{equation}\label{eq:relationsV}
    \begin{split}
        P_{Qq} (p)\cdot V_{qq \bar{q}} (p,q,k) \cdot P_{qQ}(q) ={}&0.
\end{split}
\end{equation}
This implies that the diagrams (e) must vanish at the integrand level, which is consistent with our computations.

\section{Relations between the tetradic and the metric formulations}\label{section:tetradandmetric}

The metric $ h_{\mu \nu} $ and the tetrad $ \tensor{q}{^{a}_{\mu}} $ are related by 
\begin{equation}\label{eq:relation1}
    h_{\mu \nu} =\eta_{ab } \left ( \tensor{\bar{e}}{^{a}_{\mu}} \tensor{q}{^{b}_{\nu}} + \tensor{q}{^{a}_{\mu}} \tensor{\bar{e}}{^{b}_{\nu}} + \kappa \tensor{q}{^{a}_{\mu}} \tensor{q}{^{b}_{\nu}}\right ). 
\end{equation}
From the EH action, we can obtain the quadratic form 
\begin{equation}\label{eq:relation2}
    B_{hh} = \int \mathop{d^{4} x} h_{\mu \nu} X^{\mu \nu \rho \sigma} h_{\rho \sigma}. 
\end{equation}
Using Eq.~\eqref{eq:relation1}, we find that 
\begin{equation}\label{eq:relation3}
    B_{hh} = \int \mathop{d^{4} x} \tensor{q}{^{a}_{\mu}}   \delta_{\alpha}^{\mu} \eta_{\beta a} \left ( X^{\alpha \beta \rho \sigma} + X^{\beta \alpha \rho \sigma} + X^{\alpha \beta \sigma \rho} + X^{\beta \alpha \sigma \rho}   \right )  \delta_{\rho}^{\nu} \eta_{\sigma b} \tensor{q}{^{b}_{\nu}} + O(q^{2} ).
\end{equation}
Thus, we see that the quadratic terms in the tetrad would be equal to 
\begin{equation}\label{eq:relation4}
B_{qq} =
\delta_{\alpha}^{\mu} \eta_{\beta a} \delta_{\rho}^{\nu} \eta_{\sigma b} \left ( X^{\alpha \beta \rho \sigma} + X^{\beta \alpha \rho \sigma} + X^{\alpha \beta \sigma \rho} + X^{\beta \alpha \sigma \rho}   \right ).
\end{equation}

In the first-order formalism, we have contributions due to the spin connection gauge fixing term.  When $ \zeta \to \infty $, we find that \eqref{eq:bm17} is equal to \eqref{eq:relation4}. This allows us compare the propagators in the second order formalism. 
We may also use Eq.~\eqref{eq:relation1} to compute the graviton propagator using the tetrad propagator in the first-order formalism of the EC theory obtained in appendix A.  Then, the graviton two-point Green's function
\begin{equation}\label{eq:cm1}
    \langle 0|T h_{\mu \nu}(x) h_{\rho \sigma} (y)| 0 \rangle,
\end{equation}
yields
\begin{equation}\label{eq:cm3}
    \begin{split}
        & \eta_{ab} \eta_{cd} \big[    \eta_{\mu}^{a} \eta_{\rho}^{c} \langle 0|T \tensor{q}{^{b}_{\nu}} (x) \tensor{q}{^{d}_{\sigma}} (y)| 0 \rangle
    +
    \eta_{\mu}^{a} \eta_{\sigma}^{d} \langle 0|T \tensor{q}{^{b}_{\nu}} (x) \tensor{q}{^{c}_{\rho}} (y)| 0 \rangle
    + 
    \eta_{\nu}^{b} \eta_{\rho}^{c} \langle 0|T \tensor{q}{^{a}_{\mu}} (x) \tensor{q}{^{d}_{\sigma}} (y)| 0 \rangle
    +\eta_{\nu}^{b} \eta_{\sigma}^{d} \langle 0|T \tensor{q}{^{a}_{\mu}} (x) \tensor{q}{^{c}_{\rho}} (y)| 0 \rangle 
    \\  
    & +  \eta_{\rho}^{d}  \langle 0|T \tensor{\bar{q}}{^{a}_{\mu}}(x)  \tensor{q}{^{b}_{\nu}}(y) \tensor{q}{^{c}_{\sigma}}(y)| 0 \rangle 
        + \cdots 
        \langle 0|T \tensor{\bar{q}}{^{a}_{\mu}}(x) \tensor{q}{^{b}_{\nu}}(x) \tensor{\bar{q}}{^{c}_{\rho}} (y)\tensor{q}{^{d}_{\sigma}} (y) | 0 \rangle 
        +\cdots 
        + \kappa \eta_{\mu}^{a}  \langle 0|T \tensor{q}{^{b}_{\nu}} (x) \tensor{q}{^{c}_{\rho}} (y) \tensor{q}{^{d}_{\sigma}}(y) | 0 \rangle 
 \\ & \quad + \cdots + \kappa \langle 0|T \tensor{\bar{q}}{^{a}_{\mu}} (x) \tensor{q}{^{b}_{\nu}} (x) \tensor{q}{^{c}_{\rho}} (y) \tensor{q}{^{d}_{\sigma}} (y) | 0 \rangle  
+ \eta_{ab} \eta_{cd} \kappa^{2} \langle 0|T \tensor{q}{^{a}_{\nu}} (x) \tensor{q}{^{b}_{\mu}} (x) \tensor{q}{^{c}_{\rho}} (y) \tensor{q}{^{d}_{\sigma}} (y)| 0 \rangle \big].
        \end{split} 
\end{equation}

In lowest order, the two-point function \eqref{eq:cm1} leads to the graviton propagator which correspond to 
\begin{equation}\label{eq:cm4}
\eta_{ab} \eta_{cd} \left(    \eta_{\mu}^{a} \eta_{\rho}^{c} \langle 0|T \tensor{q}{^{b}_{\nu}} (x) \tensor{q}{^{d}_{\sigma}} (y)| 0 \rangle
    +
    \eta_{\mu}^{a} \eta_{\sigma}^{d} \langle 0|T \tensor{q}{^{b}_{\nu}} (x) \tensor{q}{^{c}_{\rho}} (y)| 0 \rangle
    + 
    \eta_{\nu}^{b} \eta_{\rho}^{c} \langle 0|T \tensor{q}{^{a}_{\mu}} (x) \tensor{q}{^{d}_{\sigma}} (y)| 0 \rangle
    +\eta_{\nu}^{b} \eta_{\sigma}^{d} \langle 0|T \tensor{q}{^{a}_{\mu}} (x) \tensor{q}{^{c}_{\rho}} (y)| 0 \rangle \right)
\end{equation}
in lowest order ($ \kappa^{0} $). Using the propagator computed in Eq.~\eqref{eq:bm19}, Eq.~\eqref{eq:cm4} leads (in the de Donder-like gauge $ \sigma = 1/2 $) to 
\begin{equation}\label{eq:cm5}
-\frac{i}{p^2}\left[  \frac{2 \eta_{\sigma  \rho } \eta_{\mu  \nu }}{D-2} - \eta_{\sigma  \mu } \eta_{\nu  \rho }-\eta_{\sigma  \nu } \eta_{\mu  \rho } 
+ (1+\xi )\frac{ p_{\rho } \left(p_{\nu } \eta_{\sigma  \mu }+p_{\mu } \eta_{\sigma  \nu }\right)+p_{\sigma } \left(p_{\nu } \eta_{\mu  \rho }+p_{\mu } \eta_{\nu  \rho }\right)}{ p^2} \right],
\end{equation}
which is the graviton propagator in the de Donder gauge \cite{Brandt:2022und}.

\bibliography{RofECinFO.bib}

\begin{thebibliography}{35}%
\makeatletter
\providecommand \@ifxundefined [1]{%
 \@ifx{#1\undefined}
}%
\providecommand \@ifnum [1]{%
 \ifnum #1\expandafter \@firstoftwo
 \else \expandafter \@secondoftwo
 \fi
}%
\providecommand \@ifx [1]{%
 \ifx #1\expandafter \@firstoftwo
 \else \expandafter \@secondoftwo
 \fi
}%
\providecommand \natexlab [1]{#1}%
\providecommand \enquote  [1]{``#1''}%
\providecommand \bibnamefont  [1]{#1}%
\providecommand \bibfnamefont [1]{#1}%
\providecommand \citenamefont [1]{#1}%
\providecommand \href@noop [0]{\@secondoftwo}%
\providecommand \href [0]{\begingroup \@sanitize@url \@href}%
\providecommand \@href[1]{\@@startlink{#1}\@@href}%
\providecommand \@@href[1]{\endgroup#1\@@endlink}%
\providecommand \@sanitize@url [0]{\catcode `\\12\catcode `\$12\catcode `\&12\catcode `\#12\catcode `\^12\catcode `\_12\catcode `\%12\relax}%
\providecommand \@@startlink[1]{}%
\providecommand \@@endlink[0]{}%
\providecommand \url  [0]{\begingroup\@sanitize@url \@url }%
\providecommand \@url [1]{\endgroup\@href {#1}{\urlprefix }}%
\providecommand \urlprefix  [0]{URL }%
\providecommand \Eprint [0]{\href }%
\providecommand \doibase [0]{https://doi.org/}%
\providecommand \selectlanguage [0]{\@gobble}%
\providecommand \bibinfo  [0]{\@secondoftwo}%
\providecommand \bibfield  [0]{\@secondoftwo}%
\providecommand \translation [1]{[#1]}%
\providecommand \BibitemOpen [0]{}%
\providecommand \bibitemStop [0]{}%
\providecommand \bibitemNoStop [0]{.\EOS\space}%
\providecommand \EOS [0]{\spacefactor3000\relax}%
\providecommand \BibitemShut  [1]{\csname bibitem#1\endcsname}%
\let\auto@bib@innerbib\@empty
\bibitem [{\citenamefont {Pop{\l}awski}(2012)}]{Poplawski:2011jz}%
  \BibitemOpen
  \bibfield  {author} {\bibinfo {author} {\bibfnamefont {N.}~\bibnamefont {Pop{\l}awski}},\ }\href {https://doi.org/10.1103/PhysRevD.85.107502} {\bibfield  {journal} {\bibinfo  {journal} {Phys. Rev. D}\ }\textbf {\bibinfo {volume} {85}},\ \bibinfo {pages} {107502} (\bibinfo {year} {2012})}\BibitemShut {NoStop}%
\bibitem [{\citenamefont {Huang}\ \emph {et~al.}(2024)\citenamefont {Huang}, \citenamefont {Huang}, \citenamefont {Xu}, \citenamefont {Zhang},\ and\ \citenamefont {Chen}}]{Huang:2024ujj}%
  \BibitemOpen
  \bibfield  {author} {\bibinfo {author} {\bibfnamefont {Q.}~\bibnamefont {Huang}}, \bibinfo {author} {\bibfnamefont {H.}~\bibnamefont {Huang}}, \bibinfo {author} {\bibfnamefont {B.}~\bibnamefont {Xu}}, \bibinfo {author} {\bibfnamefont {K.}~\bibnamefont {Zhang}},\ and\ \bibinfo {author} {\bibfnamefont {H.}~\bibnamefont {Chen}},\ }\href {http://arxiv.org/abs/2404.10400} {\bibinfo {title} {Phase space analysis of the evolution of the early universe in {{Einstein-Cartan}} theory}} (\bibinfo {year} {2024}),\ \Eprint {https://arxiv.org/abs/2404.10400} {arXiv:2404.10400 [gr-qc]} \BibitemShut {NoStop}%
\bibitem [{\citenamefont {Buchbinder}\ and\ \citenamefont {Shapiro}(2021)}]{Buchbinder:2021wzv}%
  \BibitemOpen
  \bibfield  {author} {\bibinfo {author} {\bibfnamefont {I.~L.}\ \bibnamefont {Buchbinder}}\ and\ \bibinfo {author} {\bibfnamefont {I.~L.}\ \bibnamefont {Shapiro}},\ }\href {https://doi.org/10.1093/oso/9780198838319.001.0001} {\emph {\bibinfo {title} {Introduction to Quantum Field Theory with Applications to Quantum Gravity}}},\ \bibinfo {edition} {first edition}\ ed.,\ Oxford Graduate Texts\ (\bibinfo  {publisher} {Oxford University Press},\ \bibinfo {address} {Oxford},\ \bibinfo {year} {2021})\BibitemShut {NoStop}%
\bibitem [{\citenamefont {Hehl}\ \emph {et~al.}(1976)\citenamefont {Hehl}, \citenamefont {Von Der~Heyde}, \citenamefont {Kerlick},\ and\ \citenamefont {Nester}}]{Hehl:1976kj}%
  \BibitemOpen
  \bibfield  {author} {\bibinfo {author} {\bibfnamefont {F.~W.}\ \bibnamefont {Hehl}}, \bibinfo {author} {\bibfnamefont {P.}~\bibnamefont {Von Der~Heyde}}, \bibinfo {author} {\bibfnamefont {G.~D.}\ \bibnamefont {Kerlick}},\ and\ \bibinfo {author} {\bibfnamefont {J.~M.}\ \bibnamefont {Nester}},\ }\href {https://doi.org/10.1103/RevModPhys.48.393} {\bibfield  {journal} {\bibinfo  {journal} {Rev. Mod. Phys.}\ }\textbf {\bibinfo {volume} {48}},\ \bibinfo {pages} {393} (\bibinfo {year} {1976})}\BibitemShut {NoStop}%
\bibitem [{\citenamefont {Hehl}(2023)}]{Hehl:2023khc}%
  \BibitemOpen
  \bibfield  {author} {\bibinfo {author} {\bibfnamefont {F.~W.}\ \bibnamefont {Hehl}},\ }\href {http://arxiv.org/abs/2303.05366} {\bibinfo {title} {Four {{Lectures}} on {{Poincar{\`e} Gauge Field Theory}}}} (\bibinfo {year} {2023}),\ \Eprint {https://arxiv.org/abs/2303.05366} {arXiv:2303.05366 [gr-qc, physics:hep-th]} \BibitemShut {NoStop}%
\bibitem [{\citenamefont {Shapiro}(2002)}]{Shapiro:2001rz}%
  \BibitemOpen
  \bibfield  {author} {\bibinfo {author} {\bibfnamefont {I.}~\bibnamefont {Shapiro}},\ }\href {https://doi.org/10.1016/S0370-1573(01)00030-8} {\bibfield  {journal} {\bibinfo  {journal} {Physics Reports}\ }\textbf {\bibinfo {volume} {357}},\ \bibinfo {pages} {113} (\bibinfo {year} {2002})}\BibitemShut {NoStop}%
\bibitem [{\citenamefont {Brandt}\ \emph {et~al.}(2024)\citenamefont {Brandt}, \citenamefont {Frenkel}, \citenamefont {{Martins-Filho}},\ and\ \citenamefont {McKeon}}]{Brandt:2024rsy}%
  \BibitemOpen
  \bibfield  {author} {\bibinfo {author} {\bibfnamefont {F.~T.}\ \bibnamefont {Brandt}}, \bibinfo {author} {\bibfnamefont {J.}~\bibnamefont {Frenkel}}, \bibinfo {author} {\bibfnamefont {S.}~\bibnamefont {{Martins-Filho}}},\ and\ \bibinfo {author} {\bibfnamefont {D.~G.~C.}\ \bibnamefont {McKeon}},\ }\href {https://doi.org/10.1016/j.aop.2024.169607} {\bibfield  {journal} {\bibinfo  {journal} {Annals of Physics}\ }\textbf {\bibinfo {volume} {462}},\ \bibinfo {pages} {169607} (\bibinfo {year} {2024})}\BibitemShut {NoStop}%
\bibitem [{\citenamefont {Batalin}\ and\ \citenamefont {Vilkovisky}(1981)}]{Batalin:1981jr}%
  \BibitemOpen
  \bibfield  {author} {\bibinfo {author} {\bibfnamefont {I.~A.}\ \bibnamefont {Batalin}}\ and\ \bibinfo {author} {\bibfnamefont {G.~A.}\ \bibnamefont {Vilkovisky}},\ }\href {https://doi.org/10.1016/0370-2693(81)90205-7} {\bibfield  {journal} {\bibinfo  {journal} {Phys. Lett.}\ }\textbf {\bibinfo {volume} {102B}},\ \bibinfo {pages} {27} (\bibinfo {year} {1981})}\BibitemShut {NoStop}%
\bibitem [{\citenamefont {Batalin}\ and\ \citenamefont {Vilkovisky}(1983)}]{Batalin:1983ggl}%
  \BibitemOpen
  \bibfield  {author} {\bibinfo {author} {\bibfnamefont {I.~A.}\ \bibnamefont {Batalin}}\ and\ \bibinfo {author} {\bibfnamefont {G.~A.}\ \bibnamefont {Vilkovisky}},\ }\href {https://doi.org/10.1103/PhysRevD.28.2567} {\bibfield  {journal} {\bibinfo  {journal} {Phys. Rev. D}\ }\textbf {\bibinfo {volume} {28}},\ \bibinfo {pages} {2567} (\bibinfo {year} {1983})}\BibitemShut {NoStop}%
\bibitem [{\citenamefont {{'t Hooft}}\ and\ \citenamefont {Veltman}(1974)}]{tHooft:1974toh}%
  \BibitemOpen
  \bibfield  {author} {\bibinfo {author} {\bibfnamefont {G.}~\bibnamefont {{'t Hooft}}}\ and\ \bibinfo {author} {\bibfnamefont {M.~J.~G.}\ \bibnamefont {Veltman}},\ }\href {http://www.numdam.org/item/?id=AIHPA_1974__20_1_69_0} {\bibfield  {journal} {\bibinfo  {journal} {Annales Poincare Phys. Theor.}\ }\textbf {\bibinfo {volume} {A20}},\ \bibinfo {pages} {69} (\bibinfo {year} {1974})}\BibitemShut {NoStop}%
\bibitem [{\citenamefont {Abbott}(1981)}]{Abbott:1980hw}%
  \BibitemOpen
  \bibfield  {author} {\bibinfo {author} {\bibfnamefont {L.~F.}\ \bibnamefont {Abbott}},\ }\href {https://doi.org/10.1016/0550-3213(81)90371-0} {\bibfield  {journal} {\bibinfo  {journal} {Nucl.Phys.}\ }\textbf {\bibinfo {volume} {B185}},\ \bibinfo {pages} {189} (\bibinfo {year} {1981})}\BibitemShut {NoStop}%
\bibitem [{\citenamefont {Abbott}(1982)}]{abbott:1982}%
  \BibitemOpen
  \bibfield  {author} {\bibinfo {author} {\bibfnamefont {L.~F.}\ \bibnamefont {Abbott}},\ }\href {https://www.actaphys.uj.edu.pl/R/13/1/33} {\bibfield  {journal} {\bibinfo  {journal} {Acta Phys. Polon.}\ }\textbf {\bibinfo {volume} {B13}},\ \bibinfo {pages} {33} (\bibinfo {year} {1982})}\BibitemShut {NoStop}%
\bibitem [{\citenamefont {Frenkel}\ and\ \citenamefont {Taylor}(2018)}]{Frenkel:2018xup}%
  \BibitemOpen
  \bibfield  {author} {\bibinfo {author} {\bibfnamefont {J.}~\bibnamefont {Frenkel}}\ and\ \bibinfo {author} {\bibfnamefont {J.~C.}\ \bibnamefont {Taylor}},\ }\href {https://doi.org/10.1016/j.aop.2017.12.014} {\bibfield  {journal} {\bibinfo  {journal} {Annals Phys.}\ }\textbf {\bibinfo {volume} {389}},\ \bibinfo {pages} {234} (\bibinfo {year} {2018})}\BibitemShut {NoStop}%
\bibitem [{\citenamefont {Barvinsky}\ \emph {et~al.}(2018)\citenamefont {Barvinsky}, \citenamefont {Blas}, \citenamefont {{Herrero-Valea}}, \citenamefont {Sibiryakov},\ and\ \citenamefont {Steinwachs}}]{Barvinsky:2017zlx}%
  \BibitemOpen
  \bibfield  {author} {\bibinfo {author} {\bibfnamefont {A.~O.}\ \bibnamefont {Barvinsky}}, \bibinfo {author} {\bibfnamefont {D.}~\bibnamefont {Blas}}, \bibinfo {author} {\bibfnamefont {M.}~\bibnamefont {{Herrero-Valea}}}, \bibinfo {author} {\bibfnamefont {S.~M.}\ \bibnamefont {Sibiryakov}},\ and\ \bibinfo {author} {\bibfnamefont {C.~F.}\ \bibnamefont {Steinwachs}},\ }\href {https://doi.org/10.1007/JHEP07(2018)035} {\bibfield  {journal} {\bibinfo  {journal} {JHEP}\ }\textbf {\bibinfo {volume} {07}},\ \bibinfo {pages} {035}}\BibitemShut {NoStop}%
\bibitem [{\citenamefont {Brandt}\ \emph {et~al.}(2022)\citenamefont {Brandt}, \citenamefont {Frenkel},\ and\ \citenamefont {McKeon}}]{Brandt:2022und}%
  \BibitemOpen
  \bibfield  {author} {\bibinfo {author} {\bibfnamefont {F.~T.}\ \bibnamefont {Brandt}}, \bibinfo {author} {\bibfnamefont {J.}~\bibnamefont {Frenkel}},\ and\ \bibinfo {author} {\bibfnamefont {D.~G.~C.}\ \bibnamefont {McKeon}},\ }\href {https://doi.org/10.1103/PhysRevD.106.065010} {\bibfield  {journal} {\bibinfo  {journal} {Phys. Rev. D}\ }\textbf {\bibinfo {volume} {106}},\ \bibinfo {pages} {065010} (\bibinfo {year} {2022})}\BibitemShut {NoStop}%
\bibitem [{\citenamefont {{Zinn-Justin}}(1984)}]{Zinn-Justin:1984tfs}%
  \BibitemOpen
  \bibfield  {author} {\bibinfo {author} {\bibfnamefont {J.}~\bibnamefont {{Zinn-Justin}}},\ }\href {https://doi.org/10.1016/0550-3213(84)90295-5} {\bibfield  {journal} {\bibinfo  {journal} {Nuclear Physics B}\ }\textbf {\bibinfo {volume} {246}},\ \bibinfo {pages} {246} (\bibinfo {year} {1984})}\BibitemShut {NoStop}%
\bibitem [{\citenamefont {Kiriushcheva}\ and\ \citenamefont {Kuzmin}(2009)}]{Kiriushcheva:2009tg}%
  \BibitemOpen
  \bibfield  {author} {\bibinfo {author} {\bibfnamefont {N.}~\bibnamefont {Kiriushcheva}}\ and\ \bibinfo {author} {\bibfnamefont {S.~V.}\ \bibnamefont {Kuzmin}},\ }\href {http://arxiv.org/abs/0907.1553} {\bibinfo {title} {The {{Hamiltonian}} formulation of {{N-bein}}, {{Einstein-Cartan}}, gravity in any dimension: The {{Progress Report}} ({{Extended}} version of a talk given on {{CAIMS-2009}}, {{June}} 11-14, {{London}}, {{Canada}})}} (\bibinfo {year} {2009}),\ \Eprint {https://arxiv.org/abs/0907.1553} {arXiv:0907.1553 [gr-qc, physics:hep-th]} \BibitemShut {NoStop}%
\bibitem [{\citenamefont {Tseytlin}(1982)}]{Tseytlin:1981ks}%
  \BibitemOpen
  \bibfield  {author} {\bibinfo {author} {\bibfnamefont {A.~A.}\ \bibnamefont {Tseytlin}},\ }\href {https://doi.org/10.1088/0305-4470/15/3/005} {\bibfield  {journal} {\bibinfo  {journal} {J. Phys. A: Math. Gen.}\ }\textbf {\bibinfo {volume} {15}},\ \bibinfo {pages} {L105} (\bibinfo {year} {1982})}\BibitemShut {NoStop}%
\bibitem [{\citenamefont {Nakanishi}(1966)}]{Nakanishi:1966zz}%
  \BibitemOpen
  \bibfield  {author} {\bibinfo {author} {\bibfnamefont {N.}~\bibnamefont {Nakanishi}},\ }\href {https://doi.org/10.1143/PTP.35.1111} {\bibfield  {journal} {\bibinfo  {journal} {Prog. Theor. Phys.}\ }\textbf {\bibinfo {volume} {35}},\ \bibinfo {pages} {1111} (\bibinfo {year} {1966})}\BibitemShut {NoStop}%
\bibitem [{\citenamefont {Lautrup}(1967)}]{lautrup:1967}%
  \BibitemOpen
  \bibfield  {author} {\bibinfo {author} {\bibfnamefont {B.}~\bibnamefont {Lautrup}},\ }\href@noop {} {\bibfield  {journal} {\bibinfo  {journal} {Kong. Dan. Vid. Sel. Mat. Fys. Med.}\ }\textbf {\bibinfo {volume} {35}} (\bibinfo {year} {1967})}\BibitemShut {NoStop}%
\bibitem [{\citenamefont {Harrison}(2013)}]{harrison2013einstein}%
  \BibitemOpen
  \bibfield  {author} {\bibinfo {author} {\bibfnamefont {P.}~\bibnamefont {Harrison}},\ }\emph {\bibinfo {title} {Einstein-Cartan Theory and Its Formulation as a Quantum Field Theory}},\ \href {https://www.escholar.manchester.ac.uk/uk-ac-man-scw:214321a} {Master's thesis},\ \bibinfo  {school} {The University of Manchester (United Kingdom)} (\bibinfo {year} {2013})\BibitemShut {NoStop}%
\bibitem [{\citenamefont {Das}\ and\ \citenamefont {Namazie}(1981)}]{Das:1980qe}%
  \BibitemOpen
  \bibfield  {author} {\bibinfo {author} {\bibfnamefont {A.}~\bibnamefont {Das}}\ and\ \bibinfo {author} {\bibfnamefont {M.}~\bibnamefont {Namazie}},\ }\href {https://doi.org/10.1016/0370-2693(81)91180-1} {\bibfield  {journal} {\bibinfo  {journal} {Physics Letters B}\ }\textbf {\bibinfo {volume} {99}},\ \bibinfo {pages} {463} (\bibinfo {year} {1981})}\BibitemShut {NoStop}%
\bibitem [{\citenamefont {Das}(1982)}]{Das:1982rz}%
  \BibitemOpen
  \bibfield  {author} {\bibinfo {author} {\bibfnamefont {A.}~\bibnamefont {Das}},\ }\href {https://doi.org/10.1103/PhysRevD.26.2774} {\bibfield  {journal} {\bibinfo  {journal} {Phys. Rev. D}\ }\textbf {\bibinfo {volume} {26}},\ \bibinfo {pages} {2774} (\bibinfo {year} {1982})}\BibitemShut {NoStop}%
\bibitem [{\citenamefont {Weinberg}(1995)}]{weinberg:1995a}%
  \BibitemOpen
  \bibfield  {author} {\bibinfo {author} {\bibfnamefont {S.}~\bibnamefont {Weinberg}},\ }\href@noop {} {\emph {\bibinfo {title} {Quantum {{Theory}} of {{Fields II}}}}}\ (\bibinfo  {publisher} {Benjamin Cummings},\ \bibinfo {address} {Cambridge},\ \bibinfo {year} {1995})\BibitemShut {NoStop}%
\bibitem [{\citenamefont {Lavrov}\ and\ \citenamefont {Shapiro}(2019)}]{Lavrov:2019nuz}%
  \BibitemOpen
  \bibfield  {author} {\bibinfo {author} {\bibfnamefont {P.~M.}\ \bibnamefont {Lavrov}}\ and\ \bibinfo {author} {\bibfnamefont {I.~L.}\ \bibnamefont {Shapiro}},\ }\href {https://doi.org/10.1103/PhysRevD.100.026018} {\bibfield  {journal} {\bibinfo  {journal} {Phys. Rev. D}\ }\textbf {\bibinfo {volume} {100}},\ \bibinfo {pages} {026018} (\bibinfo {year} {2019})}\BibitemShut {NoStop}%
\bibitem [{\citenamefont {Lee}\ and\ \citenamefont {Ne'eman}(1990)}]{Lee:1990xq}%
  \BibitemOpen
  \bibfield  {author} {\bibinfo {author} {\bibfnamefont {C.-Y.}\ \bibnamefont {Lee}}\ and\ \bibinfo {author} {\bibfnamefont {Y.}~\bibnamefont {Ne'eman}},\ }\href {https://doi.org/10.1016/0370-2693(90)91594-2} {\bibfield  {journal} {\bibinfo  {journal} {Physics Letters B}\ }\textbf {\bibinfo {volume} {242}},\ \bibinfo {pages} {59} (\bibinfo {year} {1990})}\BibitemShut {NoStop}%
\bibitem [{\citenamefont {Brandt}\ \emph {et~al.}(2021)\citenamefont {Brandt}, \citenamefont {Frenkel}, \citenamefont {{Martins-Filho}},\ and\ \citenamefont {McKeon}}]{Brandt:2020gms}%
  \BibitemOpen
  \bibfield  {author} {\bibinfo {author} {\bibfnamefont {F.~T.}\ \bibnamefont {Brandt}}, \bibinfo {author} {\bibfnamefont {J.}~\bibnamefont {Frenkel}}, \bibinfo {author} {\bibfnamefont {S.}~\bibnamefont {{Martins-Filho}}},\ and\ \bibinfo {author} {\bibfnamefont {D.~G.~C.}\ \bibnamefont {McKeon}},\ }\href {https://doi.org/10.1016/j.aop.2021.168426} {\bibfield  {journal} {\bibinfo  {journal} {Annals of Physics}\ }\textbf {\bibinfo {volume} {427}},\ \bibinfo {pages} {168426} (\bibinfo {year} {2021})}\BibitemShut {NoStop}%
\bibitem [{\citenamefont {McKeon}\ \emph {et~al.}(2021)\citenamefont {McKeon}, \citenamefont {Brandt}, \citenamefont {Frenkel},\ and\ \citenamefont {{Martins-Filho}}}]{Brandt:2021qgh}%
  \BibitemOpen
  \bibfield  {author} {\bibinfo {author} {\bibfnamefont {D.~G.~C.}\ \bibnamefont {McKeon}}, \bibinfo {author} {\bibfnamefont {F.~T.}\ \bibnamefont {Brandt}}, \bibinfo {author} {\bibfnamefont {J.}~\bibnamefont {Frenkel}},\ and\ \bibinfo {author} {\bibfnamefont {S.}~\bibnamefont {{Martins-Filho}}},\ }\href {https://doi.org/10.1016/j.aop.2021.168659} {\bibfield  {journal} {\bibinfo  {journal} {Annals of Physics}\ }\textbf {\bibinfo {volume} {434}},\ \bibinfo {pages} {168659} (\bibinfo {year} {2021})}\BibitemShut {NoStop}%
\bibitem [{\citenamefont {Brandt}\ and\ \citenamefont {{Martins-Filho}}(2023)}]{Brandt:2022kjo}%
  \BibitemOpen
  \bibfield  {author} {\bibinfo {author} {\bibfnamefont {F.~T.}\ \bibnamefont {Brandt}}\ and\ \bibinfo {author} {\bibfnamefont {S.}~\bibnamefont {{Martins-Filho}}},\ }\href {https://doi.org/10.1016/j.aop.2023.169323} {\bibfield  {journal} {\bibinfo  {journal} {Annals of Physics}\ }\textbf {\bibinfo {volume} {453}},\ \bibinfo {pages} {169323} (\bibinfo {year} {2023})}\BibitemShut {NoStop}%
\bibitem [{\citenamefont {McKeon}\ \emph {et~al.}(2024)\citenamefont {McKeon}, \citenamefont {Brandt},\ and\ \citenamefont {{Martins-Filho}}}]{McKeon:2024psy}%
  \BibitemOpen
  \bibfield  {author} {\bibinfo {author} {\bibfnamefont {D.~G.~C.}\ \bibnamefont {McKeon}}, \bibinfo {author} {\bibfnamefont {F.~T.}\ \bibnamefont {Brandt}},\ and\ \bibinfo {author} {\bibfnamefont {S.}~\bibnamefont {{Martins-Filho}}},\ }\href {https://doi.org/10.1140/epjc/s10052-024-12764-z} {\bibfield  {journal} {\bibinfo  {journal} {The European Physical Journal C}\ }\textbf {\bibinfo {volume} {84}},\ \bibinfo {pages} {399} (\bibinfo {year} {2024})}\BibitemShut {NoStop}%
\bibitem [{\citenamefont {Mertig}\ \emph {et~al.}(1991)\citenamefont {Mertig}, \citenamefont {B{\"o}hm},\ and\ \citenamefont {Denner}}]{Mertig:1990an}%
  \BibitemOpen
  \bibfield  {author} {\bibinfo {author} {\bibfnamefont {R.}~\bibnamefont {Mertig}}, \bibinfo {author} {\bibfnamefont {M.}~\bibnamefont {B{\"o}hm}},\ and\ \bibinfo {author} {\bibfnamefont {A.}~\bibnamefont {Denner}},\ }\href {https://doi.org/10.1016/0010-4655(91)90130-D} {\bibfield  {journal} {\bibinfo  {journal} {Computer Physics Communications}\ }\textbf {\bibinfo {volume} {64}},\ \bibinfo {pages} {345} (\bibinfo {year} {1991})}\BibitemShut {NoStop}%
\bibitem [{\citenamefont {Shtabovenko}\ \emph {et~al.}(2020)\citenamefont {Shtabovenko}, \citenamefont {Mertig},\ and\ \citenamefont {Orellana}}]{Shtabovenko:2020gxv}%
  \BibitemOpen
  \bibfield  {author} {\bibinfo {author} {\bibfnamefont {V.}~\bibnamefont {Shtabovenko}}, \bibinfo {author} {\bibfnamefont {R.}~\bibnamefont {Mertig}},\ and\ \bibinfo {author} {\bibfnamefont {F.}~\bibnamefont {Orellana}},\ }\href {https://doi.org/10.1016/j.cpc.2020.107478} {\bibfield  {journal} {\bibinfo  {journal} {Computer Physics Communications}\ }\textbf {\bibinfo {volume} {256}},\ \bibinfo {pages} {107478} (\bibinfo {year} {2020})}\BibitemShut {NoStop}%
\bibitem [{\citenamefont {Passarino}\ and\ \citenamefont {Veltman}(1979)}]{Passarino:1978jh}%
  \BibitemOpen
  \bibfield  {author} {\bibinfo {author} {\bibfnamefont {G.}~\bibnamefont {Passarino}}\ and\ \bibinfo {author} {\bibfnamefont {M.}~\bibnamefont {Veltman}},\ }\href {https://doi.org/10.1016/0550-3213(79)90234-7} {\bibfield  {journal} {\bibinfo  {journal} {Nuclear Physics B}\ }\textbf {\bibinfo {volume} {160}},\ \bibinfo {pages} {151} (\bibinfo {year} {1979})}\BibitemShut {NoStop}%
\bibitem [{\citenamefont {Brandt}\ \emph {et~al.}(2020)\citenamefont {Brandt}, \citenamefont {Frenkel}, \citenamefont {{Martins-Filho}},\ and\ \citenamefont {McKeon}}]{Brandt:2020vre}%
  \BibitemOpen
  \bibfield  {author} {\bibinfo {author} {\bibfnamefont {F.~T.}\ \bibnamefont {Brandt}}, \bibinfo {author} {\bibfnamefont {J.}~\bibnamefont {Frenkel}}, \bibinfo {author} {\bibfnamefont {S.}~\bibnamefont {{Martins-Filho}}},\ and\ \bibinfo {author} {\bibfnamefont {D.~G.~C.}\ \bibnamefont {McKeon}},\ }\href {https://doi.org/10.1103/PhysRevD.102.045013} {\bibfield  {journal} {\bibinfo  {journal} {Physics Review D}\ }\textbf {\bibinfo {volume} {102}},\ \bibinfo {pages} {045013} (\bibinfo {year} {2020})}\BibitemShut {NoStop}%
\bibitem [{\citenamefont {Leibbrandt}(1975)}]{Leibbrandt:1975dj}%
  \BibitemOpen
  \bibfield  {author} {\bibinfo {author} {\bibfnamefont {G.}~\bibnamefont {Leibbrandt}},\ }\href {https://doi.org/10.1103/RevModPhys.47.849} {\bibfield  {journal} {\bibinfo  {journal} {Rev. Mod. Phys.}\ }\textbf {\bibinfo {volume} {47}},\ \bibinfo {pages} {849} (\bibinfo {year} {1975})}\BibitemShut {NoStop}%
\end{thebibliography}%
\end{document}